\documentclass{aa}  
\usepackage{subfig,graphicx}
\usepackage{txfonts}
\graphicspath{{./figures/}}
\usepackage{natbib,twoopt} 
\usepackage[breaklinks=true]{hyperref} %% to avoid \citeads line fills 
\bibpunct{(}{)}{;}{a}{}{,} %% natbib format like A&A and ApJ
\makeatletter
 \newcommandtwoopt{\citeads}[3][][]{\href{http://adsabs.harvard.edu/abs/#3}%
   {\def\hyper@linkstart##1##2{}%
    \let\hyper@linkend\@empty\citealp[#1][#2]{#3}}}    %% Rutten, 2000
 \newcommandtwoopt{\citepads}[3][][]{\href{http://adsabs.harvard.edu/abs/#3}%
   {\def\hyper@linkstart##1##2{}%
    \let\hyper@linkend\@empty\citep[#1][#2]{#3}}}      %% (Rutten 2000)
 \newcommandtwoopt{\citetads}[3][][]{\href{http://adsabs.harvard.edu/abs/#3}%
   {\def\hyper@linkstart##1##2{}%
    \let\hyper@linkend\@empty\citet[#1][#2]{#3}}}      %% Rutten (2000)
 \newcommandtwoopt{\citeyearads}[3][][]%
   {\href{http://adsabs.harvard.edu/abs/#3}%
   {\def\hyper@linkstart##1##2{}%
    \let\hyper@linkend\@empty\citeyear[#1][#2]{#3}}}   %% 2000
\makeatother
\usepackage{color}

\begin{document}

%
%   \title{Quasi-periodic oscillations in accreting magnetic white dwarfs: \\I. Observational constraints in X-rays and optical.}
 \title{Quasi-periodic oscillations in accreting magnetic white dwarfs}
 \subtitle{I. Observational constraints in X-ray and optical}
 
   \titlerunning{Search for X-ray QPOs in Polars}

%   \author{The authors\inst{1,2}
%   \and Collaborators\inst{1,2}
\author {J.M. Bonnet-Bidaud\inst{1} 
\and M. Mouchet\inst{2} 
\and  C. Busschaert\inst{2,3}
\and E. Falize\inst{3}
\and C. Michaut\inst{2}
   }
       
   \offprints{J.M. Bonnet-Bidaud}

   \institute{
 Service d'Astrophysique-Laboratoire AIM, CEA/DSM/Irfu,  91191 Gif-sur-Yvette, France.  (\email{bonnetbidaud@cea.fr)}       
%\and Laboratoire AIM, CEA-CNRS-Universit\'e Denis Diderot, 75013 Paris, France
%\and Laboratoire APC, UMR 7164, Universit\'e Paris-Diderot, 75013 Paris, France
\and LUTH-Observatoire de Paris, UMR 8102-CNRS, Universit\'e Paris-Diderot, 92190 Meudon, France
\and CEA-DAM-DIF, F-91297 Arpajon, France
} 
 
   \date{Received: 8 December 2014; Accepted: 28 May 2015}

% \abstract{}{}{}{}{} 
% 5 {} token are mandatory
 
  \abstract{
  % Normal abstract with abstrac{ }
   Quasi-periodic oscillations (QPOs) are observed in the optical flux of some polars with typical periods  of 1 to 3 s but none have been observed yet in X-rays where a significant part of the accreting energy is released. QPOs are expected and predicted from shock oscillations. 
   Most of the polars have been observed by the XMM-Newton satellite. We made use of the homogeneous set of observations of the polars by XMM-Newton to search for the presence of QPOs in the (0.5-10 keV) energy range and to set significant upper limits for the brightest X-ray polars.  
   We extracted high time-resolution X-ray light curves by taking advantage of the 0.07 sec resolution of the EPIC-PN camera. Among the 65 polars observed with XMM-Newton from 1998 to 2012, a sample of 24 sources was selected on the basis of their counting rate in the PN instrument to secure significant limits. We searched for QPOs using Fast Fourier Transform (FFT) methods and defined limits of detection using statistical tools.
 Among the sample surveyed, none shows QPOs at a significant level. Upper limits to the fractional flux in QPOs range from 7\% to 71\%. 
 These negative results are compared to the detailed theoretical predictions of numerical simulations based on a 2D hydrodynamical code presented in Paper II. Cooling instabilities in the accretion column are expected to produce shock quasi-oscillations with a maximum amplitude reaching $\sim 40\%$ in the bremsstrahlung (0.5-10 keV) X-ray emission and $\sim 20\%$ in the optical cyclotron emission. The absence of X-ray QPOs imposes an upper limit of $\sim (5-10)$\,g.cm$^{-2}$s$^{-1}$ on the specific accretion rate but this condition is found inconsistent with the value required to account for the amplitudes and frequencies of the observed optical QPOs. This contradiction outlines probable shortcomings with the shock instability model.
   }


   \maketitle
%
%________________________________________________________________

\section{Introduction}
Accreting magnetic white dwarfs were first discovered by their X-ray emission, starting with the identification of the prototype AM Herculis in 1977
(\citeads{1977ApJ...212L.125T}, %Tapia77
\citeads{1977ApJ...212L.121C}). %Cowley77
They are accreting binary systems in which material is transferred from a dwarf secondary star onto a magnetic  white dwarf through Roche lobe overflow. They are now subdivided in two major classes,  the polars (or AM Her systems), in which the magnetic field is strong enough (B $\sim 10-200$ MG) to synchronize the white dwarf rotation with the orbit, and intermediate polars (or DQ Her systems), where a suspected lower magnetic field allows the spin period of the white dwarf to be shorter than the orbital period (see 
\citeads{1995Ap&SS.232...89W}). %Warner95
Owing to the synchronization, polars are the "cleaner" systems where a stable accretion geometry allows the accreting flow to be captured from the secondary and to follow the magnetic field lines to the surface of the white dwarf via a stable accretion column. The release of the gravitational energy is made through a stand-off shock above the white dwarf with a hot post-shock region being the major source of emission in the system over a wide range of energy.  Output radiation includes X-ray bremsstrahlung emission from the hot (10-50 keV) post-shock gas, infrared-optical emission from the cyclotron emission of the electrons of this region, and soft X-rays and UV emission from the heated white dwarf surface
(\citeads{1973PThPh..49..776H}, %Hoshi
  \citeads{1979ApJ...234L.117L}, %Lamb-Master 1979
  \citeads{2006A&A...449.1129K}, %Konig2006
 see 
  \citeads{2000SSRv...93..611W} %Wu 00
 for a review).\\
The discovery of quasi-periodic oscillations (QPOs) among polars has led to questioning the stability of the post-shock region. Optical QPOs with (1-5\%) amplitude in the range (1.25-2.5s) were first detected in the systems V834 Cen and AN UMa
(\citeads{1982ApJ...257L..71M}) % Middelditch 82
and later also found in EF Eri
(\citeads{1987A&A...181L..15L}), % Larson 87
VV Pup
(\citeads{1989A&A...217..146L}) % Larson 89
and BL Hyi
(\citeads{1997ApJ...489..912M}). % Middelditch 97
These five sources were at that time a reasonable fraction of the known polars, suggesting that QPOs might be a general characteristic  among polars.
More detailed study of the QPO colours and eclipses  further suggested that they may originate in a region close to  the base of the accretion region (see 
\citeads{1995ASPC...85..311L}).  %Larsson95.  
High time resolution spectroscopy also shows that, at least in one source (AN UMa), short-lived coherent QPOs may be present in optical emission lines
(\citeads{1996A&A...306..199B}). %Bonnet-Bidaud 96
Additional QPO-type variability on longer timescales (4-10 min) is also found among several polars. Large-amplitude ($\sim$ 10-30\%), nearly coherent optical oscillations at a period of $\sim$ 4.5 min were found in AM Herculis during an intermediate state
(\citeads{1991A&A...251L..27B}). %Bonnet-Bidaud 91
In the case of the polar IGRJ14536-5522, similar ($\sim$ 4 to 5 min) QPOs were found to be present in circular polarization, thereby demonstrating their association with the cyclotron emission region
(\citeads{2010MNRAS.402.1161P}). %Potter 2010
\\
Already before the discovery of the fast QPOs, the stability of the accretion column was investigated from time-dependent hydrodynamic equations by 
\citetads{1981ApJ...245L..23L} %Langer 1981
and
\citetads{1982ApJ...258..289L}. %Langer 1982
They concluded that a thermal instability is present and gives rise to shock height oscillations with a period characterized by the post-shock cooling timescale. \\
These instabilities have been studied theoretically in more detail by considering different processes and modelisations, such as cyclotron emission  
(\citeads{1985ApJ...299L..87C}), %Chanmugam 1985
parametric cooling function
(\citeads{1982ApJ...261..543C}, %Chevalier 1982
\citeads{2005ApJ...626..373M}), %Mignone 2005
gravitational potential
(\citeads{1999MNRAS.306..684C}), %Cropper 99
unequal ion-electron temperatures
% ---- introduce mbox to resolve pb saut de page dans citation causes Latex error
\mbox{(\citeads{1987ApJ...313..298I}, %Imamura 87
\citeads{1999MNRAS.310..677S})}, %Saxton 99
% -------
different boundary conditions
(\citeads{2002PASA...19..282S}, %Saxton 02
\citeads{2005ApJ...626..373M}) %Mignone 2005
and noise-driven excitation
(\citeads{1992ApJ...398..593W}, %Wood 92
\citeads{1992ApJ...397..232W}). %Wu 92

All results point to the existence of an oscillating shock when the cooling is predominantly due to bremsstrahlung but also to a strong damping process when only a small contribution of cyclotron is included (see 
  \citeads{2000SSRv...93..611W}, %Wu 00
for a review). \\
The recent spectacular development of high-energy-density laser facilities also provides a new tool for studying the evolution and stability of radiating plasmas (see 
  \citeads{2006RvMP...78..755R}). %Remington 06
Using a scaling law approach, it has been shown that the physical and dynamical conditions of an accretion column can be reproduced in laboratory laser experiments
(\citeads{2009Ap&SS.322...71F}, %Falize 09
  \citeads{2011Ap&SS.336...81F}). %Falize 11
In this case,  the structure of the accretion column and its dynamics can be diagnosed precisely under variable experimental conditions, thus allowing a potential detailed analysis of the quasi-periodic oscillation phenomena in laboratory. Conclusive preliminary results for the building of  an accretion shock similar to what is observed in accreting magnetic white dwarfs have already been obtained in the context of the POLAR project, using  the LULI 2000 laser facility
 (\citeads{2012HEDP....8....1F}, %Falize 12
 \citeads{2012MmSAI..83..665M}. %Michaut 12
  \citeads{2013NJPh...15c5020B}). %Busschaert 13
Future more promising prospects are expected from the more powerful NIF and LMJ lasers
 (\citeads{2009PhPl...16d1006M}, %Moses 09
 \citeads{2010JPhCS.244a2003L}).\\ %Lion 10
Though the number of polars has significantly increased recently, reaching now more than one hundred identified sources thanks to the X-ray surveys by the different RXTE, Swift, XMM-Newton, and INTEGRAL satellites
 (\citepads{2012MmSAI..83..578M}, %Mouchet 12
only five sources show optical QPOs, with none discovered since 1997 and no QPOs detected yet in X-rays. QPO searches in the optical range have been strongly hampered for some time by the disappearance of photomultiplier tubes in favour of CCD cameras with lower time resolution. This is now compensated for by the recent advent of fast cameras based on frame-transfer CCDs, such as ULTRACAM
 \citepads{2007MNRAS.378..825D} %Dhillon 07
 or SALTICAM
  \citepads{2006MNRAS.372..151O}. %O'Donoghue 06
Negative optical searches were reported in  
\citetads{1982ApJ...257L..71M}, % Middelditch 82
\citetads{1993MNRAS.260..209R}, % Ramseyer 93
\citetads{1998ApJ...501..830I}, % Imamura 98
\citetads{1999ApJ...515..404S}, %Steinan 99
\citetads{2001MNRAS.324..899P}%Perryman 01
. Also in X-rays, only a sparse coverage of all known sources is still available. Upper limits in the range (4-20\%) were published for AM Her, EF Eri, and V834 Cen
\citepads{1997MNRAS.286...77B}, %Beardmore 97
V834 Cen
\citepads{2000PASP..112...18I},
BL Hyi 
\citepads{1999ASPC..157..149W}, %Wolff99
and V2301 Oph
(\citeads{1999ApJ...515..404S}, %Steinan 99
\citeads{2007MNRAS.379.1209R}), % Ramsay 07
based on observations by the GINGA, RXTE, and XMM-Newton satellites with various statistical criteria.\\
At present, interpretation of the detected QPOs is not satisfactory, and some important questions remain open. Fast (1-3 s) QPOs were only found in optical for a few sources, and none were detected in X-rays. For an accretion column with a pure bremsstrahlung cooling,  strong X-ray QPOs are expected from shock oscillation models. From the same models, optical QPOs are most likely associated with the cyclotron emission of the post-shock region. But as the cyclotron cooling is also known to damp the instabilities very efficiently, this  leads to an apparent paradox. Alternative hypotheses on the QPO origin have been proposed in terms of Alfven waves across the magnetosphere
(\citepads{1981ApJ...245..183T} %Tuohy 81
or excited magneto-acoustic waves across the white dwarf thin surface layer
(\citepads{1995MNRAS.275L..11L}. %Lou 95
\\
Before any conclusion can be drawn on the QPO physical process, more complete information has to be collected. 
In this paper (Paper I), we present a systematic search for QPOs in the (0.5-10 keV) X-ray flux of a homogeneous set of polars observed by the XMM-Newton satellite. We aimed at detecting or putting upper limits on QPOs that can be significant for the different theoretical models. Since polars are known to present extended low states, we carefully selected the sources by their intensity level according to different criteria, leading to a selection of 24 sources covered by 39 different XMM observations. This significant sample allows a wide range of parameters relevant  to the structure of the accretion column to be scanned for the first time (mass, field strength, accretion rate, etc.) and to compare with predictions of the theoretical models.\\
In an accompanying paper (Busschaert et al. 2015, Paper II), we use numerical simulations based on a 2D hydrodynamical code HADES adapted for high-Mach number flows and  high contrasts in hydrodynamic parameters
(\citepads{2011Ap&SS.336..175M} %Michaut 11
to reproduce the expected luminosity and time characteristics of QPOs and to be compared these theoretical predictions to our observational results.\\

\section{Observations}
We searched the XMM-Newton Science Archive (XSA) database for available polar observations. By 2012, 65 sources could be retrieved, but specific selection criteria had to be applied for the search of fast oscillations. Time resolution has to be better than 0.1 s for searching (1-3 s) QPOs, so we were restricted  to the use of data obtained in imaging mode for the EPIC-PN camera (73 ms) or in timing mode for EPIC-MOS (1.75 ms) and EPIC-PN (0.03 ms). A few polars were observed in EPIC-MOS timing mode, but since they were either too faint in this mode or already covered in EPIC-PN mode with better statistics, our analysis is here limited to the EPIC-PN data. \\
Polars are rather faint sources that show extended low states. From a survey of (5-10 ks) snap-shot observations of 37 polars, 16 sources were found in low states, and 6 were not detected
(\citepads{2004MNRAS.350.1373R}. %Ramsay04
From different simulations, we determined that a typical EPIC-PN mean counting rate higher than 0.3 c/s  is needed to derive a significant limit on QPOs (see below). For eclipsing polars or polars showing strong orbital modulation, we apply this limit to the restricted high level part of the light curve. Finally, irrespective of this criterion, we also include in our source sample the five sources known to show optical QPOs. \\
Table~\ref{log_obs} shows the resulting selection, including 24  sources with  39 different XMM observations. Exposure times range from 5 to 70 ks with the two asynchronous polars (BY Cam and CD Ind) extensively covered by seven XMM observations over their expected beat cycle. A high percentage of these observations were referenced in different previous publications, including a detailed X-ray flickering analysis of 20 selected polars at a 10 s time resolution 
\citepads{2010A&A...519A..69A} %Anzolin10
but no systematic study at higher time resolution was conducted.\\
The X-ray data were processed using the XMM-Newton Science AnalysisSoftware (SAS) v11.0.0. We extract EPIC-PN light curve with 0.1 s resolution using an aperture of 40 arcsec centred on the source position. Only X-ray events that were graded as PATTERN = 0-4 and FLAG = 0 were used. Background data were extracted from an adjacent source-free region and were scaled and subtracted from the source data, including dead-time and vignetting corrections using the  \em epiclccor  \rm task.  Data segments with high background due to solar activity were suppressed when necessary. Because the lower energy ($<$ 0.3-0.4 keV) band may be affected by soft flares that are caused by stack overflows generated by high energy particles
(\citepads{2004SPIE.5165..123B}, %Burwitz04
we restricted the extraction to the range (0.5-10 keV). Photon arrival times were corrected to the barycentre of the solar system via the \em barycen \rm task. \\

%__________________________________________________ TABLE 1
\begin{table*}
\caption[ ]{X-ray fast oscillations (0.1-5) Hz detection limits from XMM-PN imaging observations}
     \label{log_obs}
\begin{flushleft}
\begin{tabular}{llllrrrrrrrr}
\hline
\hline
\multicolumn{1}{c}{Source} & \multicolumn{1}{l}{Date} & 
\multicolumn{1}{l}{HJDstart} & \multicolumn{1}{c}{Obs. mode} & 
 \multicolumn{1}{c}{Exp.} & \multicolumn{1}{c}{Rate} &
  \multicolumn{1}{r}{M-FFT}  & \multicolumn{1}{r}{$f_{max}$} & 
 \multicolumn{1}{r}{$P_{max}$}  & \multicolumn{1}{r}{$P_{exceed}$} & 
 \multicolumn{1}{r}{$P_{detect}$}  & \multicolumn{1}{r}{Limit} \\
\multicolumn{1}{c}{ } & \multicolumn{1}{c}{yy-mm-dd} & 
\multicolumn{1}{c}{2450000+} & \multicolumn{1}{c}{Cam-Filt} &
\multicolumn{1}{c}{ks} & \multicolumn{1}{c}{c/s} & 
\multicolumn{1}{r}{ }  & \multicolumn{1}{r}{Hz} & 
\multicolumn{1}{r}{}  & \multicolumn{1}{r}{$2.6\,\sigma$} & 
\multicolumn{1}{r}{$2.6\,\sigma$}  & \multicolumn{1}{r}{\% rms} \\ \\
\hline
\noalign{\smallskip}
\object{AI Tri} &       05-08-15        &       $3604.9775$     &       PN FF(1)   &       $19.8$  &       0.50    &       193     &       3.46    &       2.43    &       1.68    &       2.65    &       33.8    \\
\object{AM Her} &       05-07-19        &       $3571.2170$     &       PN-TI   &       $8.24$  &       12.46   &       80      &       0.54    &       2.66    &       1.52    &       3.05    &       6.8     \\
AM Her  &       05-07-25        &       $3577.2004$     &       PN-TI   &       $8.24$  &       8.83    &       80      &       2.19    &       2.87    &       1.52    &       3.05    &       9.5     \\
AM Her  &       05-07-27        &       $3579.1948$     &       PN-TI   &       $8.24$  &       13.53   &       80      &       4.81    &       2.66    &       1.52    &       3.05    &       9.9     \\
\object{AN UMa}(*)      &       02-05-01        &       $2395.9861$     &       PN-SW(1)        &       $6.39$  &       1.29    &       62      &       1.88    &       2.93    &       1.46    &       3.22    &       25.5    \\
\object{BL Hyi}(*)      &       02-12-16        &       $2625.4282$     &       PN-FF(1)        &       $27.66$ &       0.09    &       270     &       3.04    &       2.38    &       1.73    &       2.54    &       71.1    \\
\object{BY Cam  }&      03-08-30        &       $2881.8820$     &       PN-TI   &       $9.69$  &       8.96    &       95      &       0.42    &       2.67    &       1.55    &       2.96    &       8.0     \\
BY Cam  &       03-08-31        &       $2882.9515$     &       PN-TI   &       $9.69$  &       11.06   &       95      &       2.06    &       2.75    &       1.55    &       2.96    &       7.9     \\
BY Cam  &       03-09-01        &       $2883.8366$     &       PN-TI   &       $11.69$ &       10.15   &       114     &       4.74    &       2.61    &       1.59    &       2.86    &       10.7    \\
BY Cam  &       03-09-02        &       $2885.0546$     &       PN-TI   &       $12.61$ &       3.16    &       123     &       3.71    &       2.69    &       1.60    &       2.83    &       16.7    \\
BY Cam  &       03-09-04        &       $2886.8794$     &       PN-TI   &       $13.49$ &       6.94    &       132     &       3.50    &       2.75    &       1.62    &       2.80    &       11.2    \\
BY Cam  &       03-09-05        &       $2887.8534$     &       PN-TI   &       $14.39$ &       5.75    &       140     &       1.45    &       2.48    &       1.63    &       2.77    &       9.0     \\
BY Cam  &       03-10-13        &       $2926.3050$     &       PN-TI   &       $9.69$  &       6.49    &       95      &       1.38    &       2.67    &       1.55    &       2.96    &       9.6     \\
\object{CD Ind} &       02-03-27        &       $2361.2386$     &       PN-LW(1)        &       $13.24$ &       3.72    &       129     &       3.30    &       2.52    &       1.61    &       2.81    &       13.4    \\
CD Ind  &       02-03-28        &       $2361.9267$     &       PN-LW(2)        &       $ 8.63$   &       2.96    &       84      &       4.36    &       2.63    &       1.53    &       3.03    &       19.2    \\
CD Ind  &       02-03-29        &       $2363.2364$     &       PN-LW(1)        &       $13.24$ &       2.02    &       129     &       4.97    &       2.46    &       1.61    &       2.81    &       22.7    \\
CD Ind  &       02-03-30        &       $2363.8653$     &       PN-LW(1)        &       $13.24$ &       2.58    &       129     &       0.25    &       2.42    &       1.61    &       2.81    &       12.6    \\
CD Ind  &       02-03-31        &       $2365.2292$     &       PN-LW(1)        &       $14.00$ &       1.84    &       137     &       1.63    &       2.50    &       1.62    &       2.78    &       16.2    \\
CD Ind  &       02-04-01        &       $2365.8831$     &       PN-LW(1)        &       $14.00$ &       0.43    &       137     &       4.92    &       2.56    &       1.62    &       2.78    &       51.5    \\
CD Ind  &       02-04-02        &       $2367.2095$     &       PN-LW(1)        &       $14.23$ &       0.39    &       139     &       2.60    &       2.55    &       1.63    &       2.77    &       38.7    \\
\object{DP Leo} &       00-11-22        &       $1870.7333$     &       PN-FF(1)        &       $19.95$ &       0.09    &       195     &       1.72    &       2.38    &       1.68    &       2.64    &       65.7    \\
\object{EF Eri}(*)      &       11-01-15        &       $5576.8583$     &       PN-FF(1)        &       $69.87$ &       0.06    &       682     &       0.64    &       2.22    &       1.83    &       2.33    &       58.1    \\
\object{EP Dra} &       02-10-18        &       $2565.9196$     &       PN-LW(1)        &       $17.55$ &       0.47    &       171     &       0.93    &       2.47    &       1.66    &       2.69    &       29.9    \\
\object{EU Lyn} &       02-10-31        &       $2579.4893$     &       PN-FF(1)        &       57.49   &       0.54    &       56      &       1.97    &       3.00    &       1.44    &       3.28    &       40.9    \\
\object{EV UMa} &       01-12-08        &       $2252.0817$     &       PN-LW(1)        &       $4.98$  &       1.38    &       49      &       1.89    &       3.07    &       1.40    &       3.40    &       26.3    \\
\object{GG Leo} &       02-05-13        &       $2408.2666$     &       PN-SW(1)        &       $7.00$  &       1.23    &       68      &       4.88    &       2.83    &       1.48    &       3.16    &       36.2    \\
\object{HS Cam} &       03-10-13        &       $2925.9261$     &       PN-FF(1)        &       $14.61$ &       1.90    &       143     &       4.79    &       2.48    &       1.63    &       2.76    &       22.7    \\
\object{HU Aqr} &       02-05-16        &       $2411.2192$     &       PN-SW(1)        &       $36.72$ &       0.41    &       359     &       2.02    &       2.28    &       1.76    &       2.46    &       27.0    \\
HU Aqr  &       03-05-20        &       $2779.9944$     &       PN-TI   &       $18.75$ &       0.49    &       183     &       3.51    &       2.48    &       1.67    &       2.67    &       35.7    \\
\object{QS Tel} &       06-09-30        &       $4009.3308$     &       PN-LW(1)        &       $19.26$ &       0.16    &       188     &       2.38    &       2.53    &       1.68    &       2.66    &       57.3    \\
\object{RX J1007}       &       01-12-07        &       $2250.7973$     &       PN-LW(1)        &       $4.71 $       &       0.52    &       46      &       1.12    &       2.98    &       1.38    &       3.45    &       40.3    \\
\object{SDSS 2050}      &       04-10-18        &       $3296.9280$     &       PN-FF(1)        &       $11.04$ &       0.78    &       108     &       3.67    &       2.49    &       1.58    &       2.89    &       30.6    \\
\object{UZ For} &       02-08-08        &       $2494.7505$     &       PN-SW(1)        &       $28.98$ &       0.23    &       283     &       2.47    &       2.44    &       1.73    &       2.53    &       43.6    \\
\object{V1309 Ori}      &       01-03-17        &       $1986.5165$     &       PN-FF(1)        &       $26.49$ &       0.14    &       259     &       2.29    &       2.29    &       1.72    &       2.55    &       49.6    \\
\object{V2301 Oph}      &       04-09-06        &       $3254.6370$     &       PN-LW(1)        &       $16.65$ &       3.84    &       163     &       4.07    &       2.37    &       1.65    &       2.71    &       13.0    \\
\object{V347 Pav}       &       02-03-16        &       $2350.0731$     &       PN-SW(1)        &       $5.00$  &       1.11    &       49      &       4.57    &       2.80    &       1.40    &       3.40    &       36.7    \\
\object{V834 Cen}(*)    &       07-01-30        &       $4130.9941$     &       PN-TI   &       $43.62$ &       3.79    &       426     &       1.88    &       2.32    &       1.78    &       2.42    &       9.0     \\
\object{VV Pup}(*)      &       07-10-20        &       $4393.7886$     &       PN-TI   &       $48.27$ &       0.35    &       471     &       2.40    &       2.33    &       1.79    &       2.40    &       30.8    \\
\object{WW Hor} &       00-12-04        &       $1882.6816$     &       PN-FF(1)        &       $21.12$ &       0.18    &       206     &       3.19    &       2.38    &       1.69    &       2.62    &       52.5    \\
\noalign{\smallskip}
\hline
\end{tabular}
\end{flushleft}
(*) source showing optical QPOs. RX J1007 and SDSS 2050 refer to RX J1007.5-2017 and SDSSJ205017.84-053626.8, respectively.
\end{table*}
%______________________________________________________________
%

\section{Light curve analysis}
To judge from the overall level and variability of the sources, the orbital light curve for each observation was first built  using the ephemerides as given in Table~\ref{ephemeris}. 
The resulting mean (0.5-10 keV) light curves folded with the orbital period are shown in  Fig.~\ref{lightcurves_1} and, separately, in Fig.~\ref{lightcurves_BYCam}\, for BY Cam and Fig.~\ref{lightcurves_CDInd}\, for CD Ind. The orbital curve is repeated twice for clarity, and for easy comparison, the different light curves have been normalized by dividing by the (0.5-10 keV) mean counting rate as listed in Table~\ref{log_obs}. This overall picture of the bright polar light curves clearly outlines the different geometries of accretion in polars and shows the close similarity of several sources. The most represented light curve shape is a bright phase covering a significant part of the cycle and including a sharp and narrow eclipse (
DP Leo, EP Dra, EV UMa, HS Cam, HU Aqr, SDSS 2050, UZ For, V2301 Oph, WW Hor). The second class includes sources with a strong regular modulation (AM Her, BL Hyi, EF Eri, GG Leo, QS Tel, V347 Pav, VV Pup). The remaining sources show more complex and less regular variations  (AI Tri, AN UMa, EU Lyn, RX J1007, V1309 Ori, V834 Cen). The two asynchronous polars, BY Cam and CD Ind, exhibit spectacular and very similar shape changes in the different observations as expected from their beat cycle. In the case of BY Cam, the first six observations cover half the (14.6 d) beat period, the last one being at a phase close to the second observation. For CD Ind, the seven observations cover about one full (6.3 d) beat cycle. 

According to our level criteria, most sources are in what can be considered as  high states except for two sources with known optical QPOs  that are unfortunately in low states: BL Hyi and EF Eri. The low state for VV Pup in 2002 was not considered because the level was very low (< 0.03 c/s).
Different sources have already been analysed for their overall spectral characteristics using the same XMM observations confirming their level. We note, however, that several observations were not yet published, including BL Hyi, BY Cam, EF Eri (2011), HS Cam, UZ For, V834 Cen (2007), and VV Pup (2007).
This paper shows their XMM light curves for the first time and allows some of the ephemerides to
be checked.\\

%______________________________________________ FIGURE 1a
%
\onlfig{
\begin{figure*}
  \begin{center}
   \begin{tabular}{ccc}
     % \resizebox{55mm}{!}{\includegraphics*[trim=0 100 0 0]{AITri_05_lccor_ph200.pdf}} & 
      \resizebox{58mm}{!}{\includegraphics*[trim=20 80 50 0]{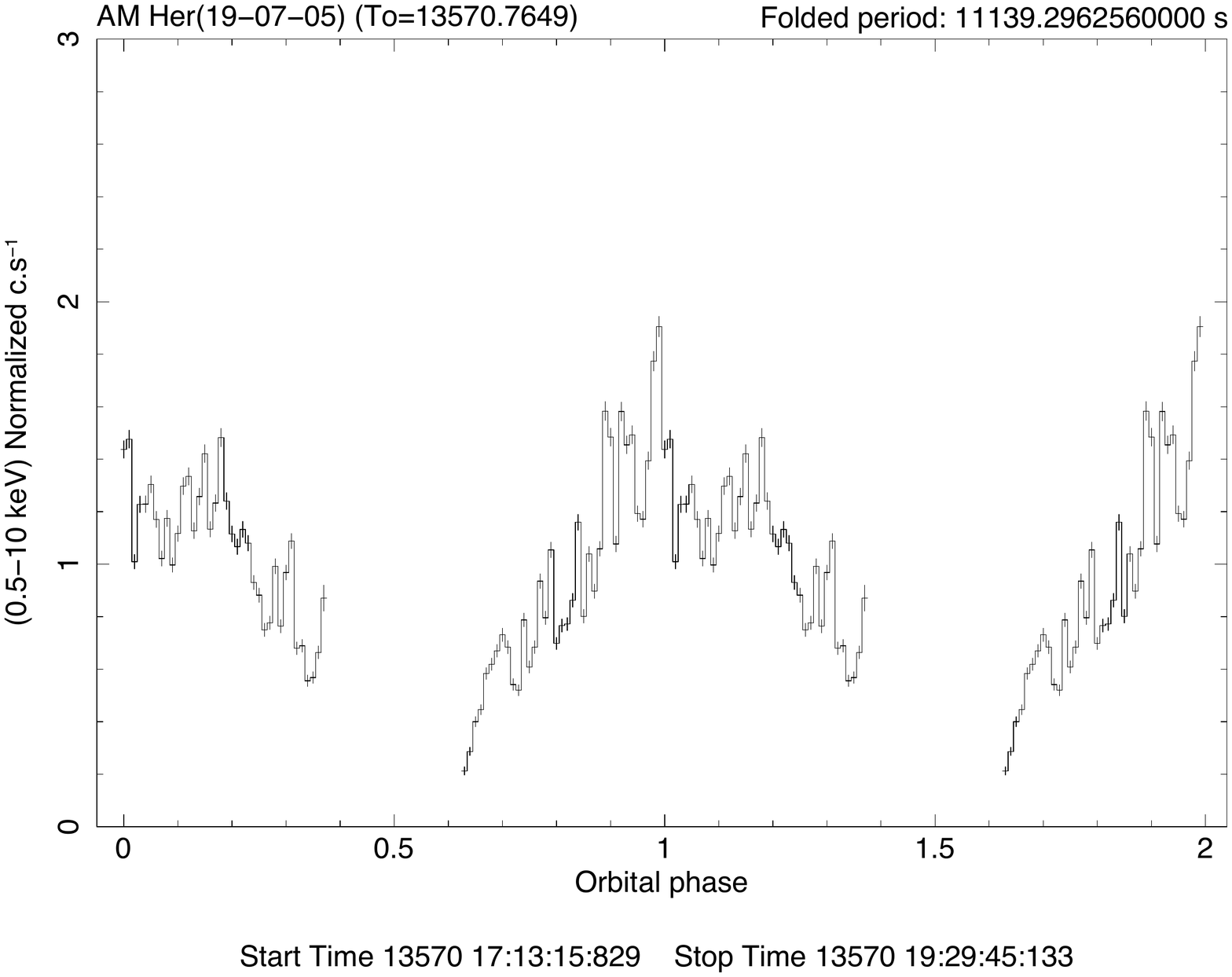}} &
      \resizebox{58mm}{!}{\includegraphics*[trim=20 80 50 0]{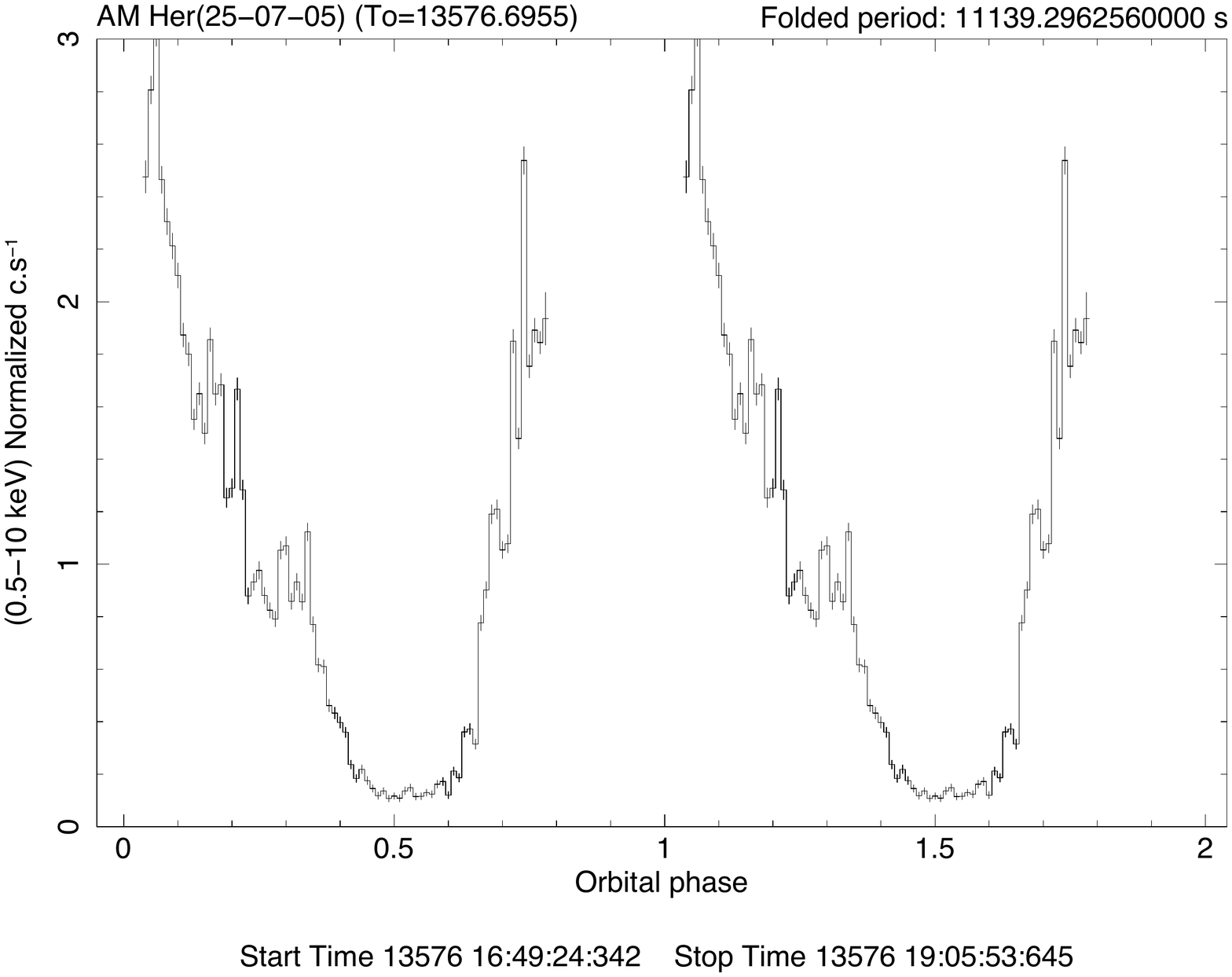}} &
      \resizebox{58mm}{!}{\includegraphics*[trim=20 80 50 0]{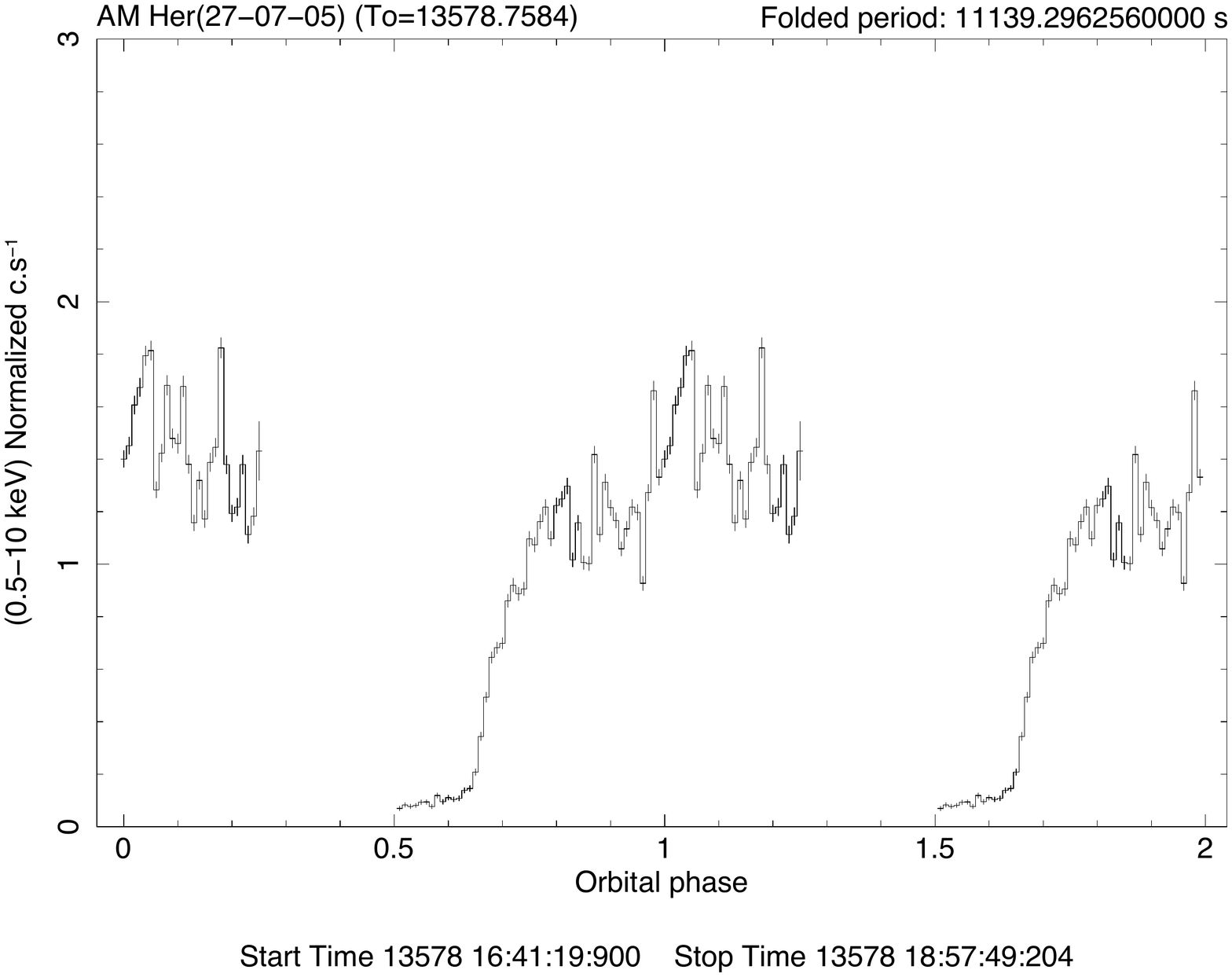}} \\
      \resizebox{58mm}{!}{\includegraphics*[trim=20 80 50 0]{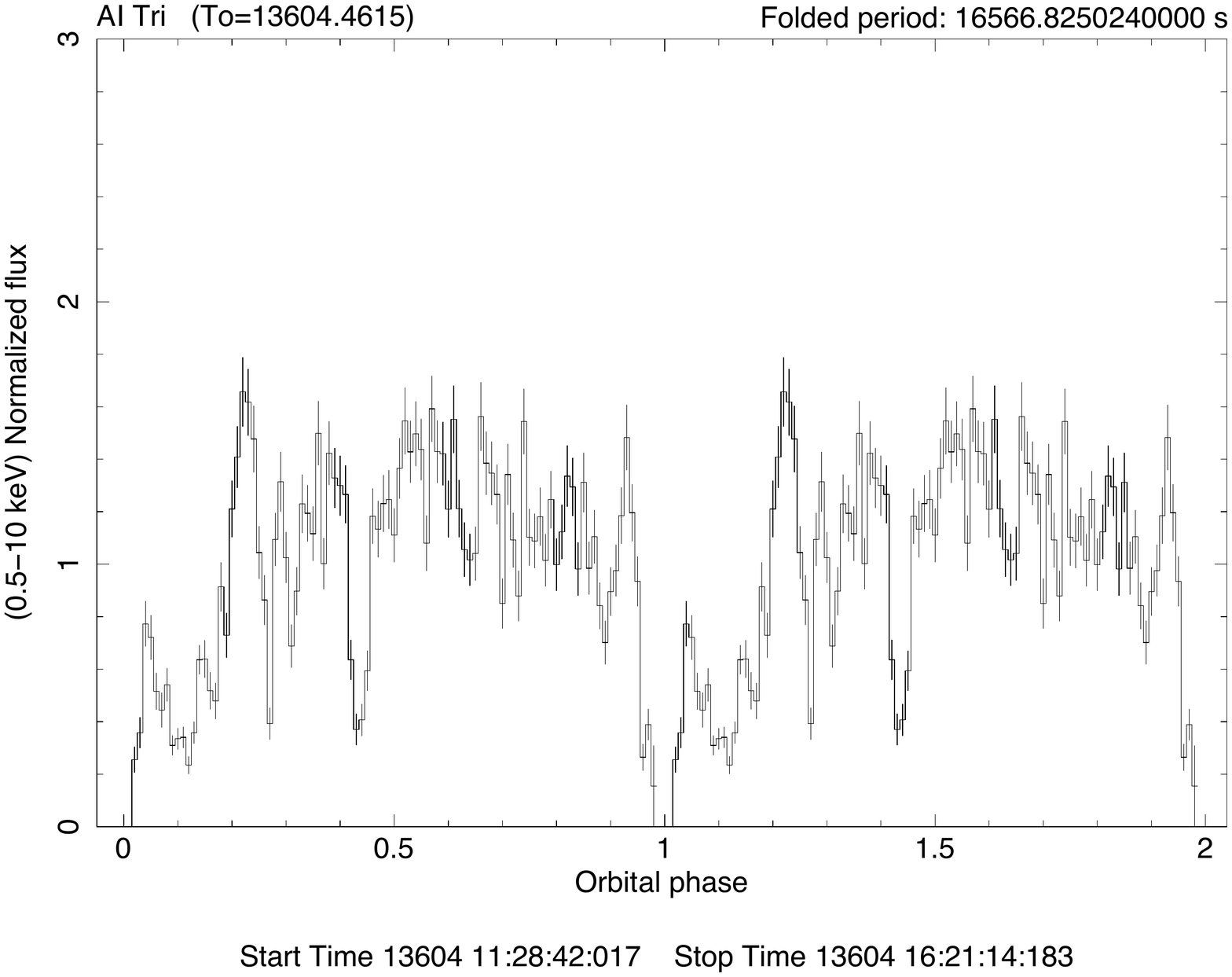}} &
      \resizebox{58mm}{!}{\includegraphics*[trim=20 80 50 0]{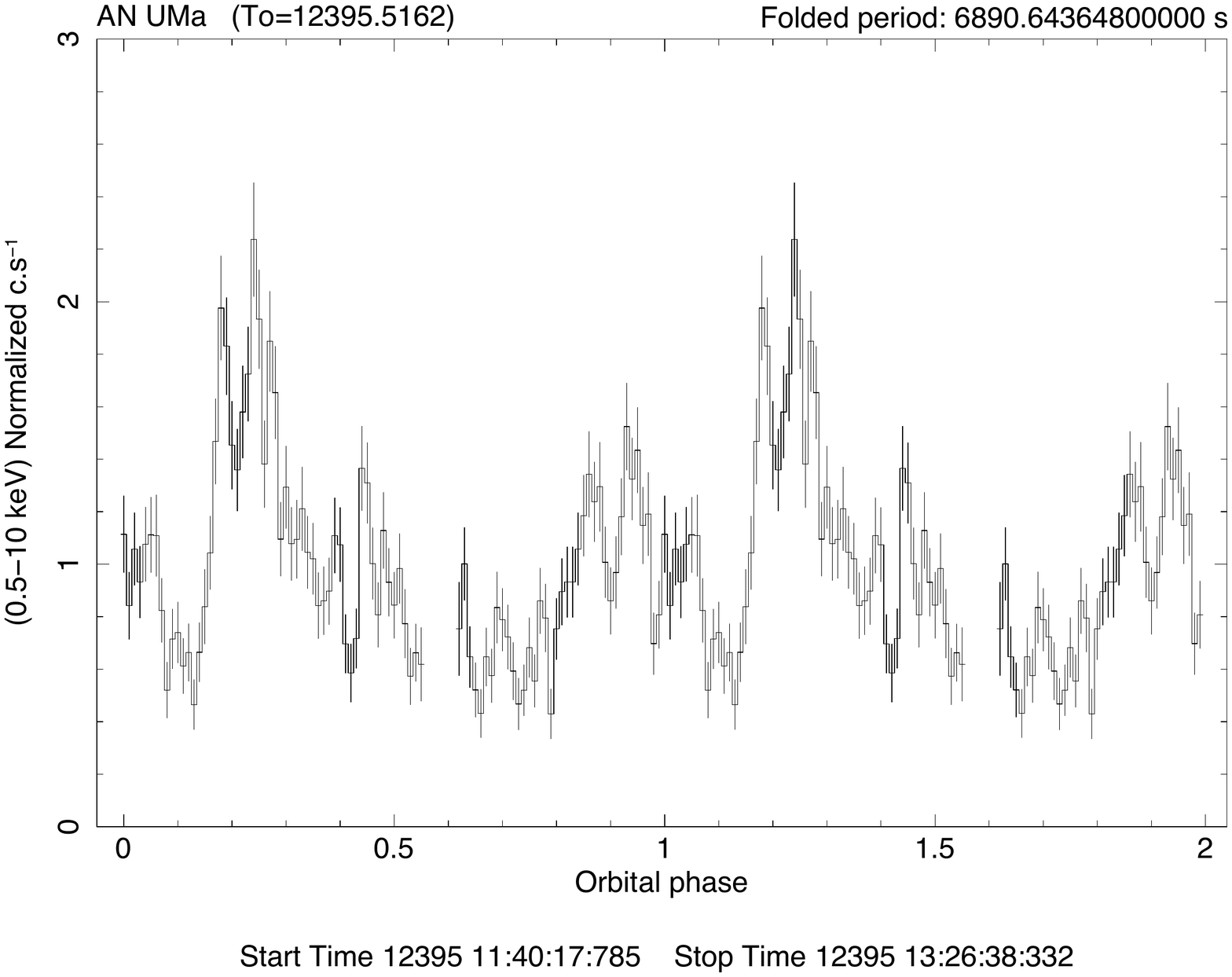}} &  
      \resizebox{58mm}{!}{\includegraphics*[trim=20 80 50 0]{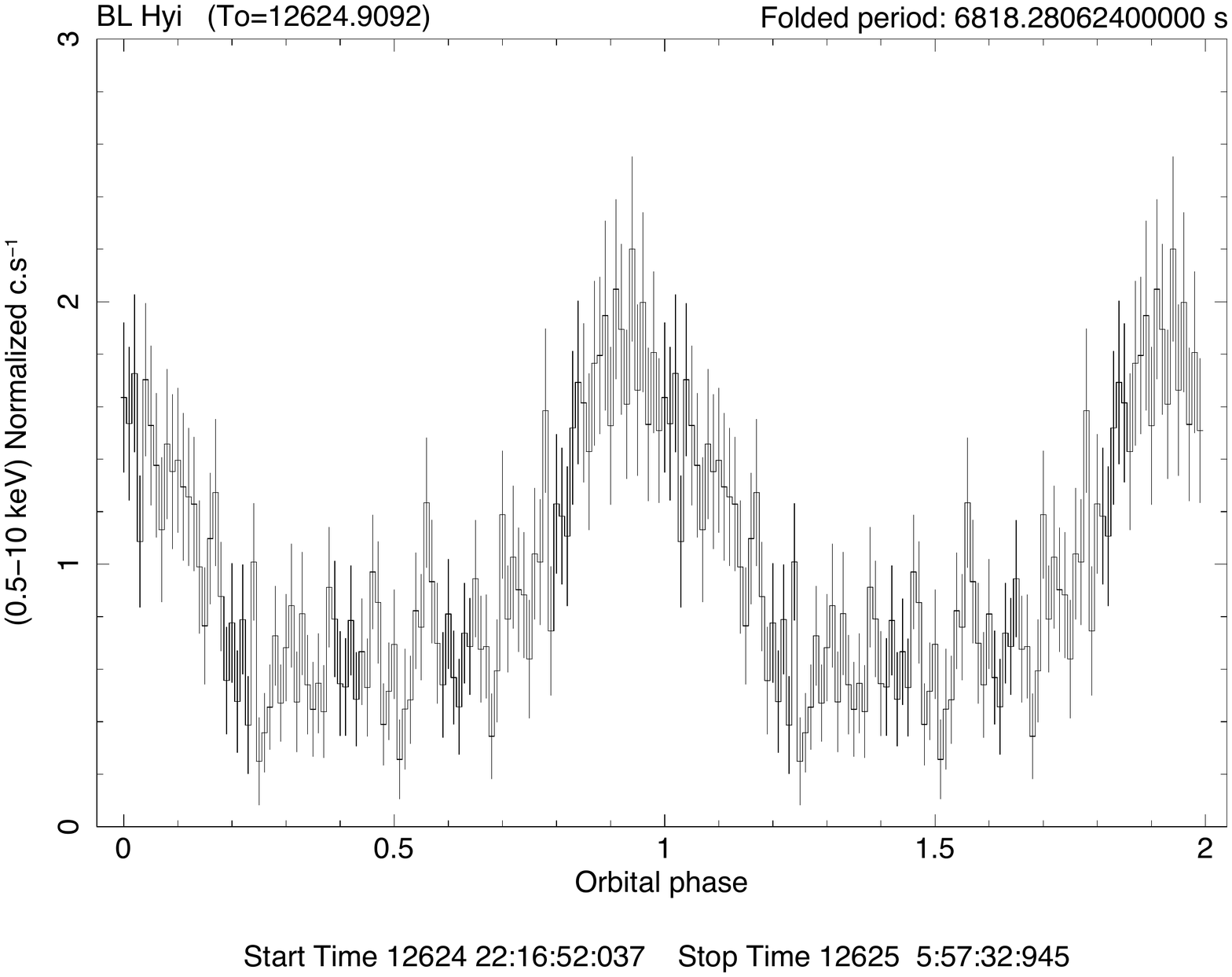}}  \\
       \resizebox{58mm}{!}{\includegraphics*[trim=20 80 50 0]{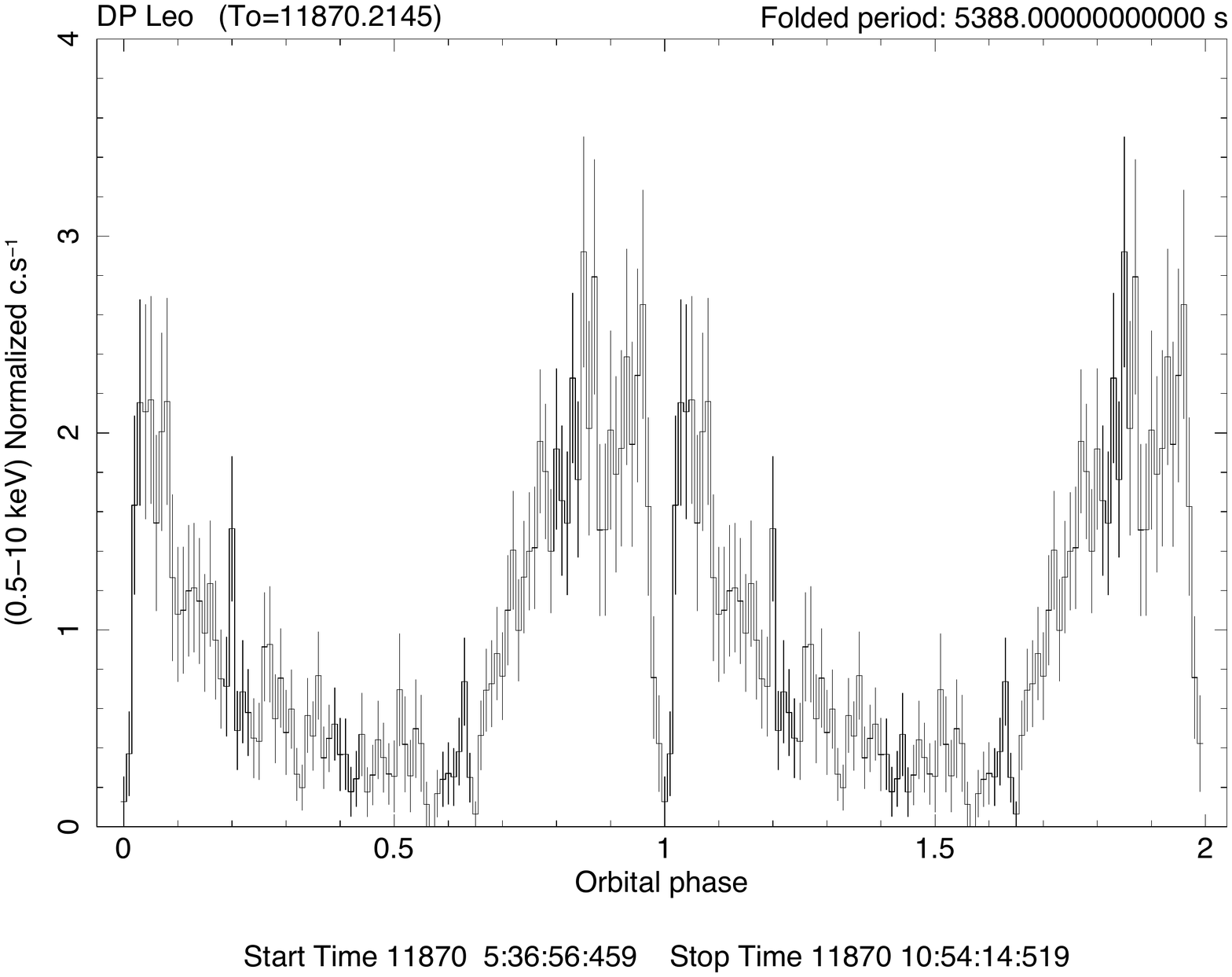}} &     
      \resizebox{58mm}{!}{\includegraphics*[trim=20 80 50 0]{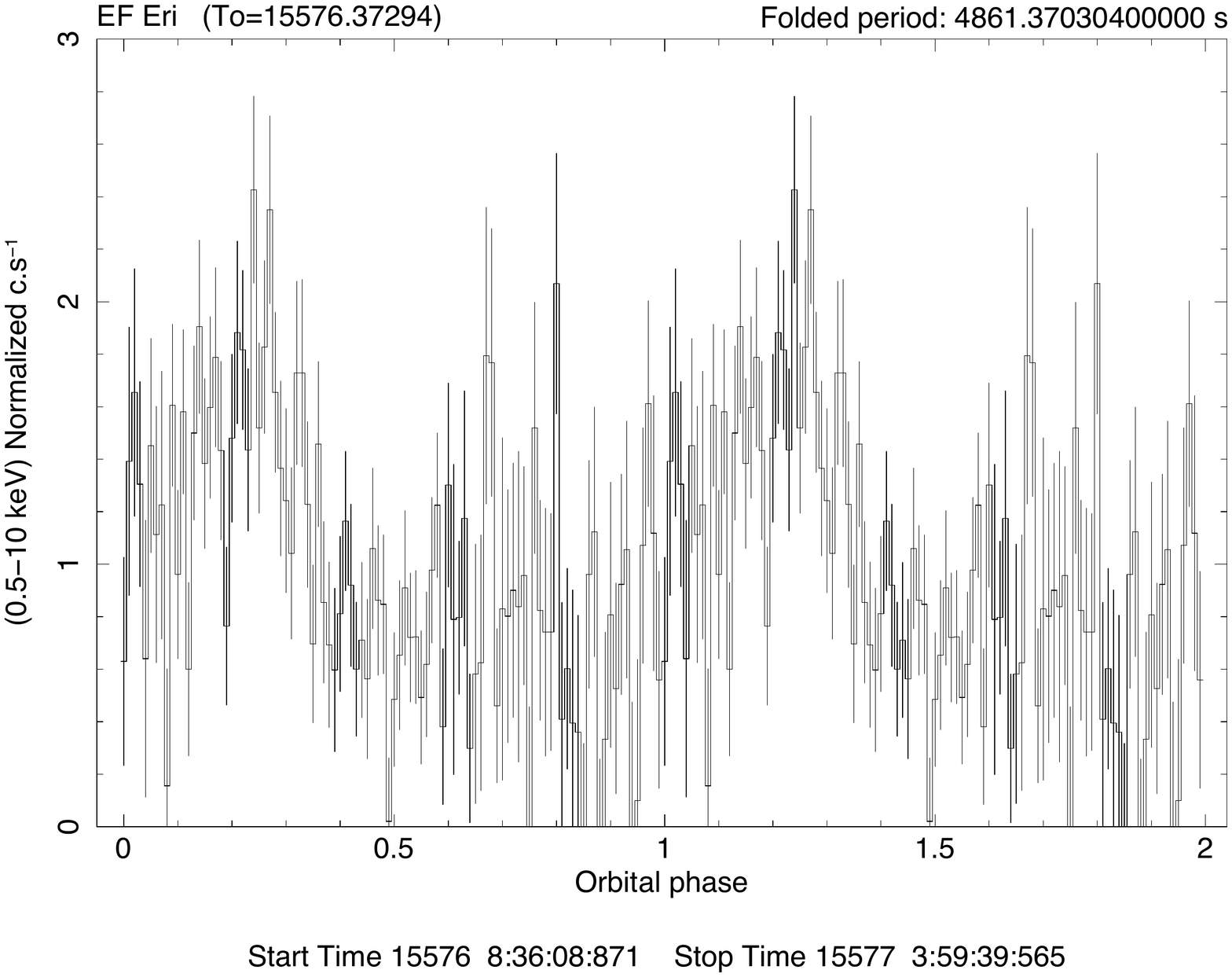}} & 
      \resizebox{58mm}{!}{\includegraphics*[trim=20 80 50 0]{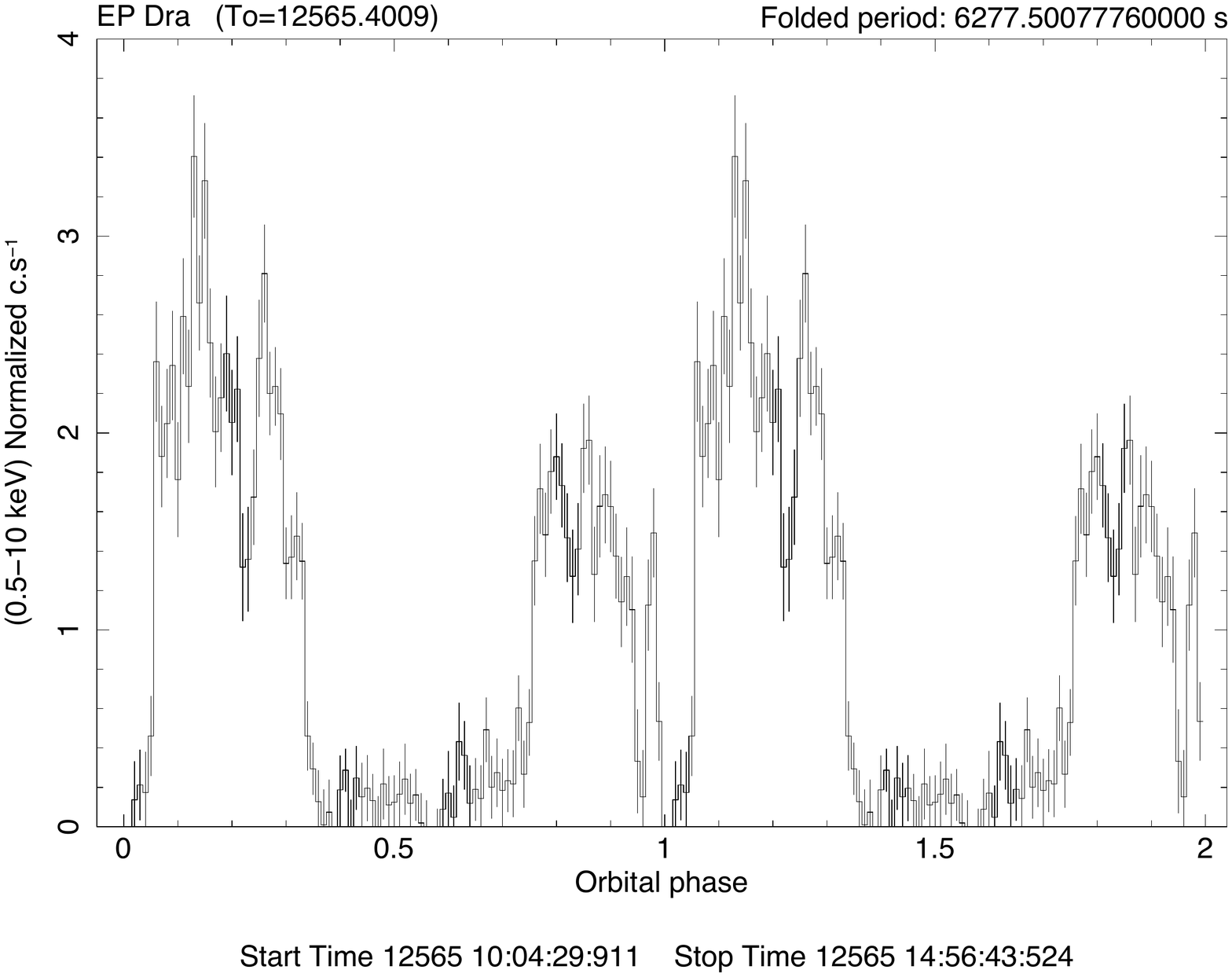}}      \\      
       \resizebox{58mm}{!}{\includegraphics*[trim=20 80 50 0]{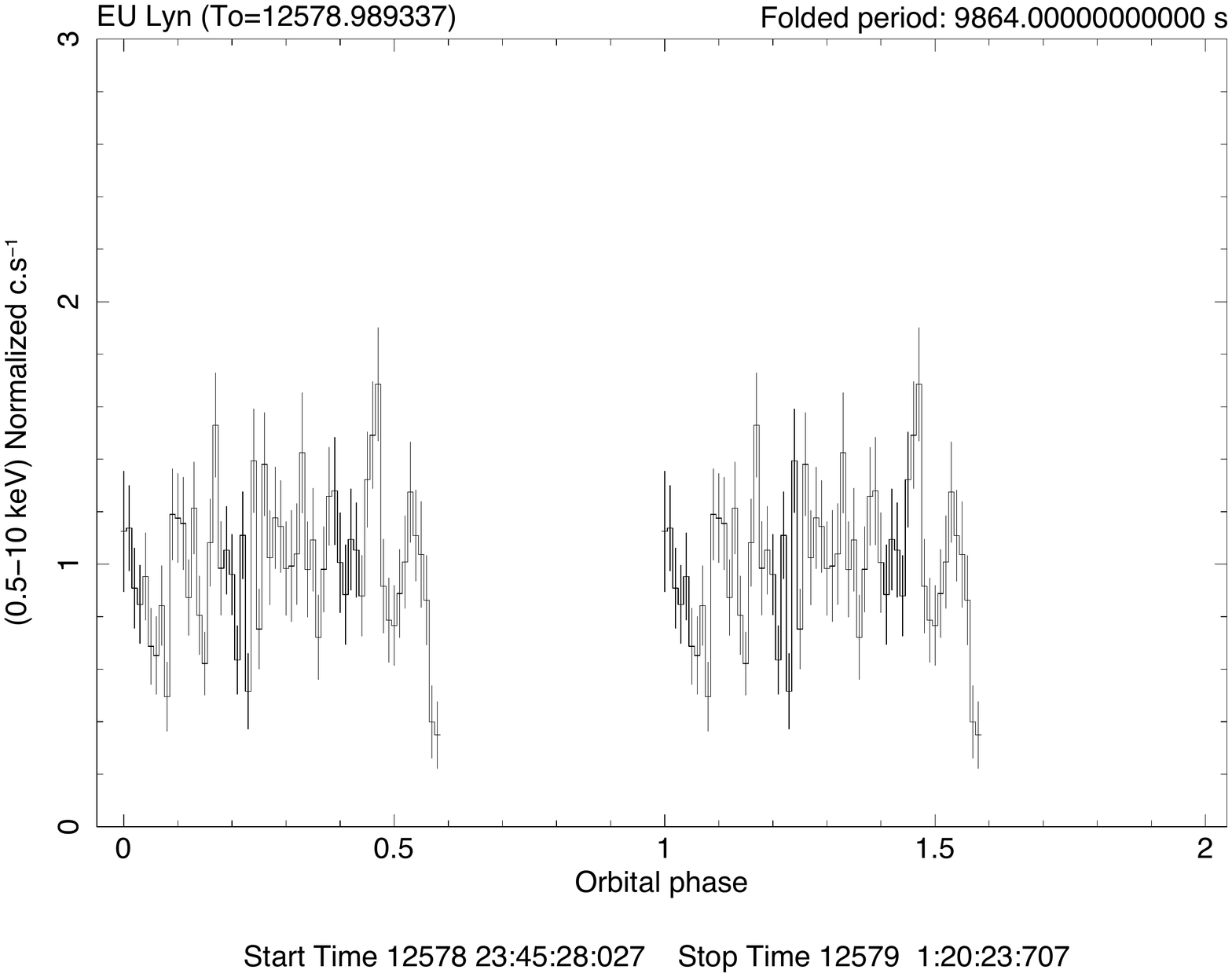}} & 
       \resizebox{58mm}{!}{\includegraphics*[trim=20 80 50 0]{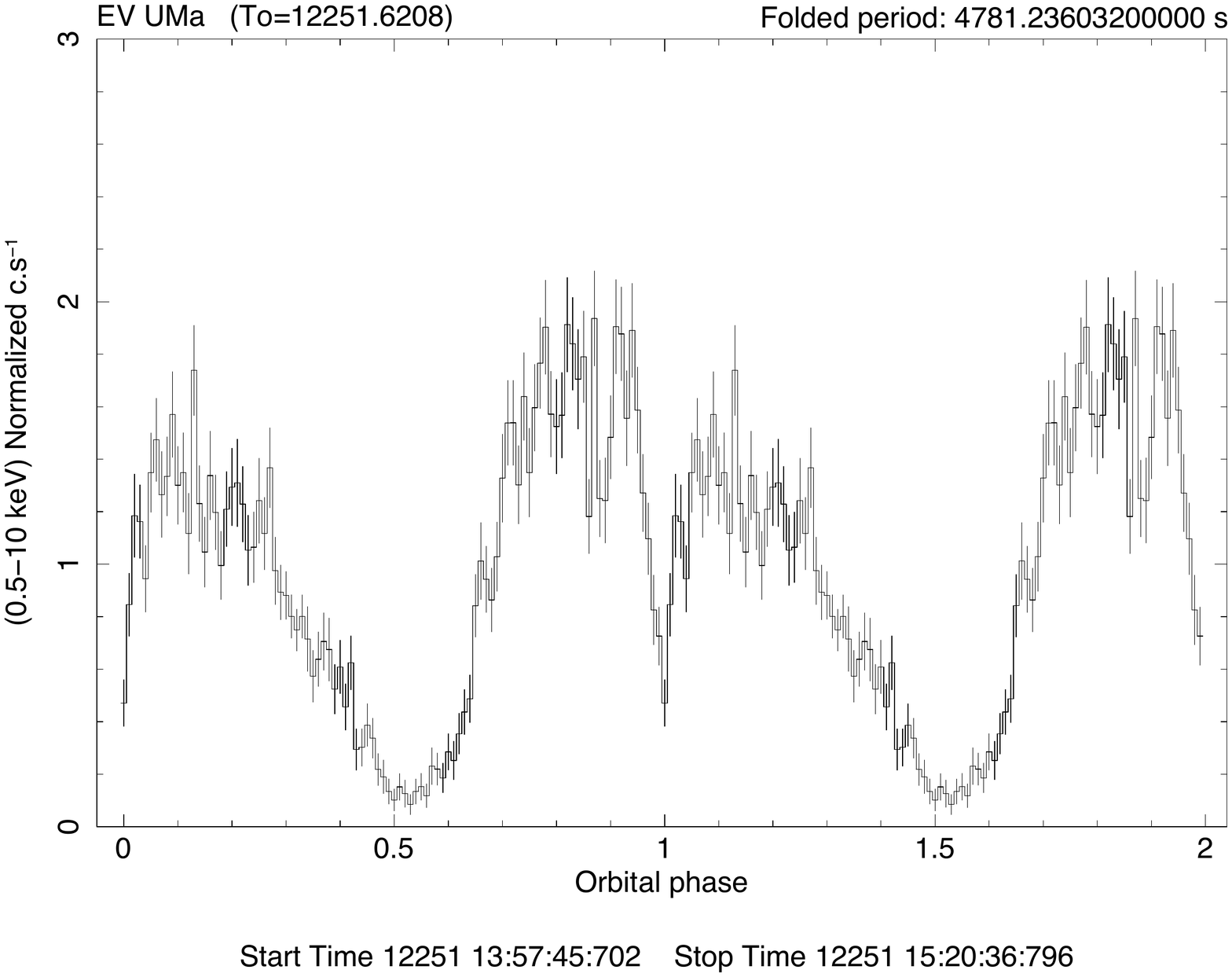}} &   
       \resizebox{58mm}{!}{\includegraphics*[trim=20 80 50 0]{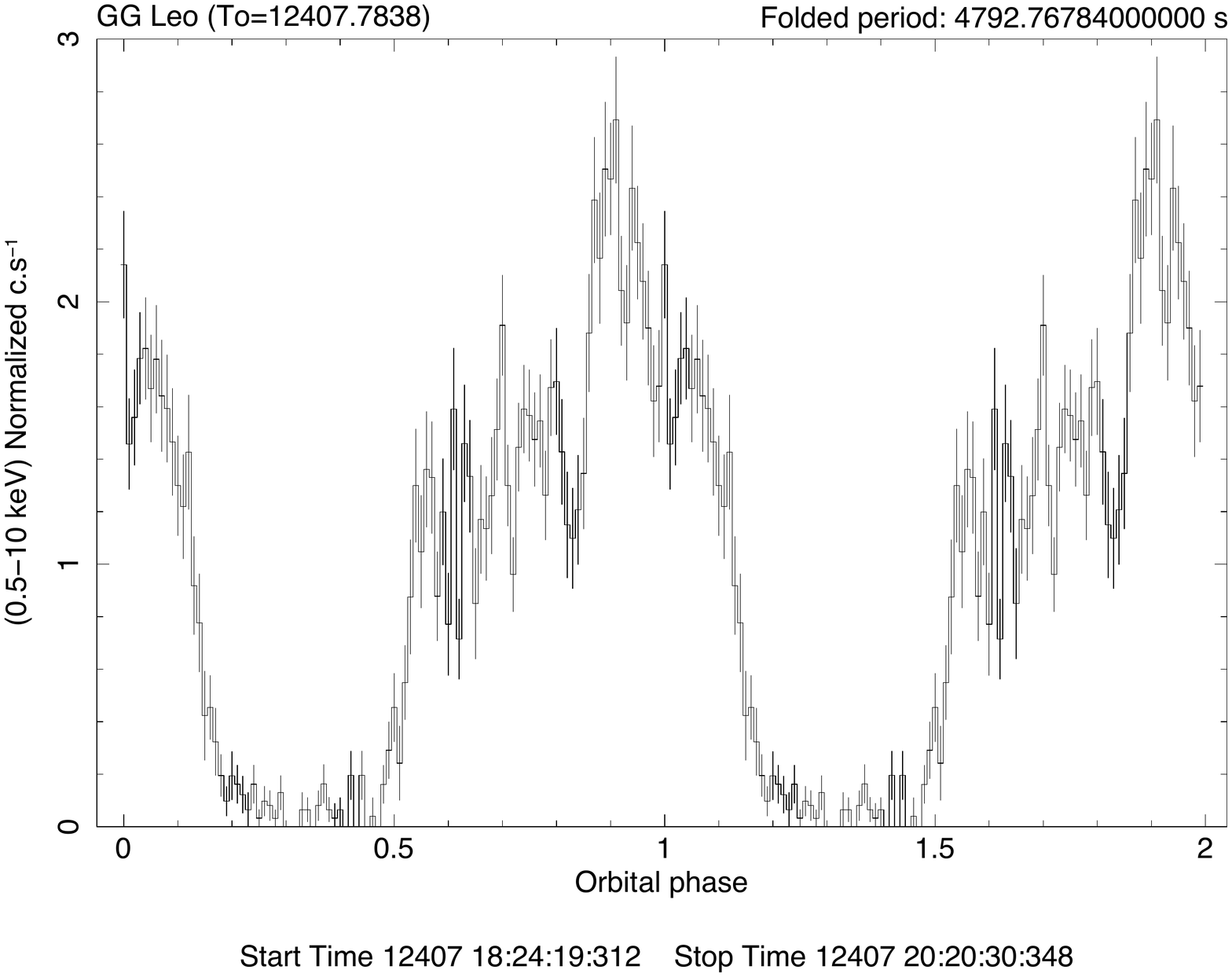}}  \\
       \resizebox{58mm}{!}{\includegraphics*[trim=20 80 50 0]{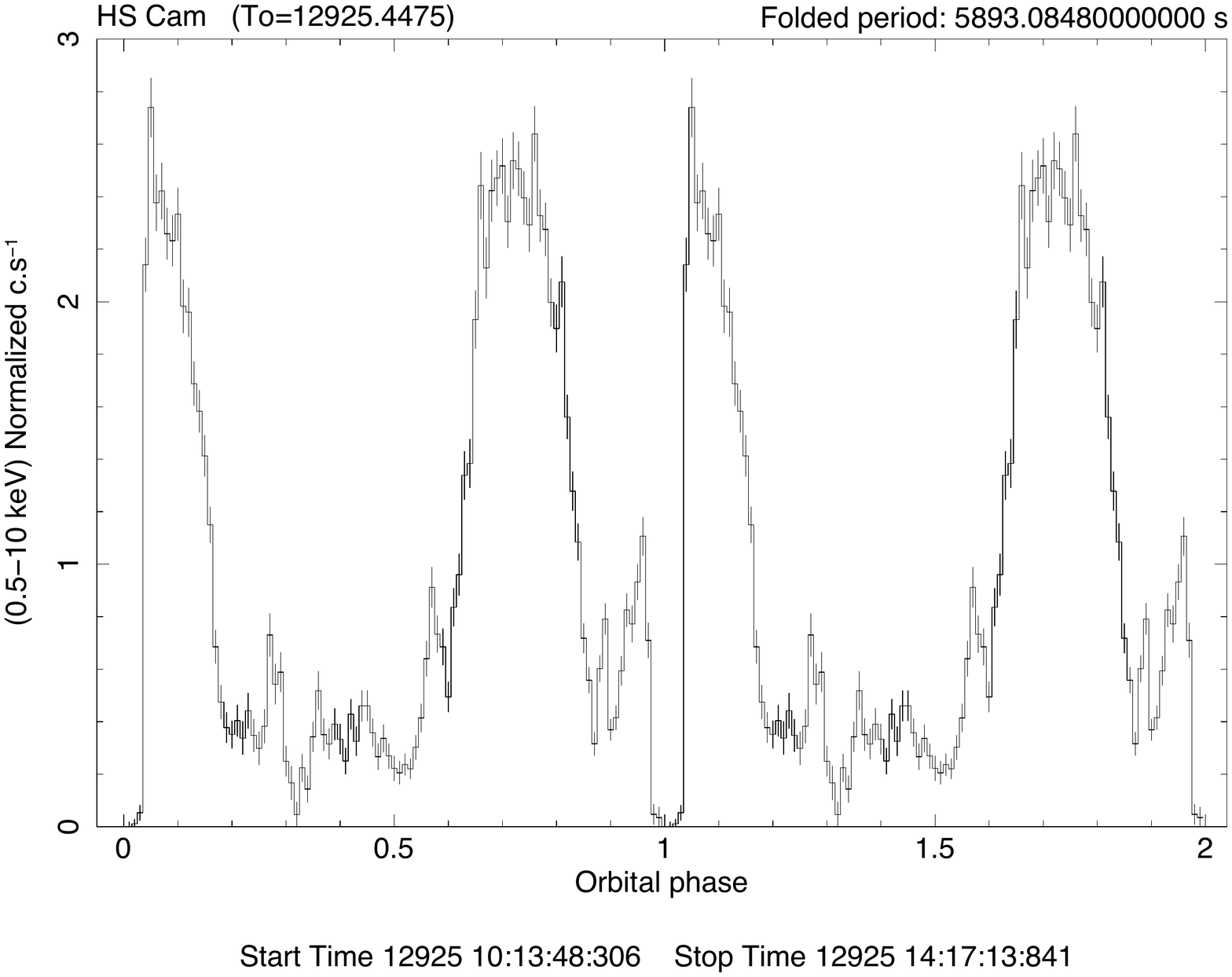}} &           
       \resizebox{58mm}{!}{\includegraphics*[trim=20 80 50 0]{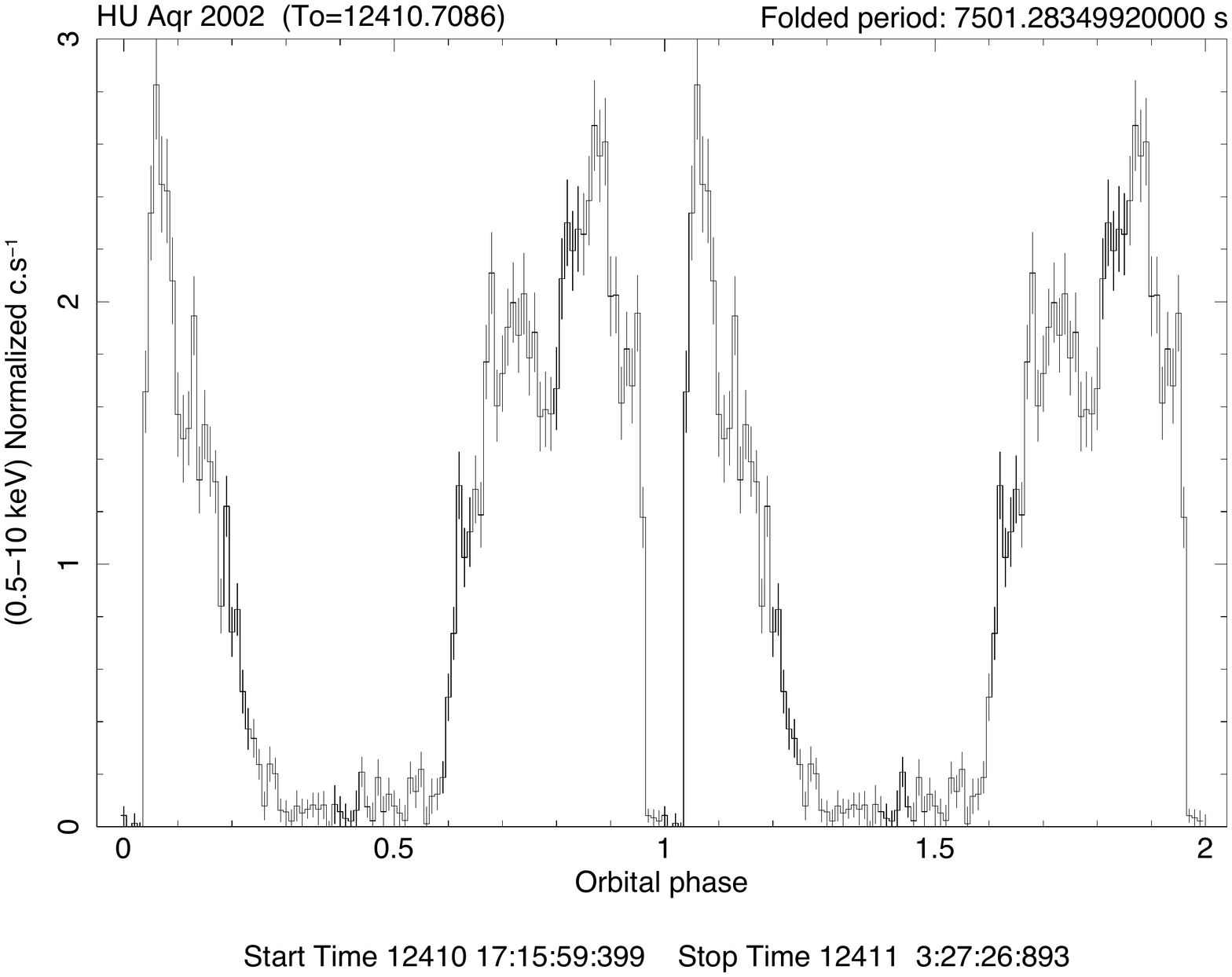}} &         
       \resizebox{58mm}{!}{\includegraphics*[trim=20 80 50 0]{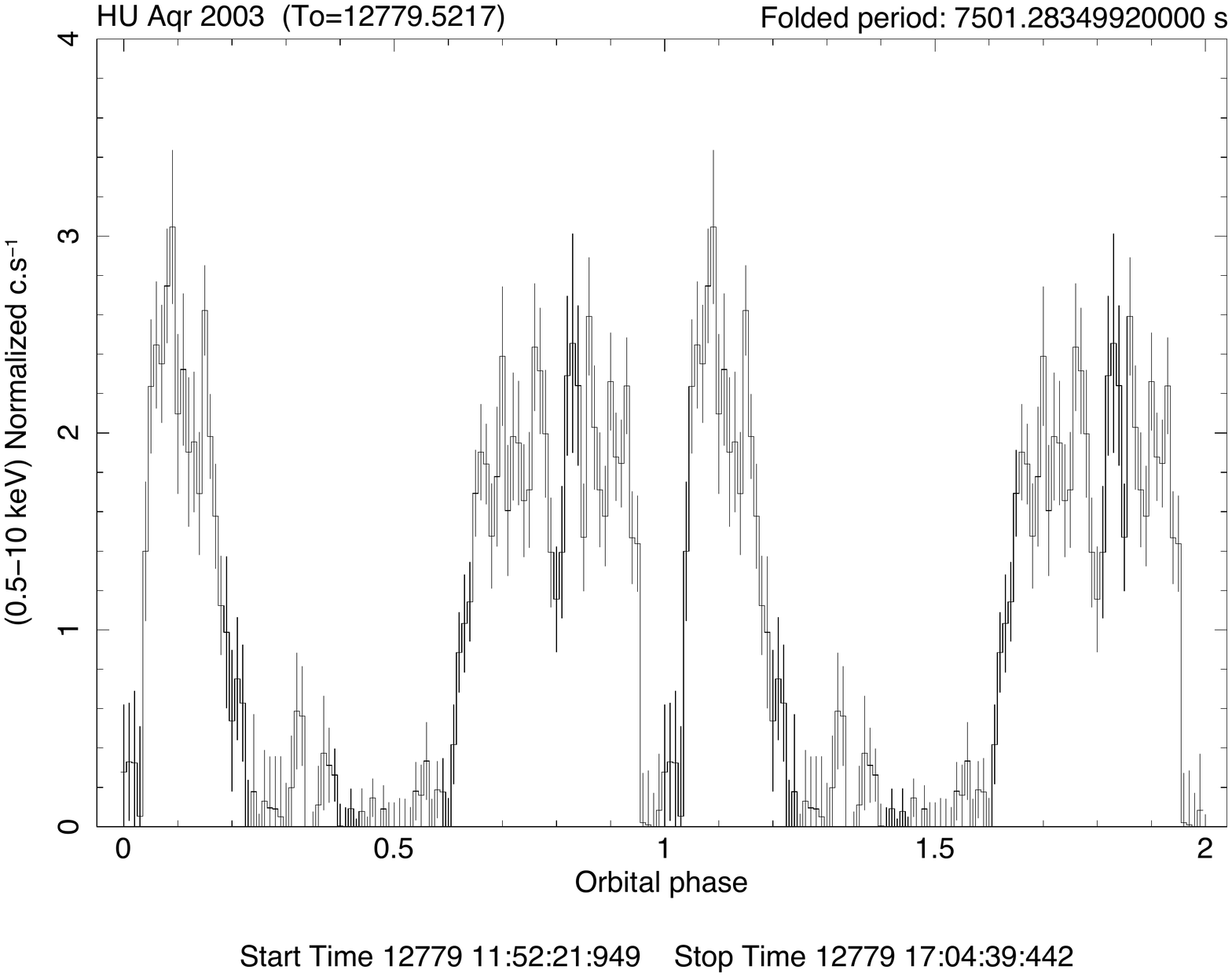}} \\ 
   \end{tabular}
    \caption{The EPIC-PN (0.5-10) keV normalized X-ray light curve of polars. The ephemeris used to fold the light curve is indicated at the top of the figure (with T$_{0}$ = HJD-2440000.5 computed at the date of the observations). The folded curved is repeated twice for clarity.}
    \label{lightcurves_1}
  \end{center}
\end{figure*}
}

%______________________________________________________________
%
%______________________________________________ FIGURE 1b
%
\addtocounter{figure}{-1}
\onlfig{
\begin{figure*}
% \ContinuedFloat
  \begin{center}
   \begin{tabular}{ccc}
          \resizebox{58mm}{!}{\includegraphics*[trim=20 80 50 0]{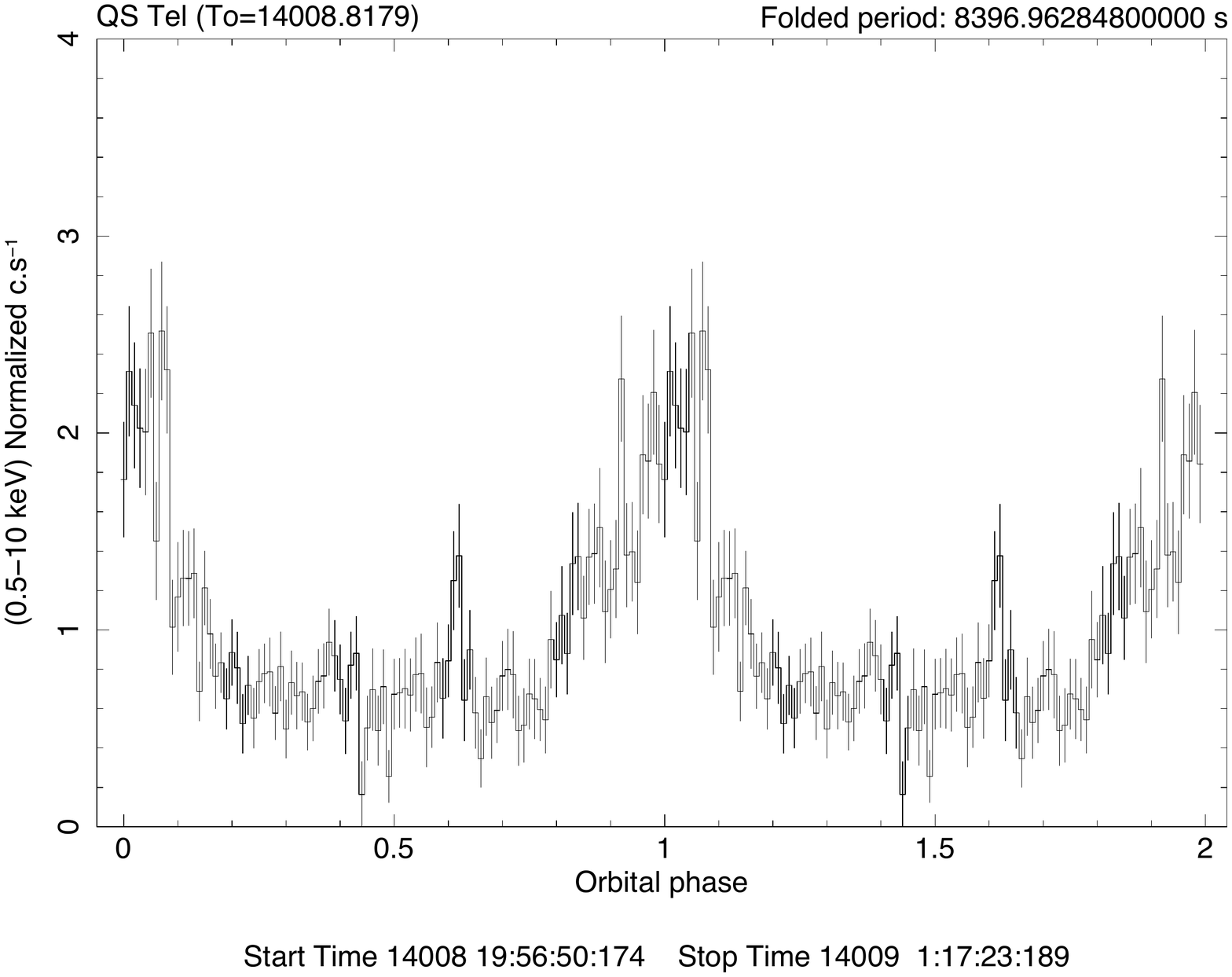}}  &
        \resizebox{58mm}{!}{\includegraphics*[trim=20 80 50 0]{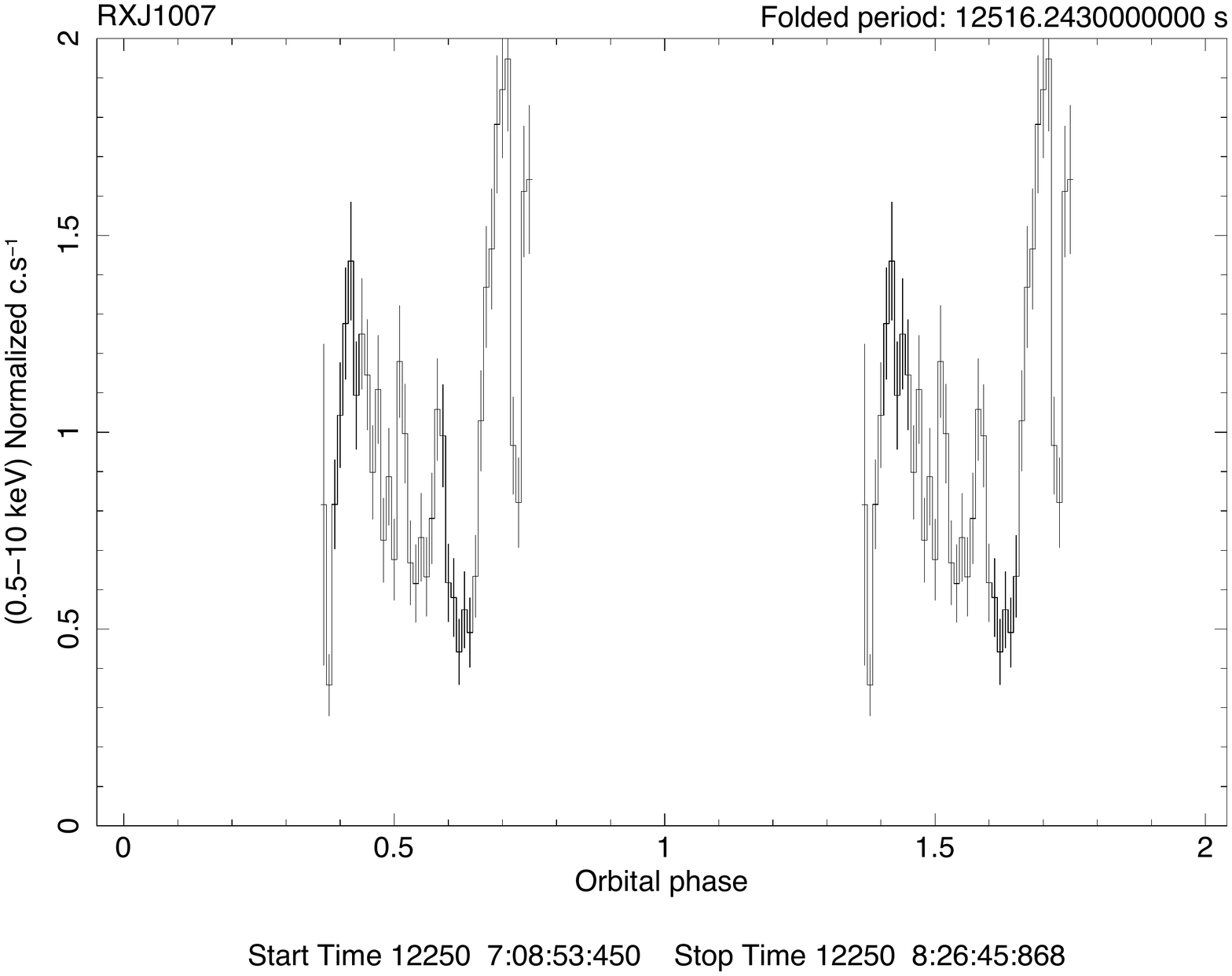}} &
         \resizebox{58mm}{!}{\includegraphics*[trim=20 80 50 0]{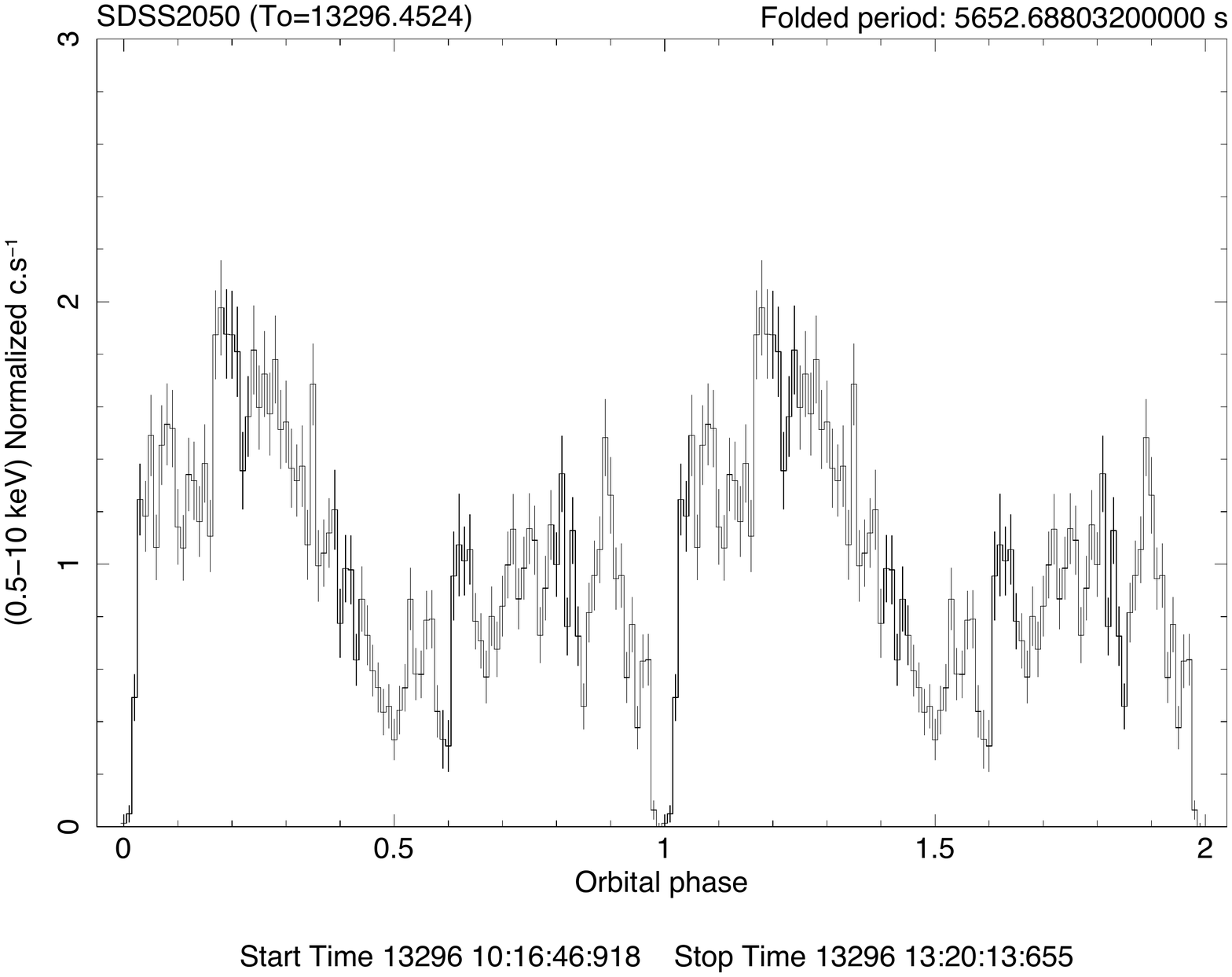}}    \\      
       \resizebox{58mm}{!}{\includegraphics*[trim=20 80 50 0]{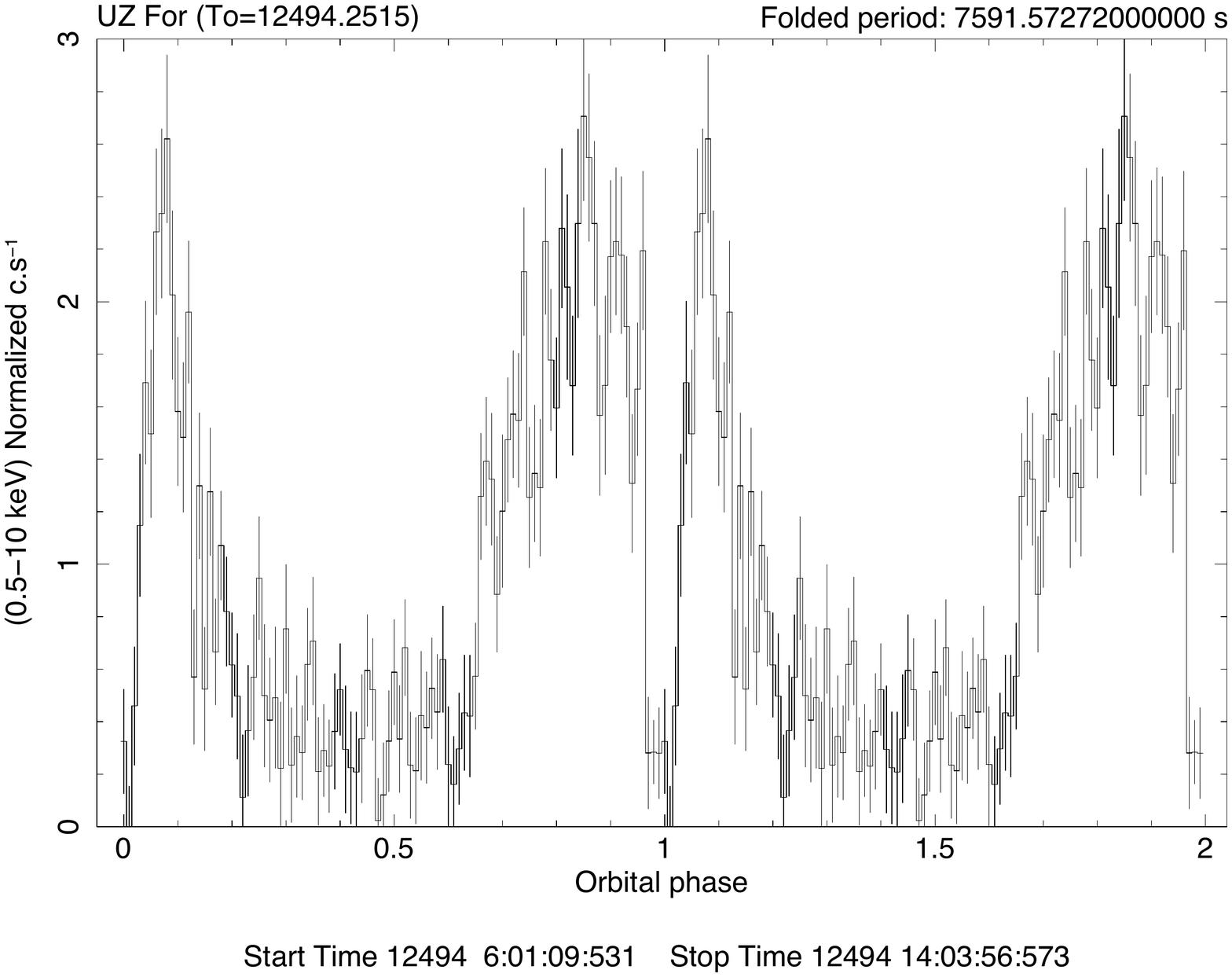}} &        
       \resizebox{58mm}{!}{\includegraphics*[trim=20 80 50 0]{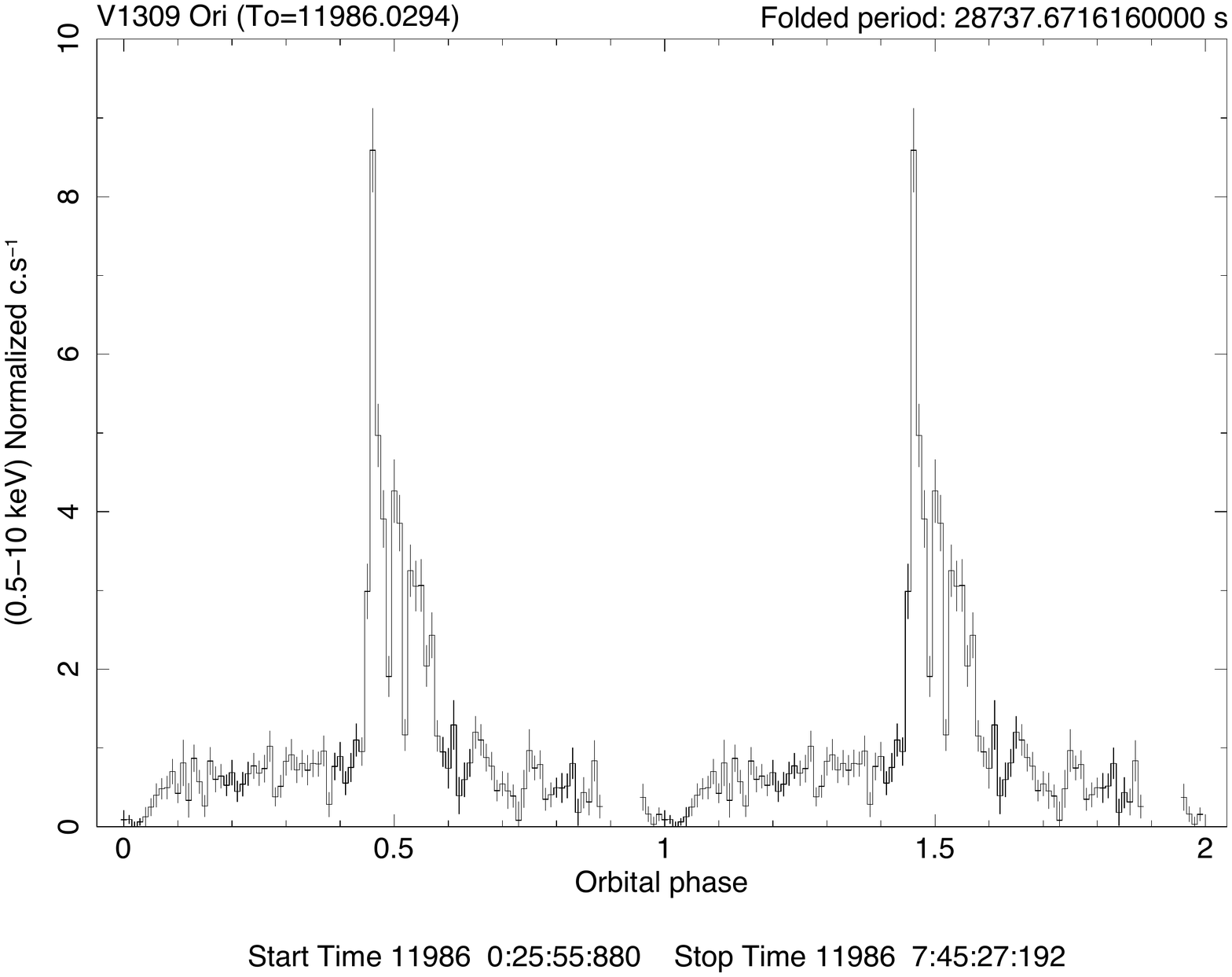}} &
       \resizebox{58mm}{!}{\includegraphics*[trim=20 80 50 0]{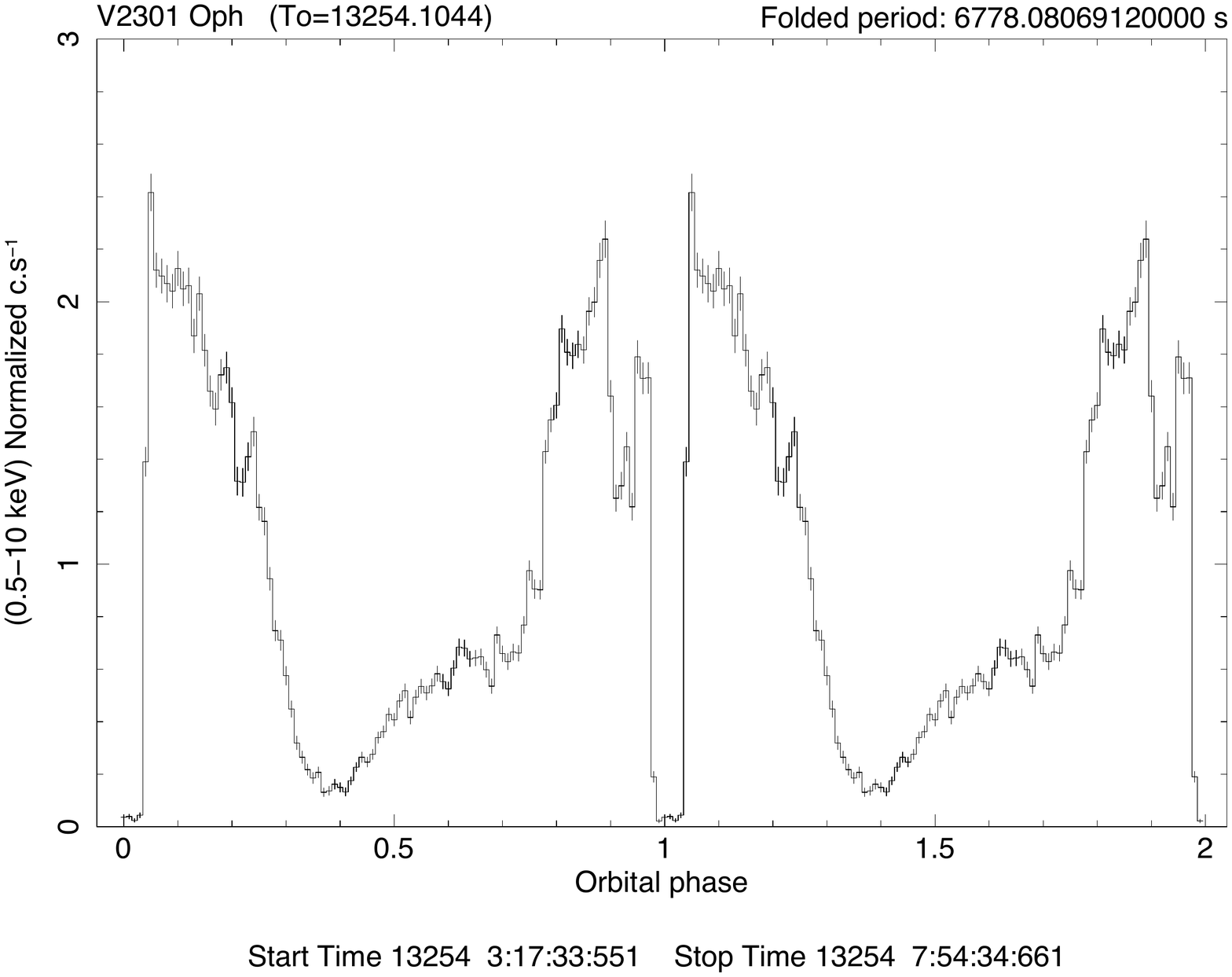}}  \\   
       \resizebox{58mm}{!}{\includegraphics*[trim=20 80 50 0]{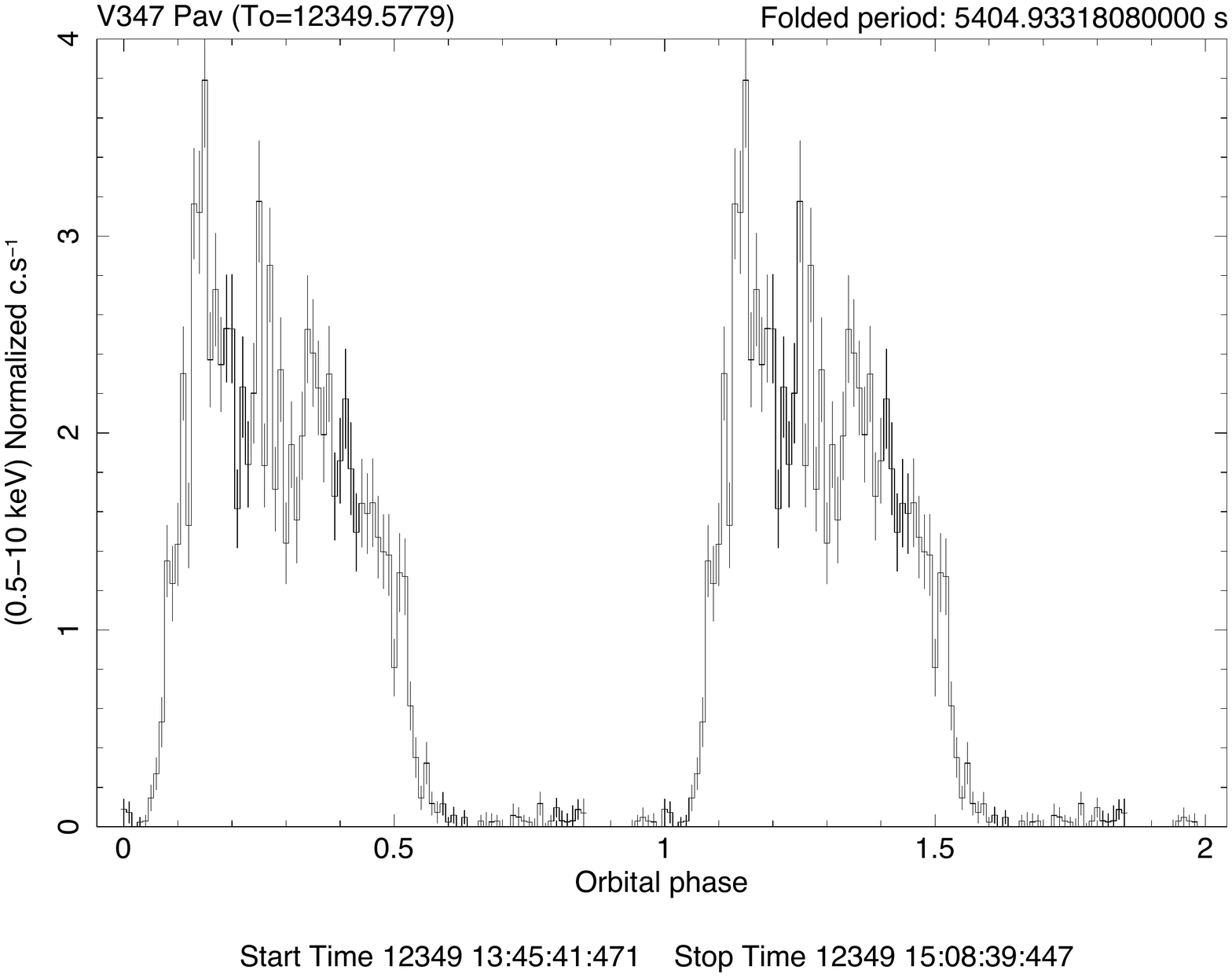}}  &
       \resizebox{58mm}{!}{\includegraphics*[trim=20 80 50 0]{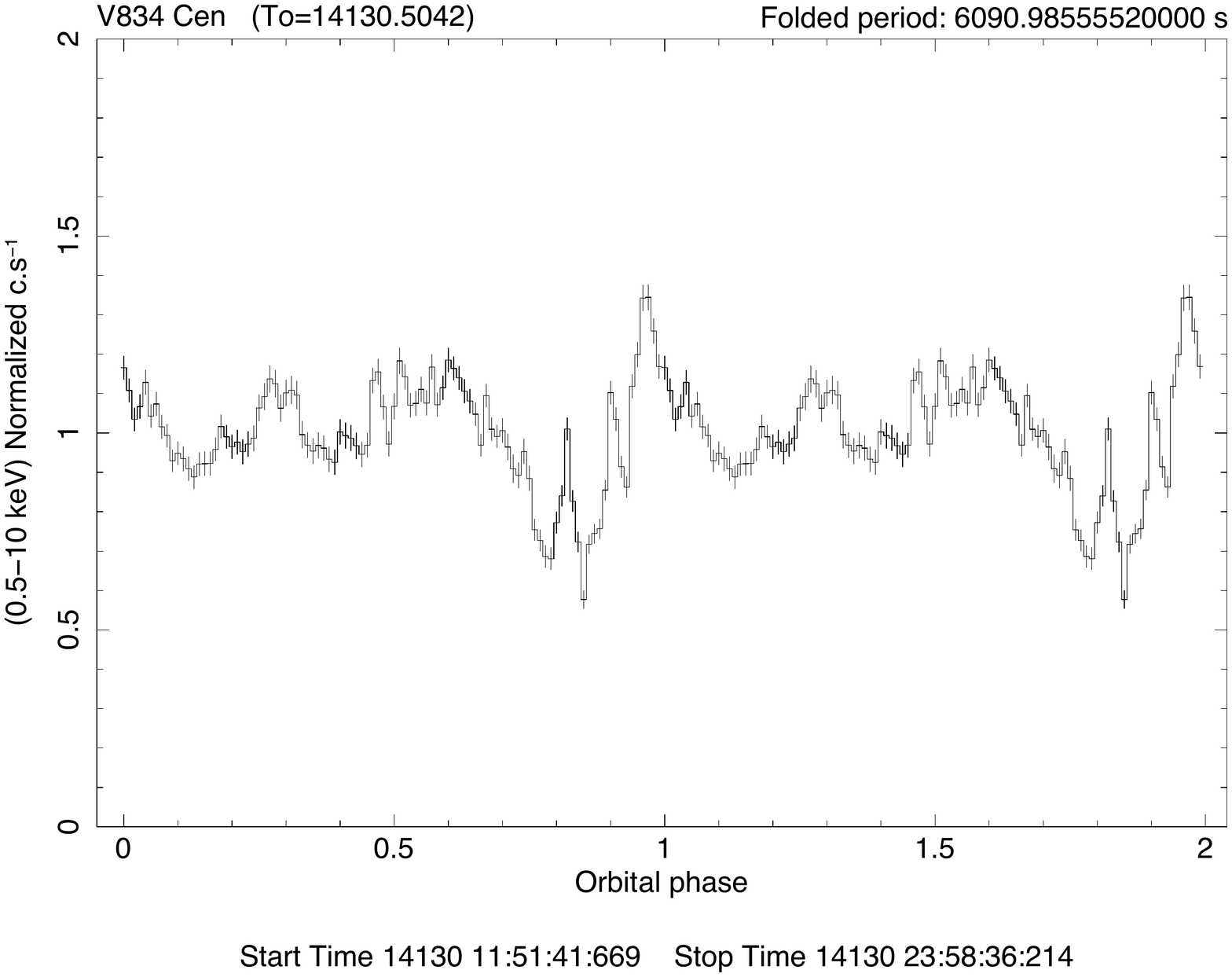}} &
       \resizebox{58mm}{!}{\includegraphics*[trim=20 80 50 0]{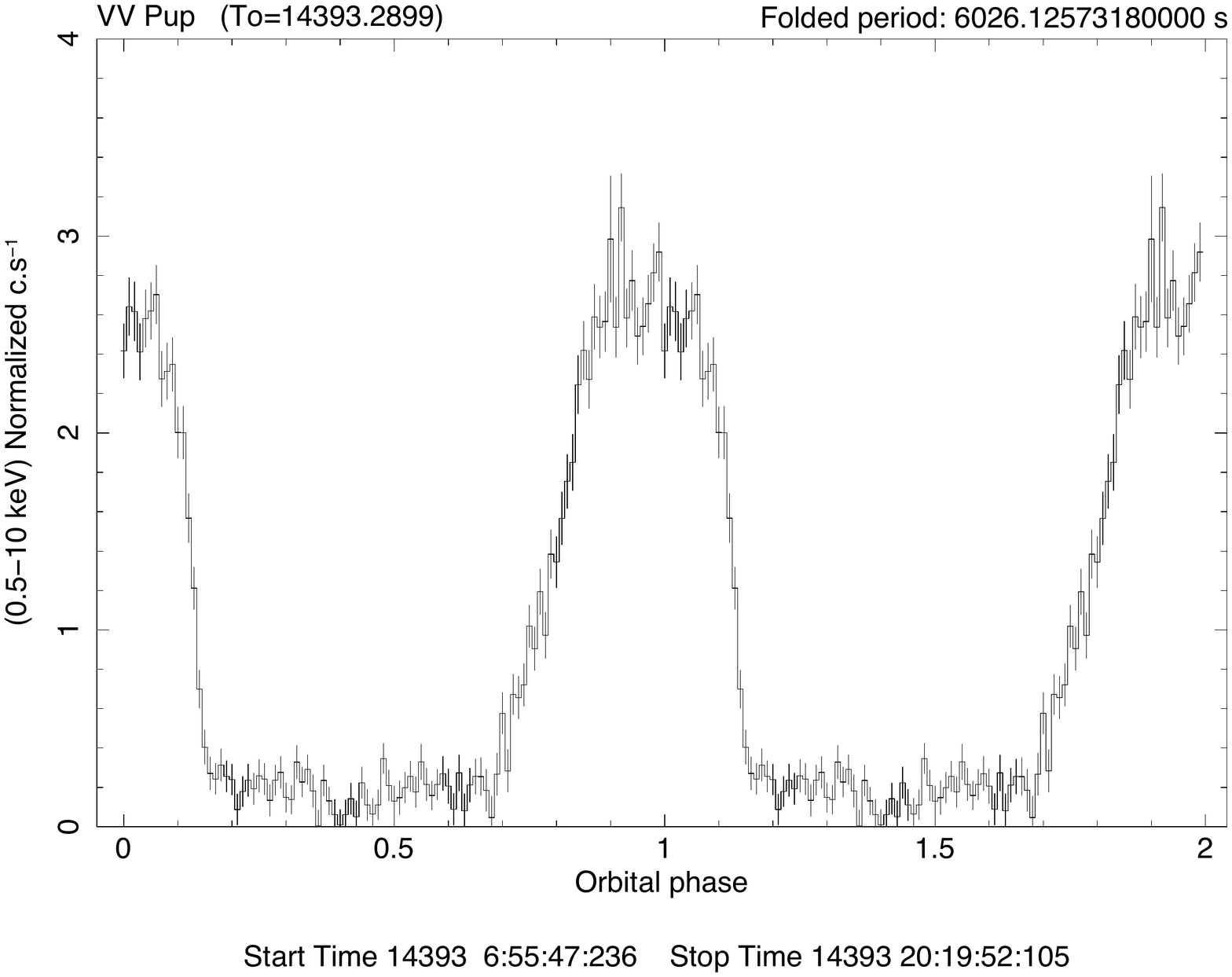}}  \\   
       \resizebox{58mm}{!}{\includegraphics*[trim=20 80 50 0]{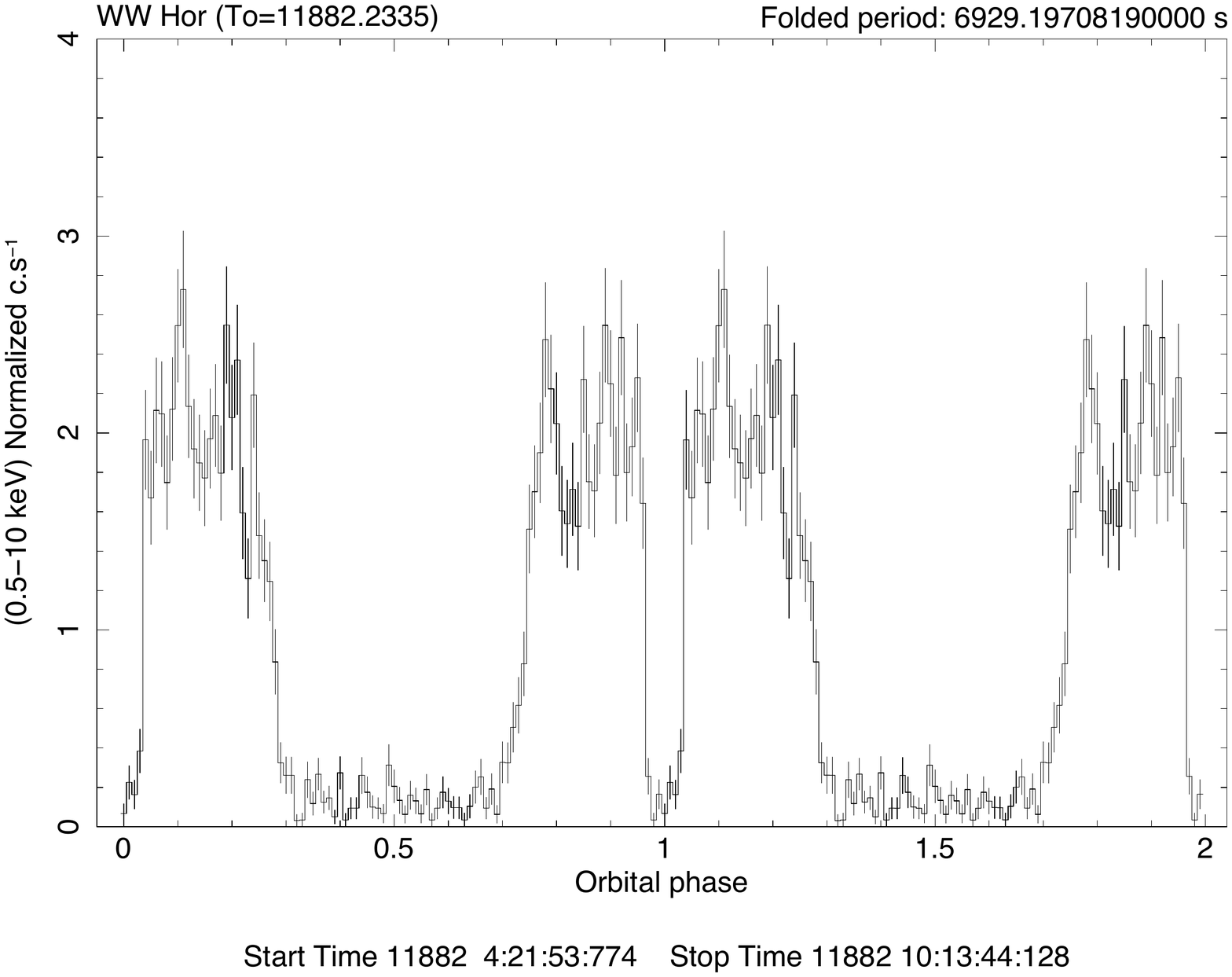}}   \\      
   \end{tabular}
    \caption{The EPIC-PN (0.5-10) keV normalized X-ray light curve of polars. The ephemeris used to fold the light curve is indicated at the top of the figure (with T$_{0}$ = HJD-2440000.5 computed at the date of the observations). The folded curved is repeated twice for clarity.}
    \label{lightcurves_2}
  \end{center}
\end{figure*}
}
%______________________________________________________________
%

%______________________________________________ FIGURE 2
%
\onlfig{
\begin{figure*}
  \begin{center}
   \begin{tabular}{ccc}
       \resizebox{58mm}{!}{\includegraphics*[trim=20 80 50 0]{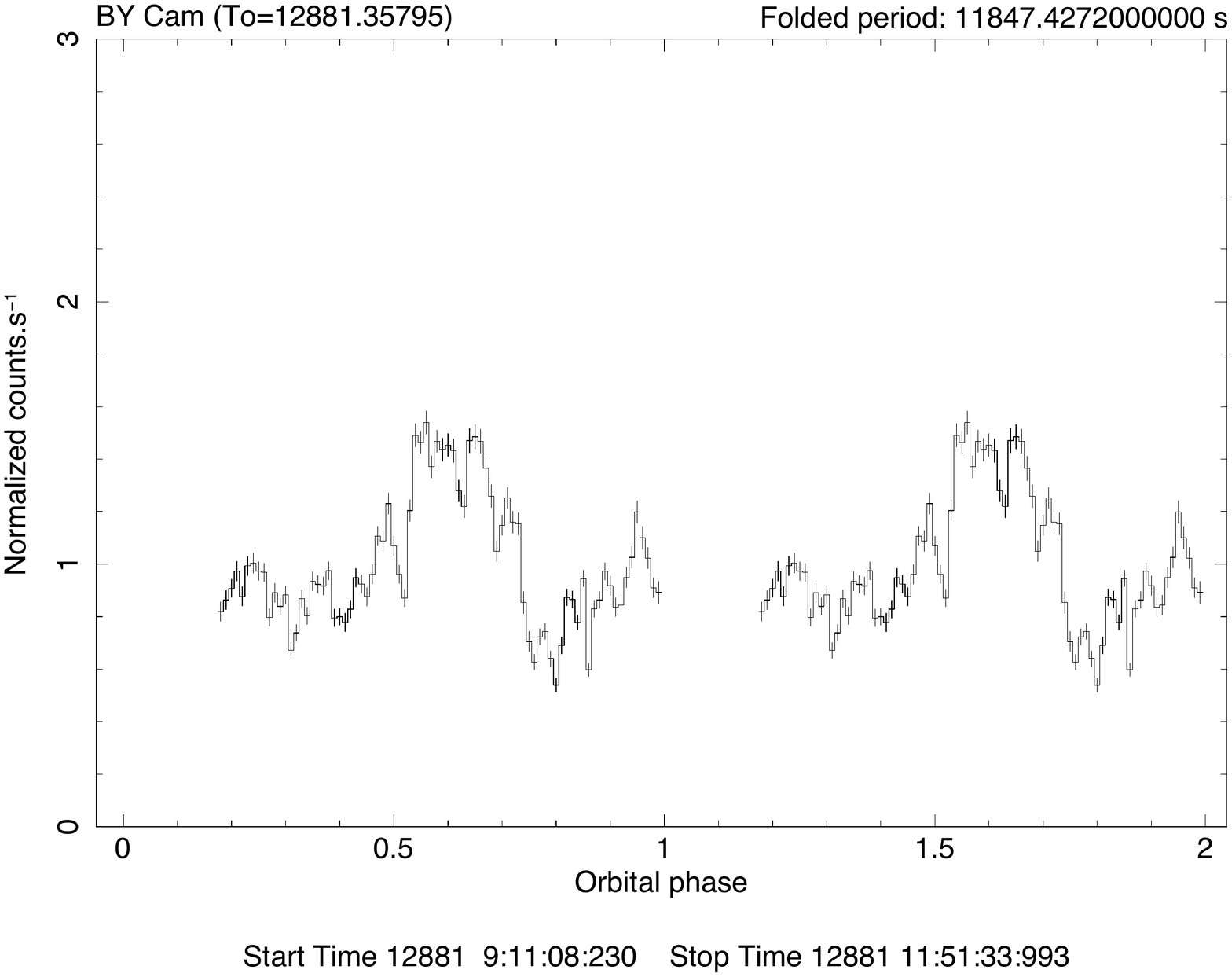}} &     
       \resizebox{58mm}{!}{\includegraphics*[trim=20 80 50 0]{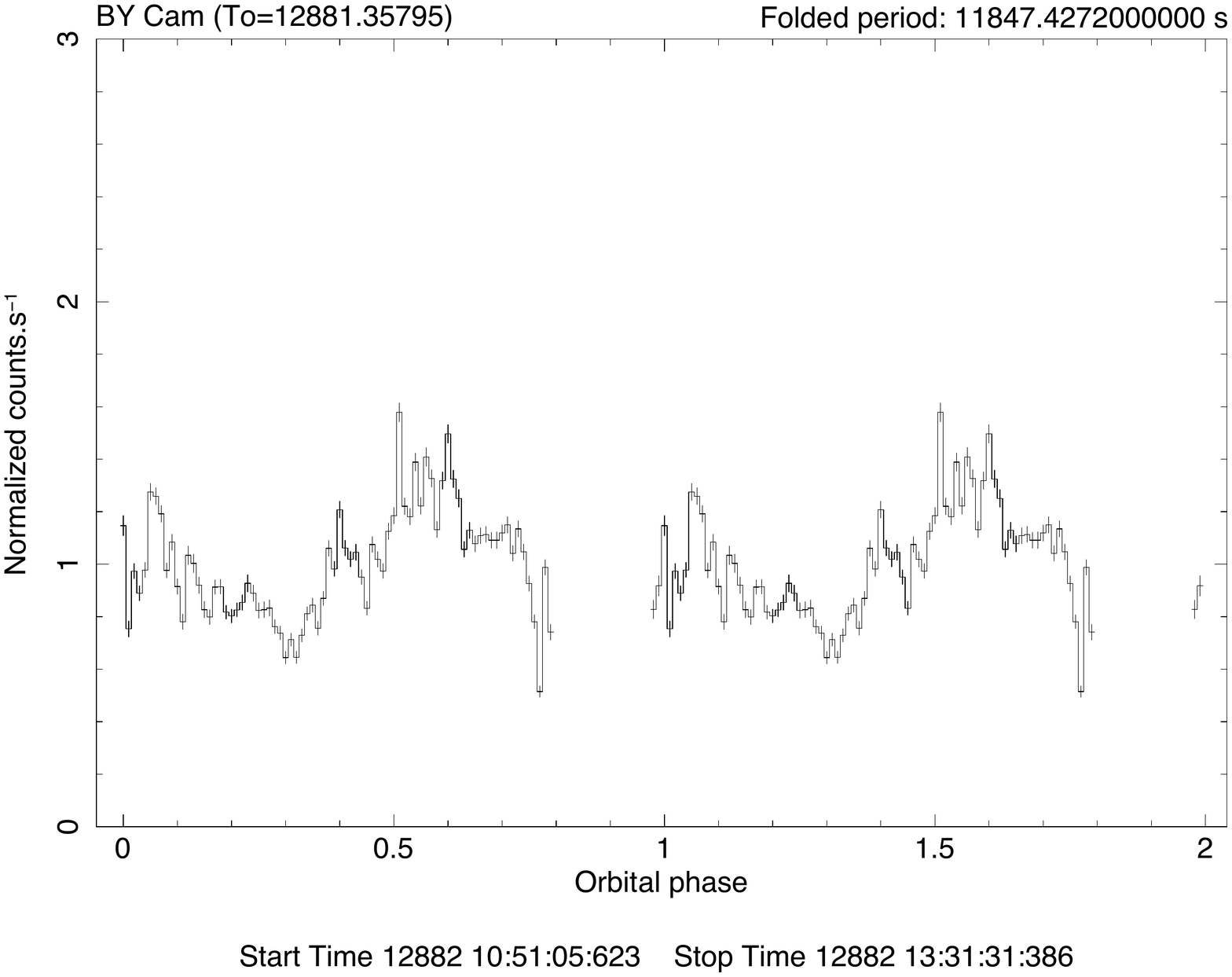}} & 
       \resizebox{58mm}{!}{\includegraphics*[trim=20 80 50 0]{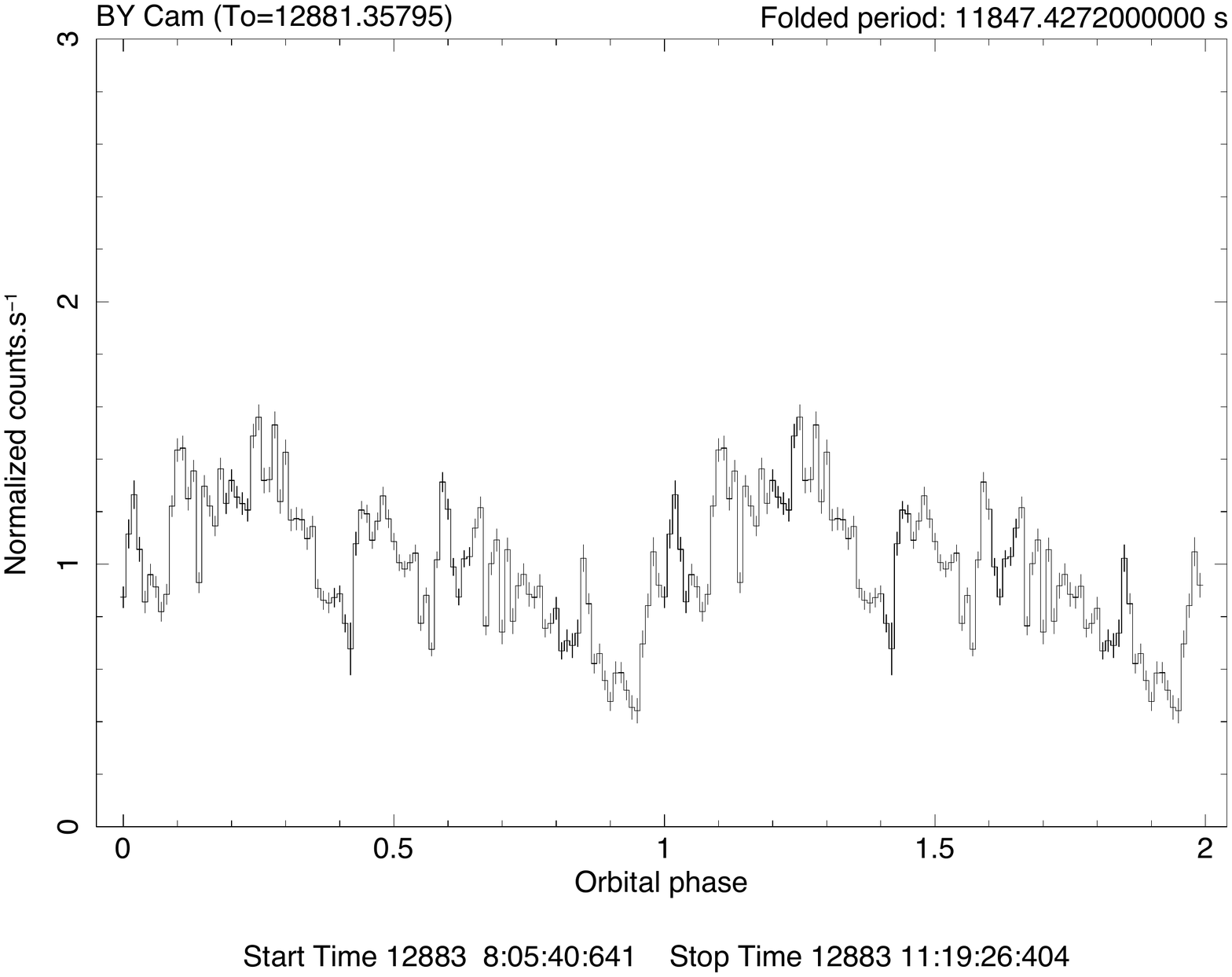}}  \\      
       \resizebox{58mm}{!}{\includegraphics*[trim=20 80 50 0]{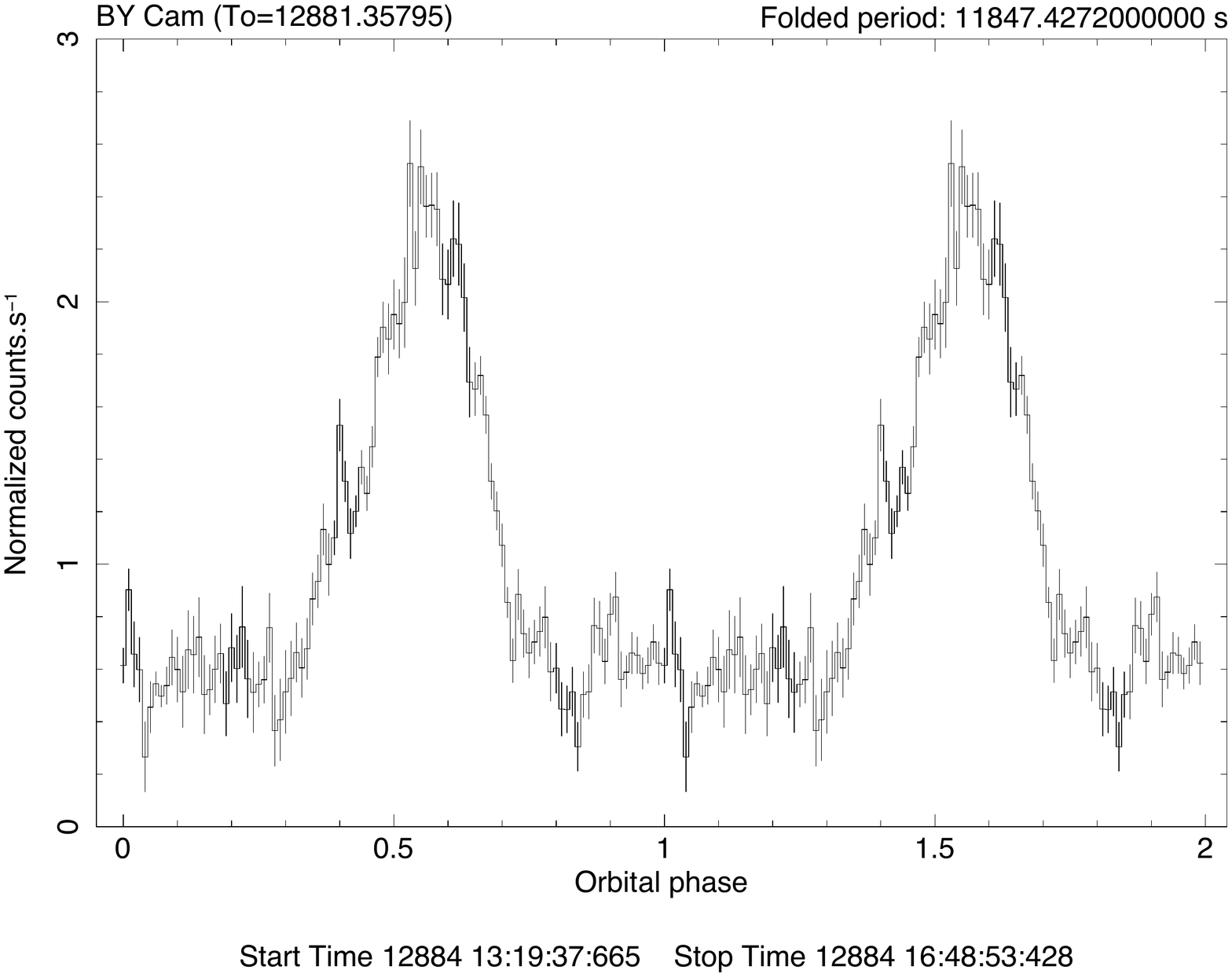}} &     
       \resizebox{58mm}{!}{\includegraphics*[trim=20 80 50 0]{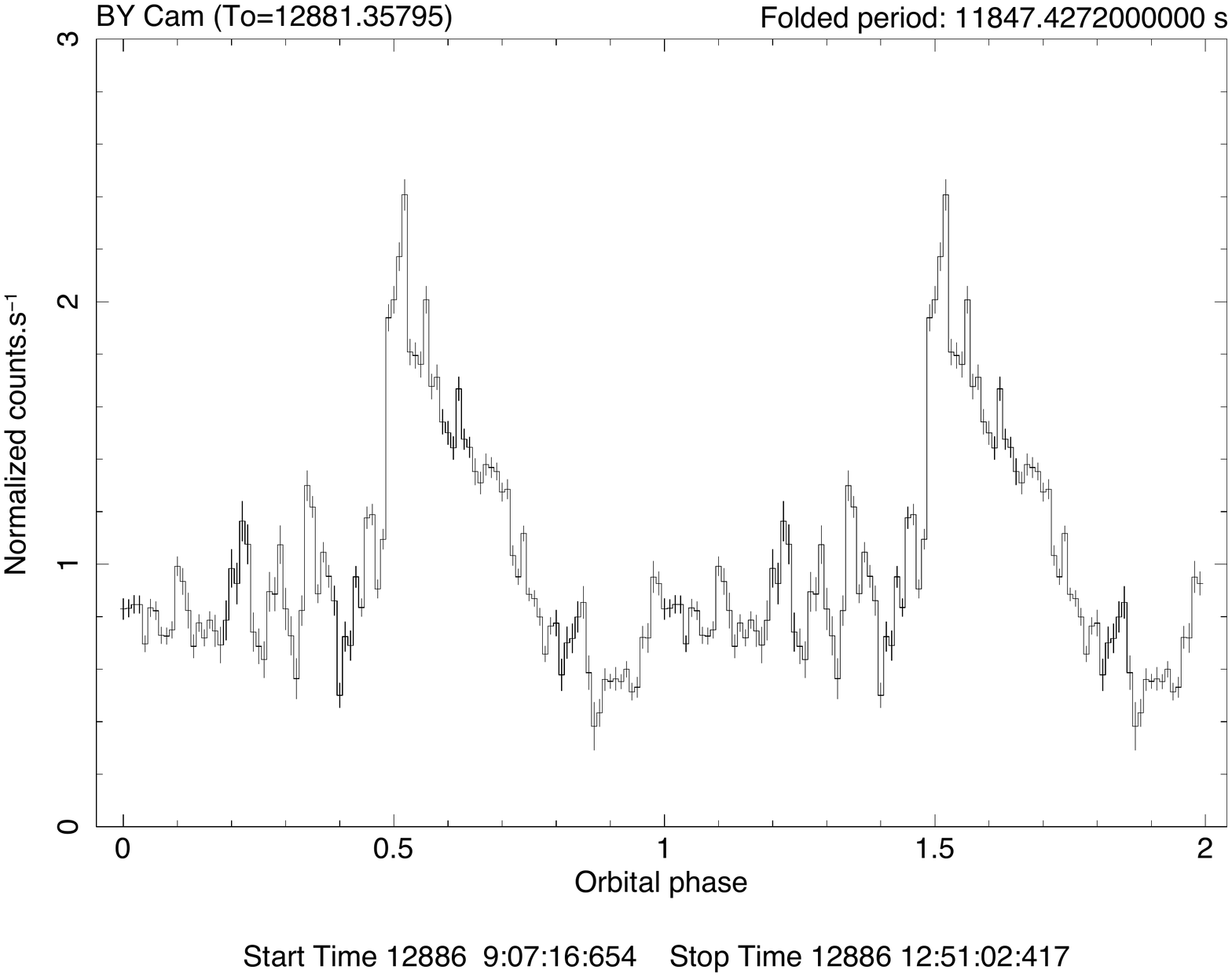}} & 
       \resizebox{58mm}{!}{\includegraphics*[trim=20 80 50 0]{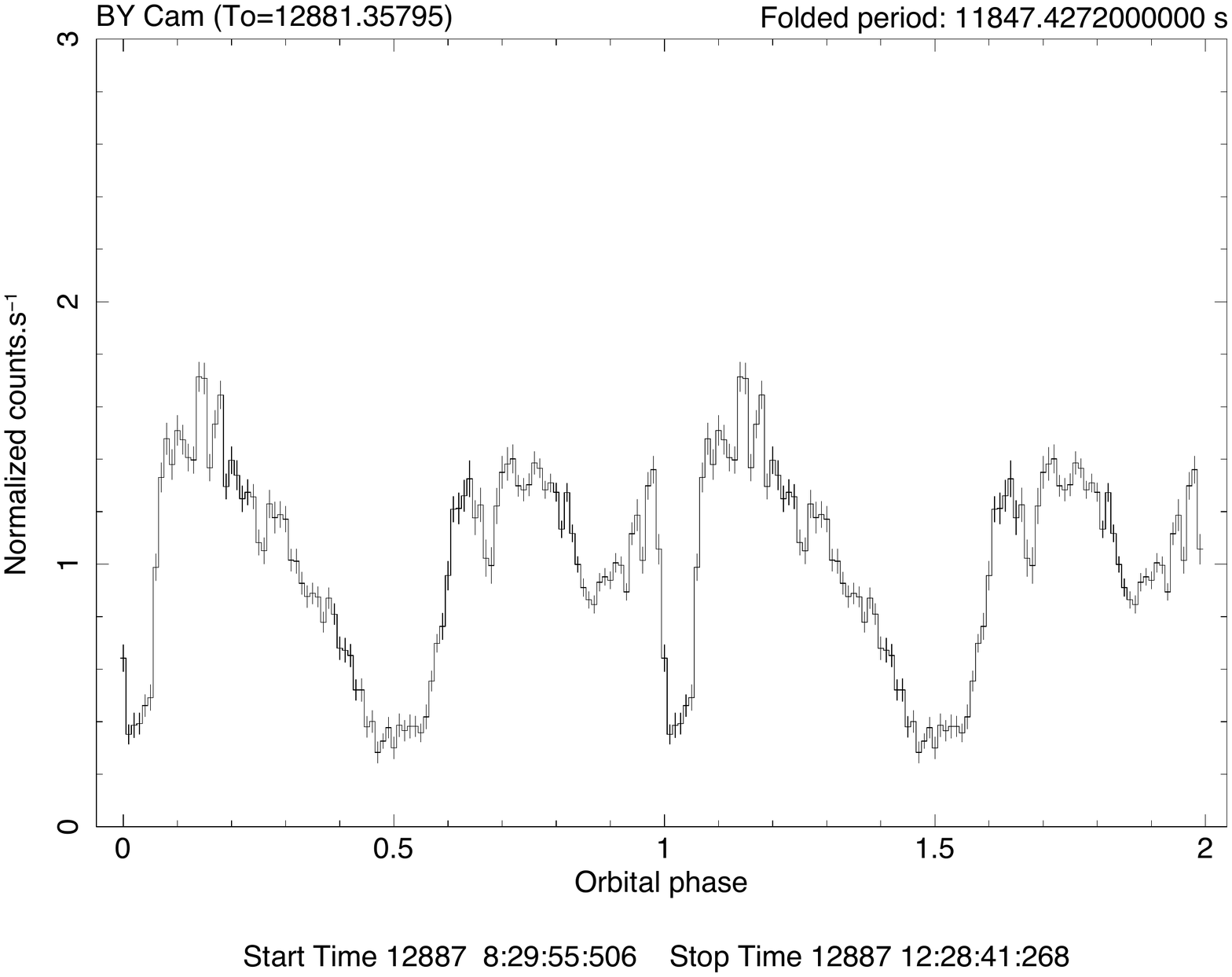}}  \\      
       \resizebox{58mm}{!}{\includegraphics*[trim=20 80 50 0]{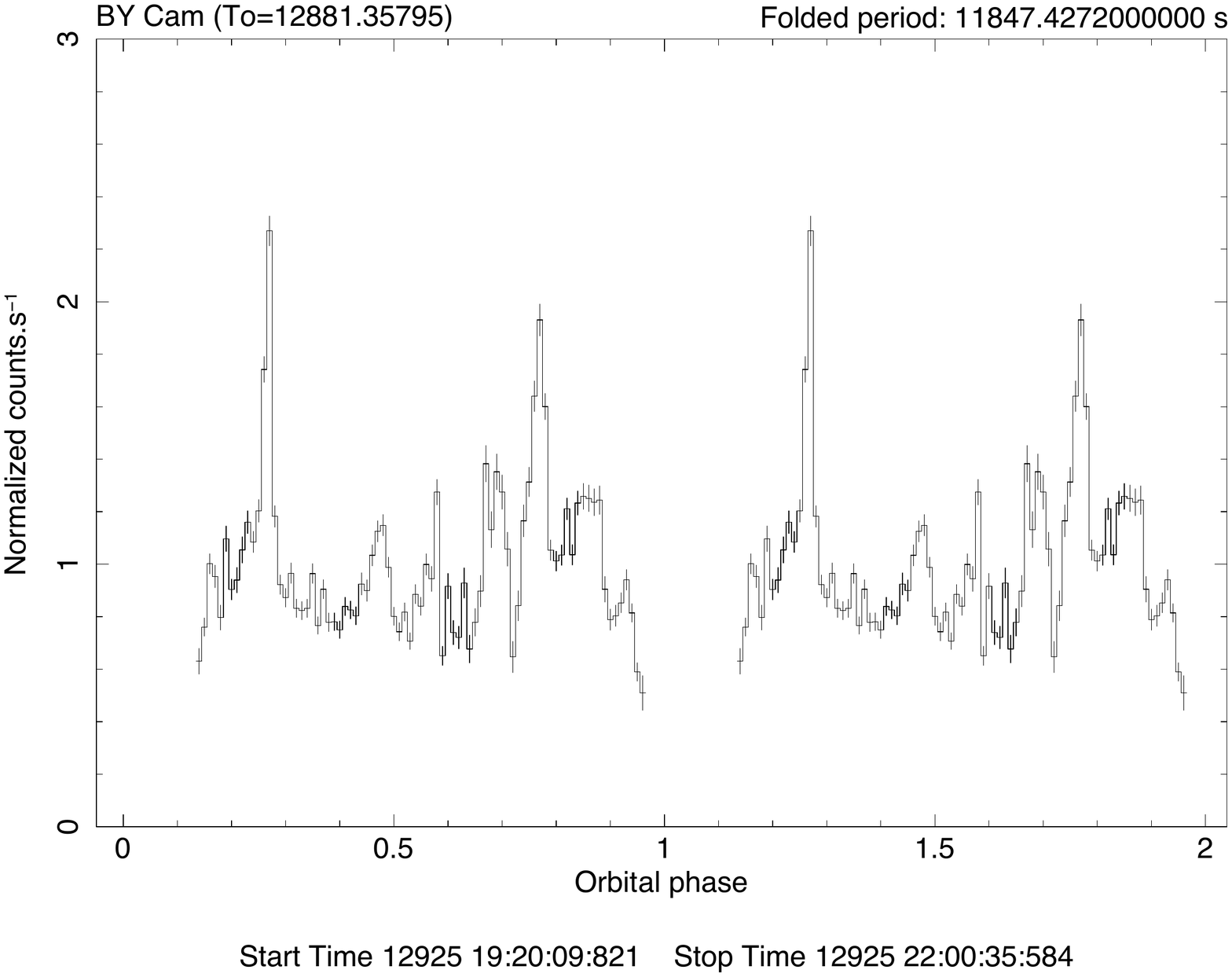}} \\  
   \end{tabular}
    \caption{The EPIC-PN (0.5-10) keV normalized X-ray light curve of the polar BY Cam with same caption as in Fig.~\ref{lightcurves_1}. Observations are shown (left to right, top to bottom) in the same chronological order as in Table~\ref{log_obs}.}
    \label{lightcurves_BYCam}
  \end{center}
\end{figure*}
}

%______________________________________________________________
%
%______________________________________________ FIGURE 3
%
\onlfig{
\begin{figure*}
  \begin{center}
   \begin{tabular}{ccc}
       \resizebox{58mm}{!}{\includegraphics*[trim=20 80 50 0]{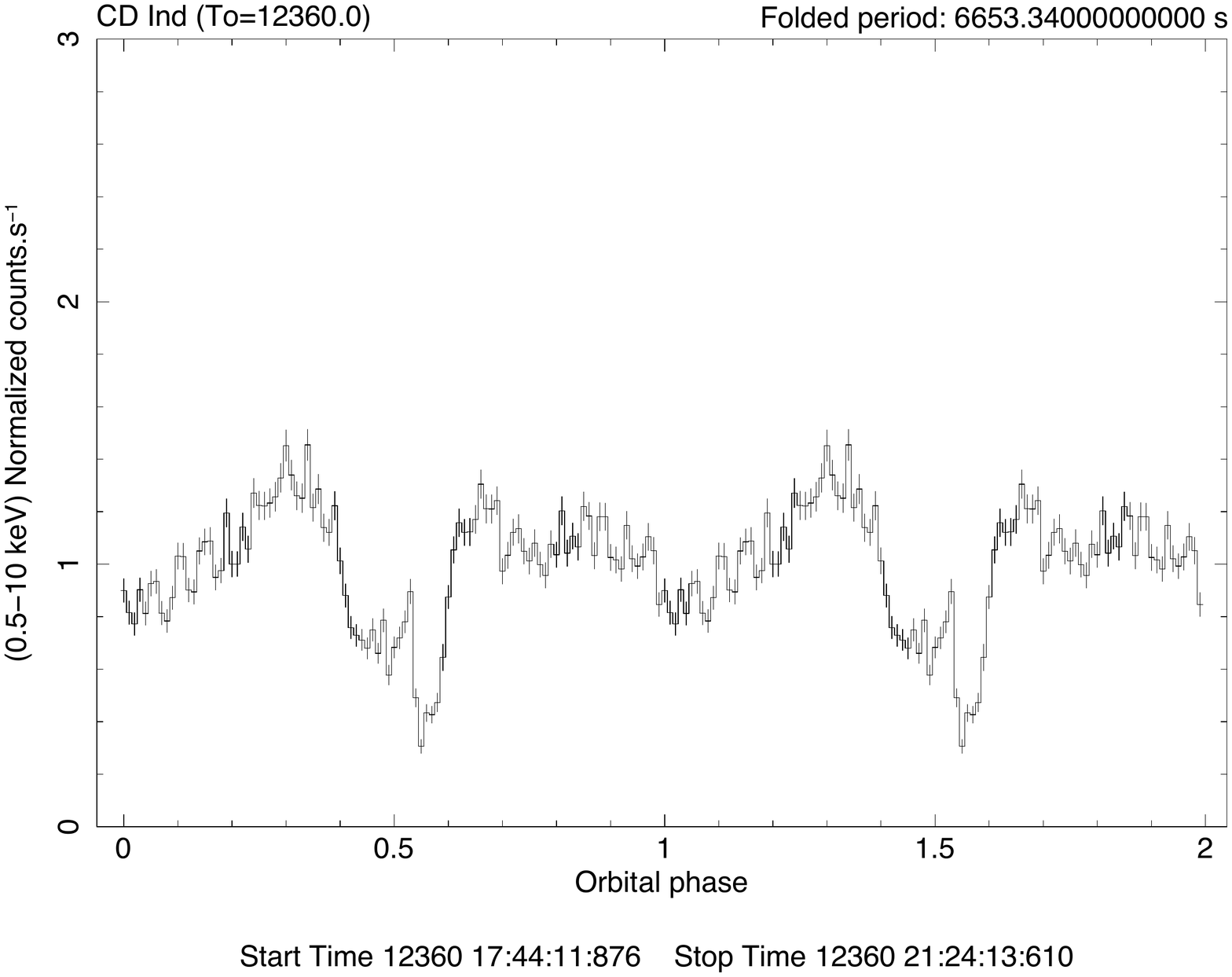}} &     
       \resizebox{58mm}{!}{\includegraphics*[trim=20 80 50 0]{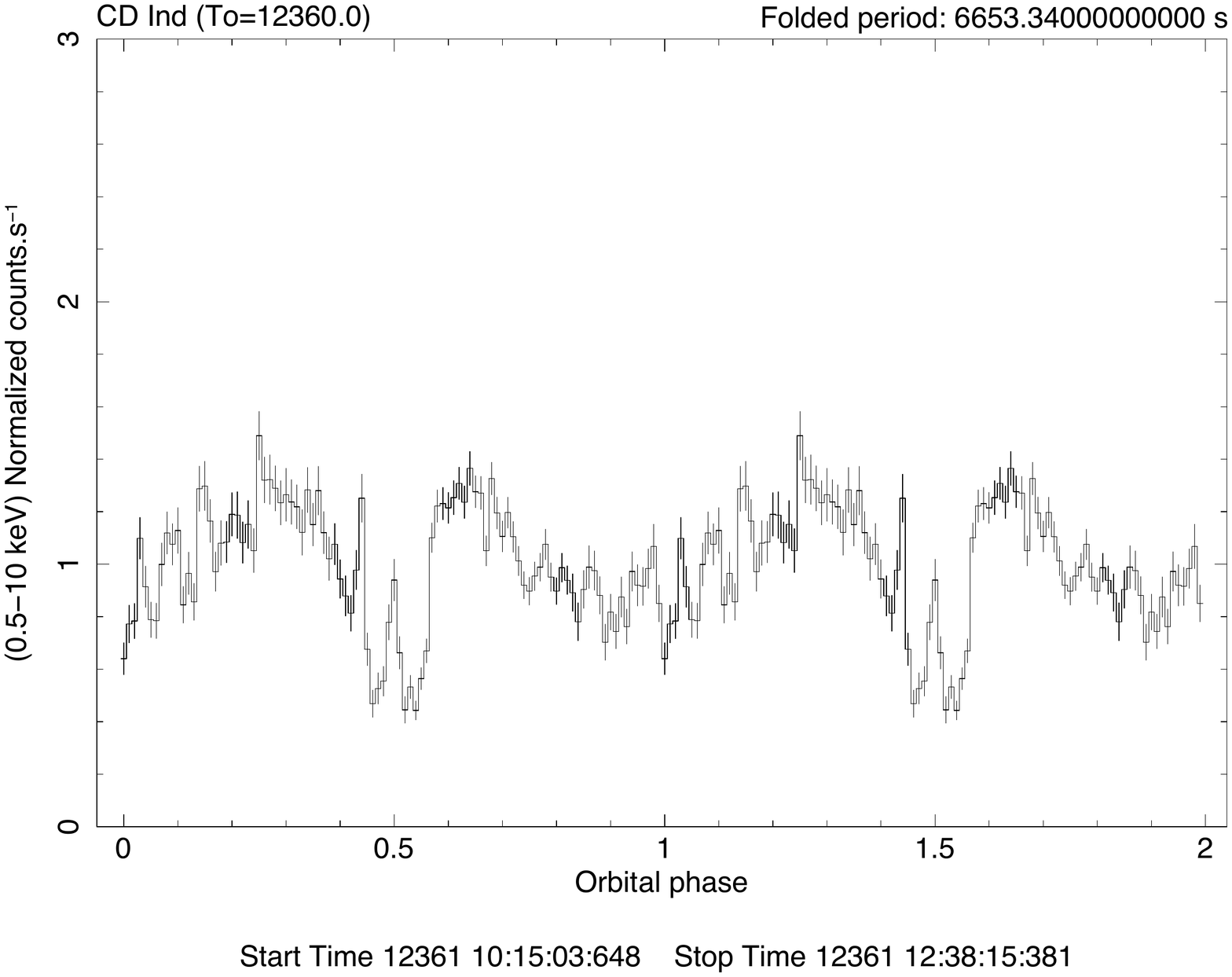}} & 
       \resizebox{58mm}{!}{\includegraphics*[trim=20 80 50 0]{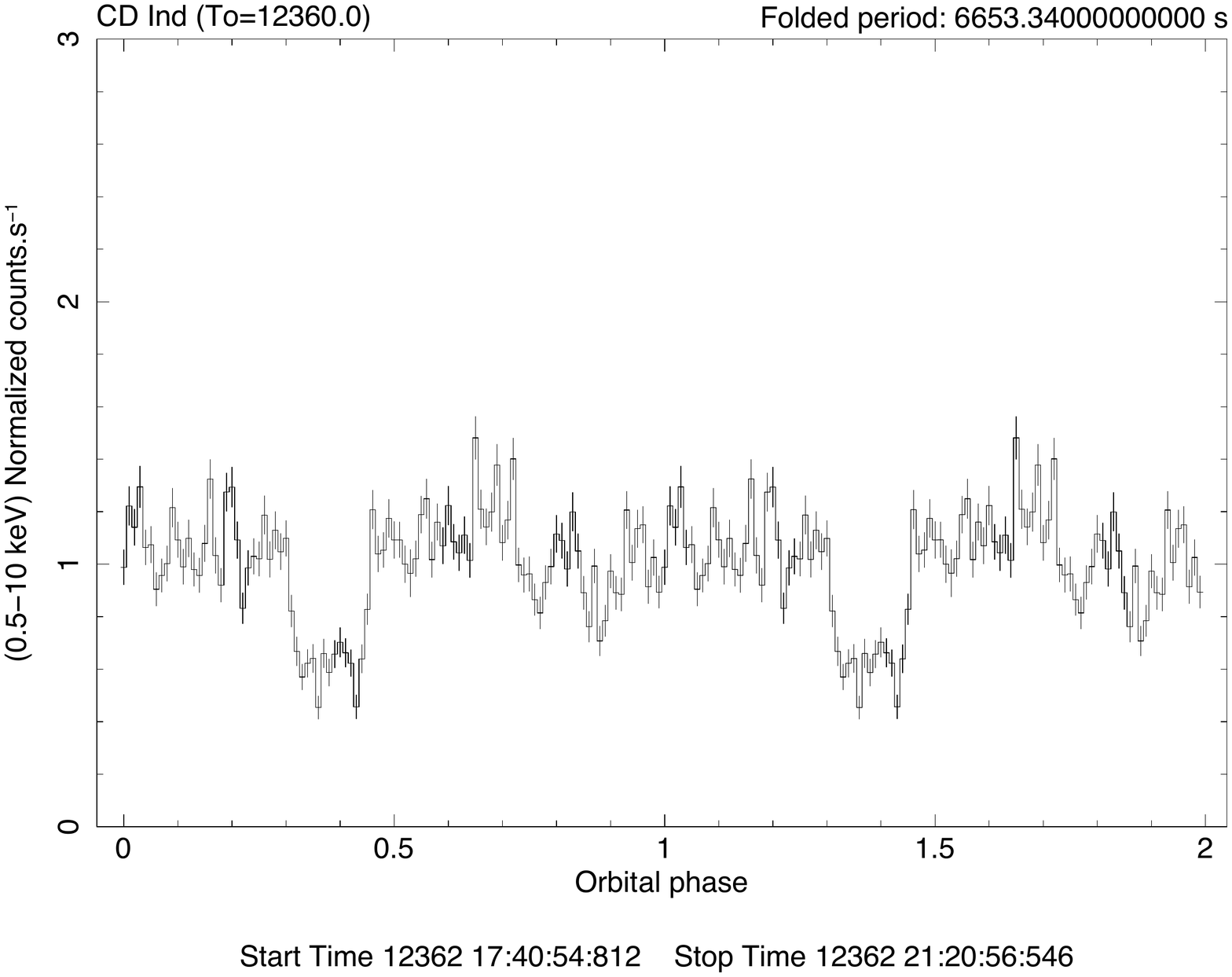}}  \\      
       \resizebox{58mm}{!}{\includegraphics*[trim=20 80 50 0]{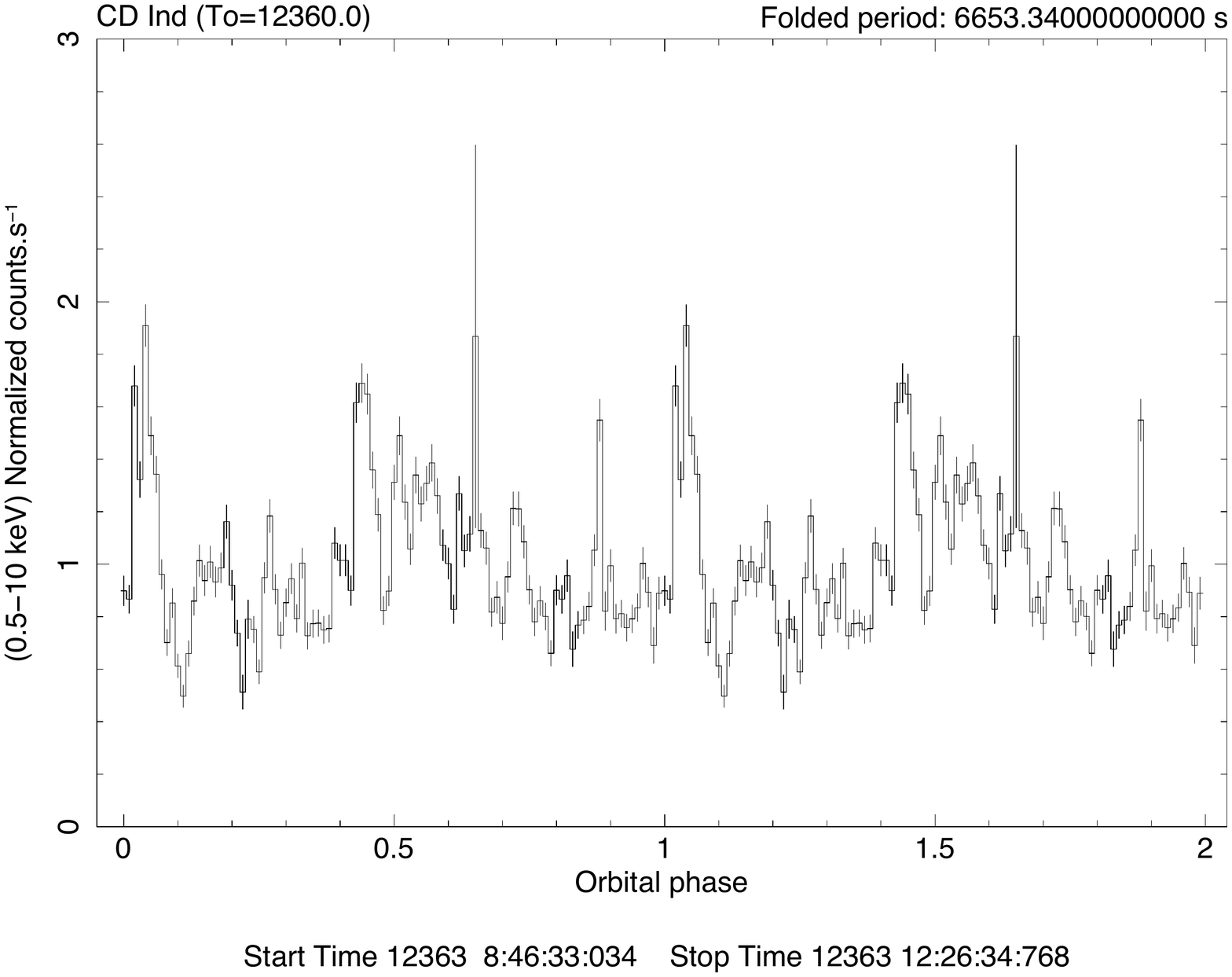}} &     
       \resizebox{58mm}{!}{\includegraphics*[trim=20 80 50 0]{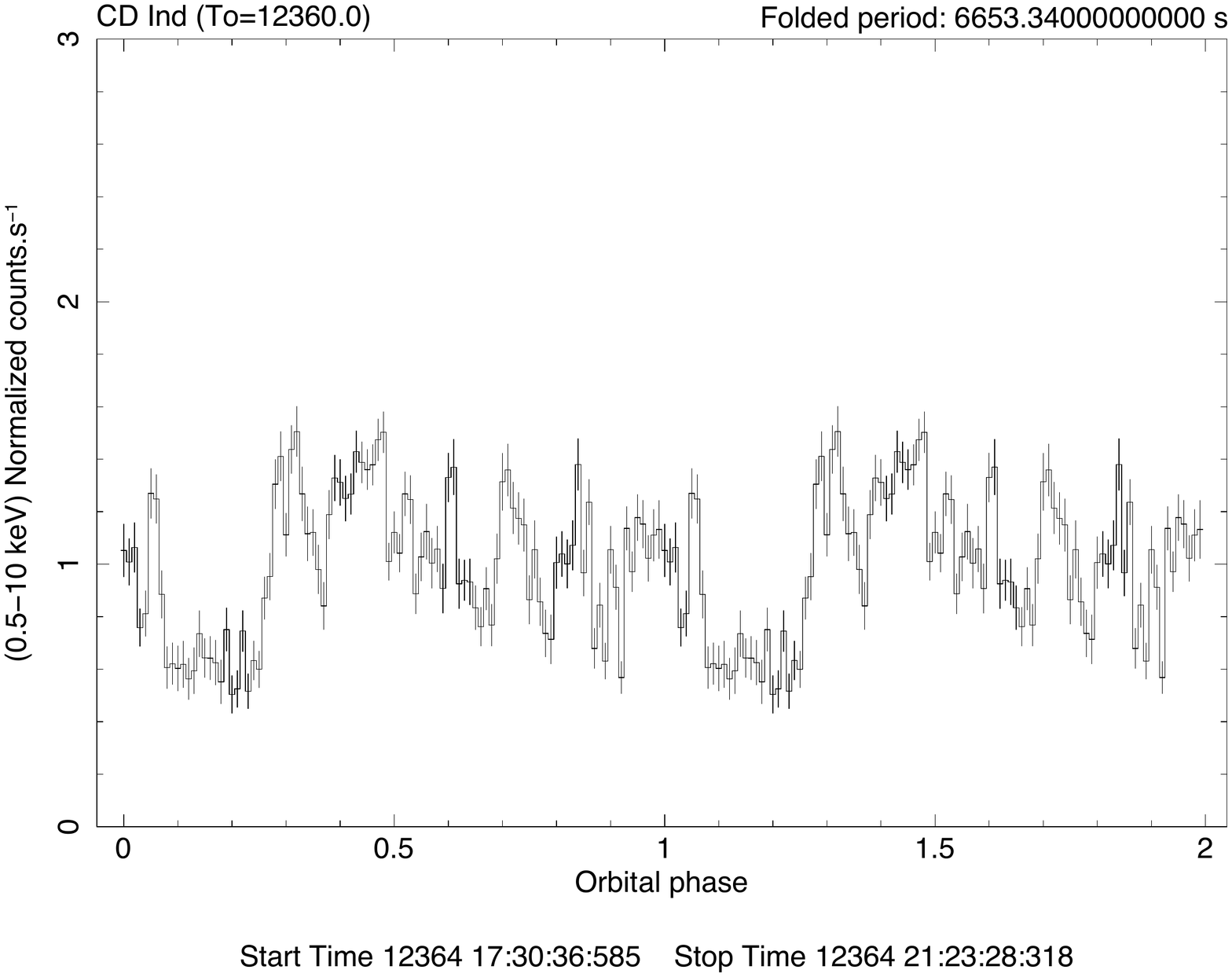}} & 
       \resizebox{58mm}{!}{\includegraphics*[trim=20 80 50 0]{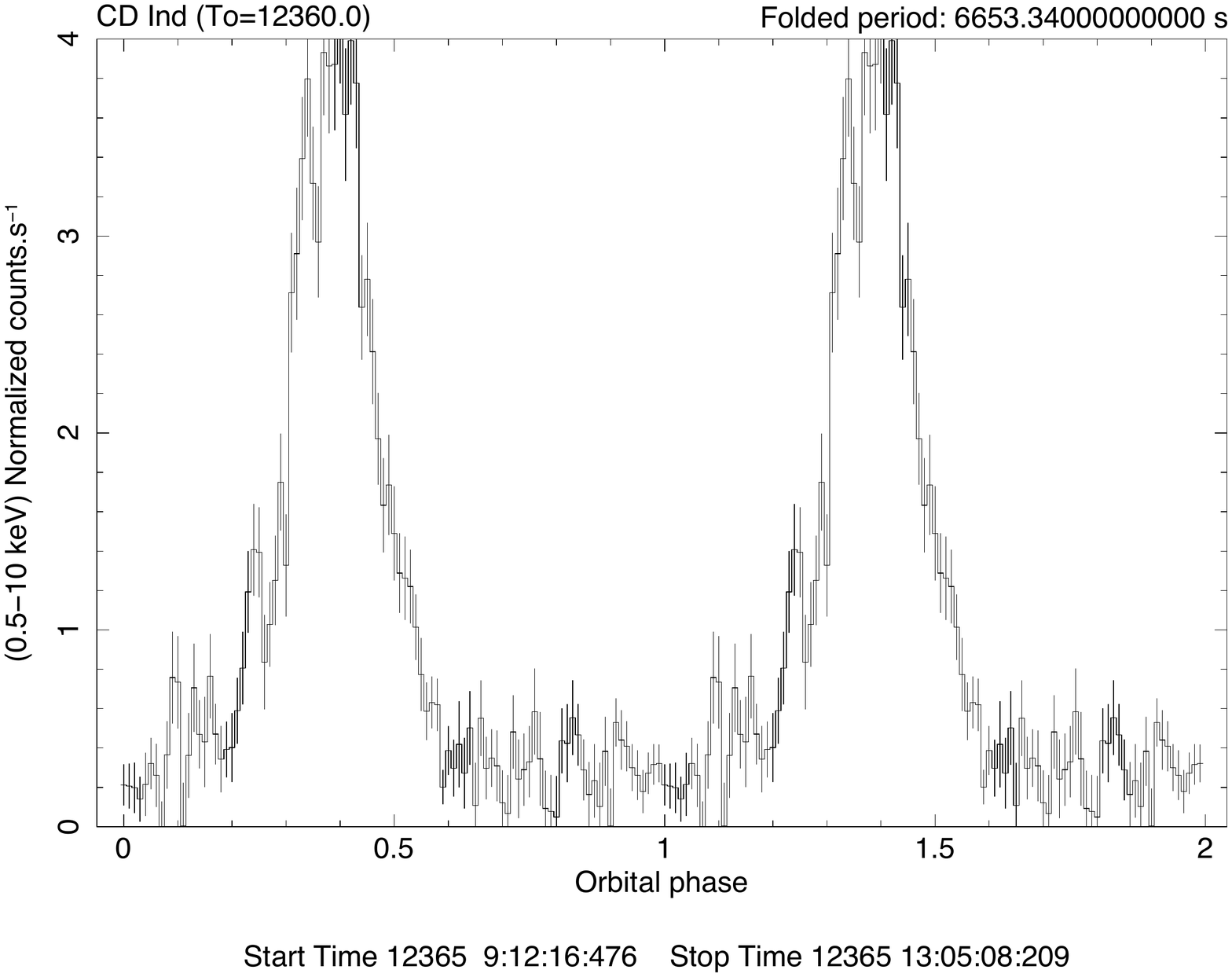}}  \\      
       \resizebox{58mm}{!}{\includegraphics*[trim=20 80 50 0]{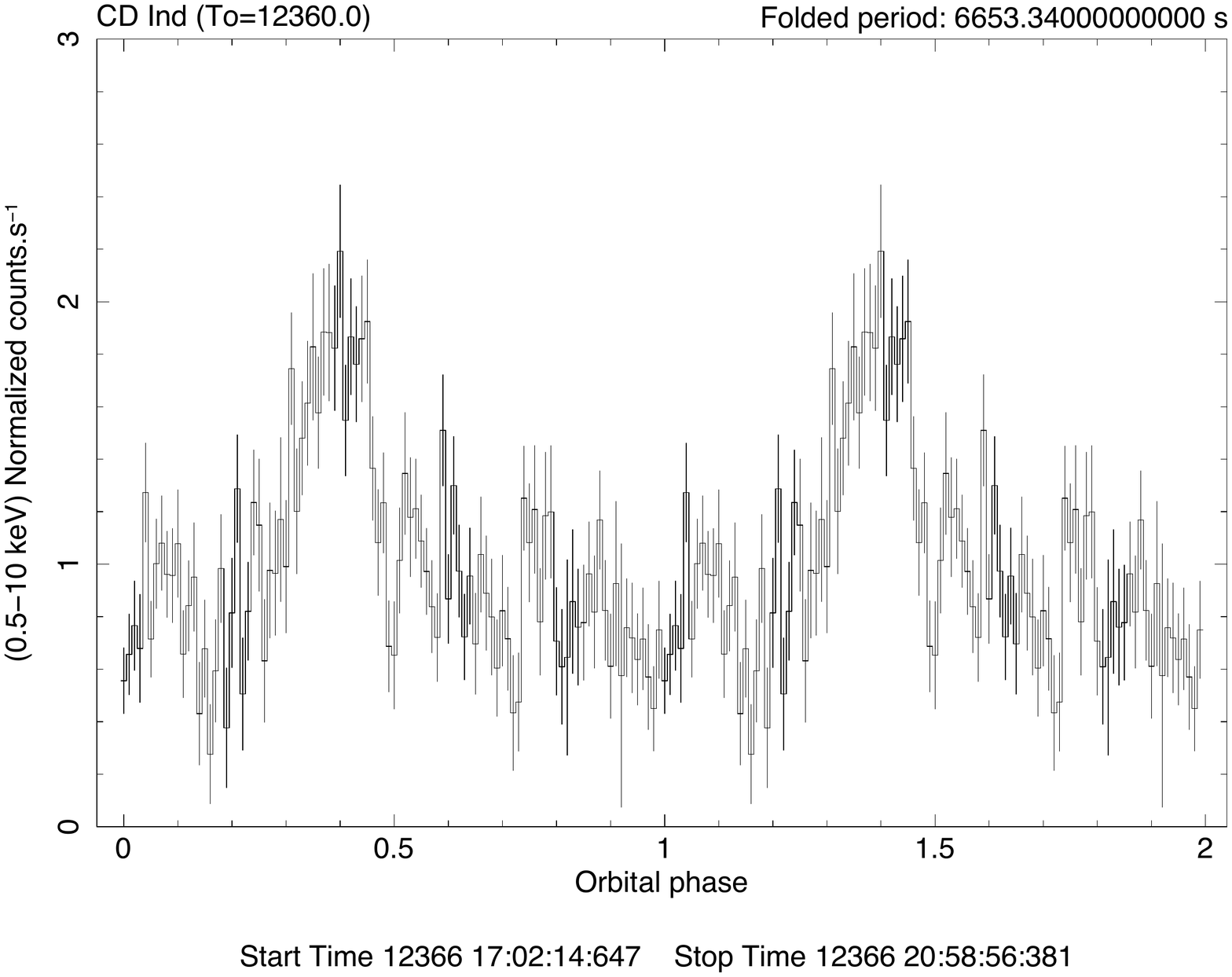}} \\  
   \end{tabular}
    \caption{The EPIC-PN (0.5-10) keV normalized X-ray light curve of the polar CD Ind with same caption as in Fig.~\ref{lightcurves_1}. Observations are shown (left to right, top to bottom) in the same chronological order as in Table~\ref{log_obs}.}
    \label{lightcurves_CDInd}
  \end{center}
\end{figure*}
}
%______________________________________________________________

\subsection{Ephemerides}

Table~\ref{ephemeris} gives the ephemeris used for each source. When the source ephemerides were already updated using the same XMM observation, we used the available ephemeris. For three sources not yet published (HS Cam, UZ For, and BL Hyi),  the XMM light curve can be used to update the ephemeris. \\  
For HS Cam, a mid-eclipse time can be computed at the time of the XMM observation, and when combining with data obtained by
\citetads{1997A&A...328..571T}, %Tovmassian97
a refined ephemeris can be derived as  
%-----------
\begin{equation}
\rm T_{ecl}= HJD\,2452925.94753(3) + 0.06820748(17)\,E
.\end{equation}
%-----------
The uncertainty in the period determination includes possible aliasing owing to the inaccuracy of the original period value.\\
For UZ For, the XMM mid-eclipse time is determined as $\rm T_{ecl}$= HJD\,2452494.751768(44). A detailed ephemeris has been given by 
\citetads{2011MNRAS.416.2202P}, %Potter11
combining different sets of data mainly from optical observations. They include the eclipse time determined from the XMM optical monitor (OM) during the same observation. We note that the X-ray eclipse time is sightly but significantly different by 0.00044(9) d.\\
For BL Hyi, the quadratic ephemeris by 
\citetads{1999ASPC..157..149W} %Wolff99
 was used. The shape of the XMM light curve is comparable to previous X-ray observations by RXTE
 \citepads{1999ASPC..157..149W} %Wolff99
and ASCA/BeppoSAX
\citepads{1998A&A...334..857M}. %Matt98
The phase of the light curve is consistent with the ephemeris but the noisy light curve and the lack of a significant sharp feature for the phasing prevent deriving a more precise ephemeris. 

%
%
%__________________________________________________ TABLE 2
\begin{table*}
\caption[ ]{Polar ephemerides and system characteristics}
     \label{ephemeris}
\begin{flushleft}
\begin{tabular}{llllrrrr}
\hline
\hline
\multicolumn{1}{l}{Source} & \multicolumn{1}{l}{To (HJD)} & 
\multicolumn{1}{l}{Porb (d)} & \multicolumn{1}{l}{Ref.} & 
\multicolumn{1}{l}{B(MG)} & \multicolumn{1}{r}{M$_{wd}$\,(M$_{\odot})$}  &
\multicolumn{1}{r}{d(pc)} & \multicolumn{1}{r}{$\rm \dot{M}_{16}$} \\
\hline
\noalign{\smallskip}
AI Tri          &       2451439.0391(10)        &       0.19174566(9)                   &       (1)     &       38              &       1*              &       600     &       0.205   \\
AM Her  &       2446603.403(5)  &       0.12892704(1)                   &       (2)     &       14              &       0.77-0.97       &       85      &       0.271   \\
AN UMa  &       2443191.0255(24)        &       0.07975282(4)                   &       (3)     &       29-36   &       1               &$ >$120   &       0.021   \\
BL Hyi  &       2444884.2199(45)        &       0.0789149644(94)                &       (4)     &       12-23   &       1               &       132     &       0.176   \\
BY Cam  &       2453213.010(3)  &       0.137123(3)                     &       (5)     &       28-41   &       1*              &       190     &       0.124   \\
CD Ind  &       -                               &       0.07700625(-)                   &       (6)     &       9-13            &                       &$ >$250   &       0.145   \\
DP Leo  &       2448773.215071(18)      &       0.06236283691(70)               &       (7)     &       30-59   &       0.71            &       260     &       0.043   \\
EF Eri  &       2453716.61108(5)        &       0.05626586(80)          &       (8)     &       16-21   &       0.6             &       150     &       0.637   \\
EP Dra  &       2447681.72918(6)        &       0.072656259(5)          &       (9)     &       16              &       0.43            &       150     &       0.887   \\
EU Lyn  &       -                               &       0.1142(21)                      &       (10)    &                       &                       &               &                       \\
EV UMa  &       2448749.4421(5) &       0.05533838(26)          &       (11)    &       30-40   &       1               &       705     &       0.731   \\
GG Leo  &       2449488.023703(61)      &       0.055471850(46)         &       (12)    &       23              &       1.13            &$ >$100   &       0.023   \\
HS Cam  &       2452925.94753(3)        &       0.06820748(17)          &       (13)    &                       &       0.85            &$ >$100   &                       \\
HU Aqr  &       2449102.9200839 (6)     &       0.08682041087 (3)       &       (14)    &       36              &       0.61            &       180     &       1.523   \\
QS Tel  &       2448894.5568(15)        &       0.09718707(16)          &       (15)    &       47-75   &       1*              &       170     &       0.022   \\
RX J1007        &       2455215.96256(48)       &       0.144863923(36)         &       (16)    &       92              &       1*              &       700     &       0.102   \\
SDSS 2050       &       2453296.29816(6)        &       0.06542463(1)           &       (17)    &                       &                       &               &                       \\
UZ For  &       2453405.30086(3)        &       0.087865425(2)          &       (18)    &       53-75   &       1*              &       220     &       0.074   \\
V1309 Ori       &       2450339.4343(8) &       0.33261194(8)                   &       (19)    &       61              &       0.7             &       500     &       4.719   \\
V2301 Oph       &       2448071.02014(7)        &       0.078450008(10) &       (20)    &       7               &       1.05            &       150     &       0.073   \\
V347 Pav        &       2448475.2913(5) &       0.062557097(36)         &       (21)    &       15-20   &       1*              & $ >$150 &       0.095   \\
V834 Cen        &       2445048.9500(5) &       0.070497518(26)         &       (22)    &       23              &       0.66            &       116     &       0.140   \\
VV Pup  &       2427889.6474(-)         &       0.0697468256(-)                 &       (23)    &       31-54   &       0.73            &       146     &       0.410   \\
WW Hor  &       2451882.73354(5)        &       0.0801990403(9)         &       (24)    &       15              &       0.9-1.3 &       430     &       0.006   \\
\noalign{\smallskip}
\hline
\end{tabular}
\end{flushleft}
(1) 
\citetads{2010A&A...516A..76T}, %Traulsen2010
(2) 
\citetads{2005AJ....130.2852K}, %Kafka2005
(3) 
\citetads{1996A&A...306..199B}, %Bonnet-Bidaud1996), 
(4) 
\citetads{1999ASPC..157..149W}, %Wolff1999), 
(5) 
\citetads{2008CEJPh...6..385A}, %Andronov et al. (2008), 
(6) 
\citetads{1999MNRAS.303...96R}, %Ramsay et al. (1999), 
(7) 
\citetads{2002A&A...392..541S}, %Schwope et al. (2002), 
(8) 
\citetads{2006ApJ...652..709H}, %Howell et al. (2006), 
(9) 
\citetads{1997AN....318...25S}, %Schwope et al. (1997), 
(10) 
\citetads{2005ApJ...620..929H}, %Homer et al. (2005), 
(11) 
\citetads{1994MNRAS.270..650O}, %Osborne et al. (1994), 
(12) 
\citetads{1998A&A...331..262B}, %Burwitz et al. (1998), 
(13) this work, 
(14) 
\citetads{2009A&A...496..833S}, %Schwarz et al. (2009), 
(15) 
\citetads{1995A&A...293..764S}, %Schwope et al. (1995), 
(16) 
\citetads{2012A&A...546A.104T}, %Thomas et al. (2012), 
(17) 
\citetads{2006AJ....132.2743H}, %Homer et al. (2006), 
(18) 
\citetads{2011MNRAS.416.2202P}, %Potter et al.  (2011), 
(19) 
\citetads{2001A&A...374..588S}, %Staude et al. (2001), 
(20) 
\citetads{1994A&A...288..204B}, %Barwig et al. (1994), 
(21) 
\citetads{2004MNRAS.347...95R}, %Ramsay et al. (2004), 
(22) 
\citetads{1993A&A...267..103S}, %Schwope et al. (1993), 
(23) 
\citetads{1965CoKon..57....1W}, %Walker (1965), 
(24) 
\citetads{2002MNRAS.332..116P}, %Pandel et al. (2002). 
Distances are from 
\citetads{1999ASPC..157..180B}, %Barrett et al. (1999). 
White dwarf masses and magnetic fields are from 
\citetads{2012MNRAS.422.1601K} %Kalomeni et al. (2012),
where (*) marks an assumed value. B values for V347 Pav and WW Hor are respectively from 
\citetads{2000MNRAS.315..423P} %Potter et al. (2000) 
and 
\citetads{2008PASP..120.1171I}. %Imamura et al.(2008)
See text for $\rm \dot{M}_{16}$.
 \end{table*}
%______________________________________________________________
%

%
%__________________________________________________ TABLE 3
\begin{table*}
\caption[ ]{X-ray fast oscillations  (0.1-5) Hz detection limits for the source bright phases}
     \label{limit_bright}
\begin{flushleft}
\begin{tabular}{lllcrrrrrrrr}
\hline
\hline
\multicolumn{1}{c}{Source} & \multicolumn{1}{l}{Date} & 
\multicolumn{1}{l}{Bright} & \multicolumn{1}{c}{Eclipse} & 
 \multicolumn{1}{c}{Exp.} & \multicolumn{1}{c}{Rate} &
 \multicolumn{1}{r}{M-FFT}  & \multicolumn{1}{r}{$f_{max}$} & 
 \multicolumn{1}{r}{$P_{max}$}  & \multicolumn{1}{r}{$P_{exceed}$} & 
 \multicolumn{1}{r}{$P_{detect}$}  & \multicolumn{1}{r}{Limit} \\
\multicolumn{1}{c}{ } & \multicolumn{1}{c}{yy-mm-dd} & 
\multicolumn{1}{c}{phase} & \multicolumn{1}{c}{phase} &
\multicolumn{1}{c}{ks} & \multicolumn{1}{c}{(c/s)} & 
\multicolumn{1}{r}{ }  & \multicolumn{1}{r}{Hz} & 
\multicolumn{1}{r}{}  & \multicolumn{1}{r}{$2.6\,\sigma$} & 
\multicolumn{1}{r}{$2.6\,\sigma$}  & \multicolumn{1}{r}{\% rms} \\ \\
\hline
\noalign{\smallskip}
AM Her          &       05-07-19        &       0.70-1.30       &       -               &       6.69            &       13.62   &       66      &       0.54    &       2.93    &       1.47    &       3.18    &       7.4     \\
AM Her          &       05-07-25        &       0.70-1.30       &       -               &       3.80            &       14.75   &       38      &       3.55    &       3.06    &       1.32    &       3.62    &       9.6     \\
AM Her          &       05-07-27        &       0.70-1.30       &       -               &       6.09            &       17.22   &       60      &       4.81    &       2.84    &       1.45    &       3.24    &       9.7     \\
BL Hyi  (*)     &       02-12-16        &       0.70-1.20       &       -               &       13.67   &       0.12            &       137     &       3.15    &       2.51    &       1.62    &       2.78    &       72.4    \\
DP Leo          &       00-11-22        &       0.70-1.20       &       0.98-1.02       &       6.78            &       0.16            &       69      &       2.99    &       2.77    &       1.48    &       3.15    &       74.3    \\
EP Dra          &       02-10-18        &       0.75-1.35       &       0.95-1.05       &       7.80            &       0.88            &       83      &       0.58    &       2.85    &       1.53    &       3.03    &       27.8    \\
EV UMa          &       01-12-08        &       0.70-1.30       &       0.95-1.05       &       2.45            &       1.99            &       25      &       1.88    &       3.41    &       1.19    &       4.08    &       25.3    \\
HS Cam          &       03-10-13        &       0.61-1.16       &       0.84-1.04       &       5.82            &       3.51            &       62      &       1.35    &       2.70    &       1.46    &       3.22    &       13.8    \\
HU Aqr          &       02-05-16        &       0.61-1.17       &       0.96-1.04       &       17.43   &       0.77            &       142     &       4.92    &       2.50    &       1.63    &       2.77    &       37.0    \\
HU Aqr          &       03-05-20        &       0.61-1.17       &       0.96-1.04       &       10.36   &       0.81            &       103     &       2.02    &       2.63    &       1.57    &       2.91    &       27.6    \\
QS Tel          &       06-09-30        &       0.95-1.07       &       -               &       5.04            &       0.34            &       20      &       3.23    &       3.53    &       1.11    &       4.39    &       71.7    \\
SDSS 2050       &       04-10-18        &      0.80-1.40        &      0.97-1.02        &       9.83            &       0.96            &       62      &       3.51    &       2.79    &       1.46    &       3.22    &       32.8    \\
UZ For          &       02-08-08        &       0.65-1.13       &       0.97-1.03       &       11.28   &       0.42            &       116     &       1.76    &       2.57    &       1.59    &       2.86    &       36.2    \\
V1309 Ori               &       01-03-17        &       0.44-0.58       &       -               &       4.00            &       0.45            &       40      &       0.58    &       3.18    &       1.34    &       3.57    &       45.8    \\
V2301 Oph       &       04-09-06        &       0.77-1.27       &       0.97-1.04       &       6.50            &       7.01            &       67      &       3.92    &       2.85    &       1.48    &       3.17    &       13.0    \\
V347 Pav                &       02-03-16        &       0.10-0.50       &       -               &       2.36            &       2.21            &       24      &       3.93    &       3.31    &       1.17    &       4.13    &       28.9    \\
VV Pup (*)      &       07-10-20        &       0.75-1.15       &       -               &       19.15   &       0.74            &       193     &       3.37    &       2.49    &       1.68    &       2.65    &       28.6    \\
WW Hor          &       00-12-04        &       0.74-1.28       &       0.96-1.03       &       9.65            &       0.39            &       96      &       4.68    &       2.87    &       1.56    &       2.95    &       61.1    \\
\noalign{\smallskip}
\hline
\end{tabular}
\end{flushleft}
(*) source showing optical QPOs
\end{table*}
%
%__________________________________________________ TABLE 4
\begin{table*}
\caption[ ]{X-ray fast oscillations  (5-50) Hz detection limits from high resolution XMM-PN timing observations}
     \label{limit_hires}
\begin{flushleft}
\begin{tabular}{llrrrrrr}
\hline
\hline
\multicolumn{1}{c}{Source} & \multicolumn{1}{l}{Date} & 
 \multicolumn{1}{r}{M-FFT}  & \multicolumn{1}{r}{$f_{max}$} & 
 \multicolumn{1}{r}{$P_{max}$}  & \multicolumn{1}{r}{$P_{exceed}$} & 
 \multicolumn{1}{r}{$P_{detect}$}  & \multicolumn{1}{r}{Limit} \\
\multicolumn{1}{c}{ } & \multicolumn{1}{c}{yy-mm-dd} & 
\multicolumn{1}{r}{ }  & \multicolumn{1}{r}{Hz} & 
\multicolumn{1}{r}{}  & \multicolumn{1}{r}{$2.6\,\sigma$} & 
\multicolumn{1}{r}{$2.6\,\sigma$}  & \multicolumn{1}{r}{\% rms} \\ \\
\hline
\noalign{\smallskip}
AM Her  &       05-07-19        &       805     &       27.05   &       2.25    &       1.84    &       2.30    &       14.6    \\
AM Her  &       05-07-25        &       805     &       34.08   &       2.19    &       1.84    &       2.30    &       17.3    \\
AM Her  &       05-07-27        &       806     &       22.95   &       2.24    &       1.84    &       2.30    &       13.3    \\
BY Cam  &       03-08-30        &       946     &       14.94   &       2.17    &       1.85    &       2.28    &       13.9    \\
BY Cam  &       03-08-31        &       946     &       14.84   &       2.26    &       1.85    &       2.28    &       14.1    \\
BY Cam  &       03-09-01        &       1141    &       16.21   &       2.19    &       1.87    &       2.25    &       13.2    \\
BY Cam  &       03-09-02        &       1232    &       20.61   &       2.17    &       1.87    &       2.24    &       23.6    \\
BY Cam  &       03-09-04        &       1315    &       35.06   &       2.17    &       1.87    &       2.24    &       18.1    \\
BY Cam  &       03-09-05        &       1405    &       40.92   &       2.13    &       1.88    &       2.23    &       19.8    \\
BY Cam  &       03-10-13        &       946     &       29.88   &       2.21    &       1.85    &       2.28    &       19.5    \\
HU Aqr  &       03-05-20        &       1833    &       29.20   &       2.14    &       1.89    &       2.20    &       58.1    \\
V834 Cen (*)    &       07-01-30        &       4256    &       14.06   &       2.11    &       1.93    &       2.13    &       16.0    \\
VV Pup (*)      &       07-10-20        &       4679    &       33.20   &       2.09    &       1.93    &       2.12    &       56.9    \\
\noalign{\smallskip}
\hline
\end{tabular}
\end{flushleft}
(*) source showing optical QPOs
\end{table*}
%______________________________________________________________

\section{Fast oscillation searches}
The search for fast oscillations  was first done on the extracted background-subtracted (0.5-10 keV) light curves binned into 0.1\,s intervals. To be able to monitor possible variable QPOs, individual FFTs were computed on consecutive 102.4 s segments and summed up to longer time intervals. The covered frequency range was therefore (0-5) Hz with a 9.766 mHz resolution.

%______________________________________________ FIGURE 4
%
  \begin{figure} 
  \centering
   %\includegraphics[width=8.9cm,angle=-0]{EVUma_01_lccor_ph200.pdf}
   % to exclude lower part of image
   %  \includegraphics*[width=8.9cm,angle=-0,trim=0 0 0 0]{V834Cen_fft_2D.jpg}
    \includegraphics*[width=8.9cm,angle=-0,trim=40 220 130 0]{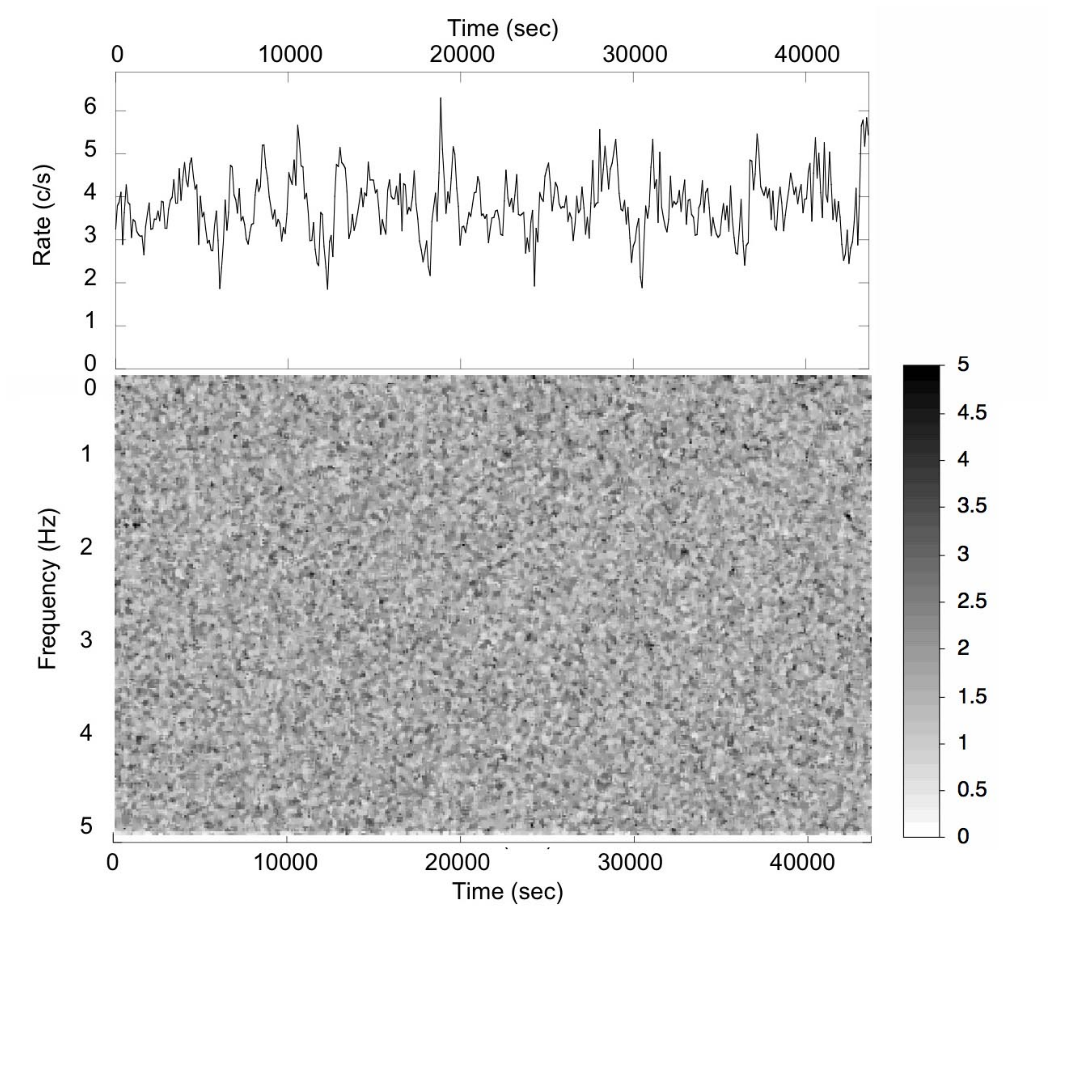}
      \caption{Variation in the power spectrum (bottom) across the V834 Cen (0.5-10 keV) XMM-PN light curve (top). Individual 102.4 s FFT with frequency range (0-5Hz) are shown vertically over the 43.6 ks observation with power value according to the vertical colour scale.  No significant excess is seen across the observation with a level well below the expected 99\% detection limit at $P_{detect}$= 21.7. 
                    }
       \label{power_2D}
   \end{figure}
%______________________________________________________________

Consecutive 102.4s FFT were built up into a 2D image to scan for a power excess in the time-frequency domain. To lower the noise, a mean FFT was also built by averaging the 102.4 s FFTs over the full observing interval as well as over typical  27.3 min intervals corresponding to 16 averaged FFTs.\\
Figure ~\ref{power_2D} shows the time variation of the FFT over the full 43.6 ks observation of the representative source V834 Cen. The trailed FFT reveals no significant excess (see full discussion below). Inspection of the averaged FFTs for the different sources in our list also reveals no conspicuous power excess in the (0-5Hz) frequency range.\\
To evaluate the detection limit for periodic phenomena, we follow the approach described by 
\citetads{1988tns..conf...27V}. %vdk88
The individual FFTs are normalized according to 
\citetads{1983ApJ...272..256L} %Leahy83
 with 
%-----------
\begin{equation}
%\rm P_{j}= \frac{2}{N_{\textem{ph}}}\, |a_{j}|^{2}
\rm P_{j}= \frac{2}{N_{ph}}\, |a_{j}|^{2}
\label{eq:norm}
\end{equation}
%-----------
where N$_{\rm ph}$ is the total number of photons per Fourier transform and $a_{j}$  the discrete Fourier amplitude at the frequency $\nu_{j}= j/\rm T$ with T the duration of the time series. 
With this convention, for a pure sinusoidal signal at a frequency $\nu_{j}$, a power P$_{\rm j}$ in the FFT will correspond to a signal of amplitude A$_{\rm j}$ 
%-----------
\begin{equation}
\rm A_{j} = { \left( \frac{2\,N_{ph}\,P_{j}}{N_{bin}^{2}}\right) }^{\frac{1}{2}}
\end{equation}
%-----------
where N$_{\rm bin}$ = T/$t_{bin}$ is the number of input points for a light curve with resolution $t_{bin}$. \\
When no signal is detected, an upper limit on the amplitude $A_{UL}$ can be derived by examining the  $P_{j}$ probability distribution. 
The limiting power for detection, $P_{UL}$,  is defined as $P_{UL} = P_{max} - P_{exceed}$, where $P_{max}$ is the highest observed power in the selected frequency range, and $P_{exceed}$  the excess power that  will correspond to a significance level of (1-$\delta$) in the power probability distribution, where $\delta$ is the low probability of being exceeded by a noise power. The amplitude upper limit is then computed as 
%-----------
\begin{equation}
\rm A_{UL} = { \left( \frac{2\,N_{ph}\,P_{UL}}{N_{bin}^{2}}\right) }^{\frac{1}{2}}
.\end{equation}
%-----------
For an individual FFT normalized as in equation \ref{eq:norm}, the power distribution is a $\chi{^2}$ distribution with n=2 degrees of freedom (dof). Owing to the additivity of the $\chi^2$ distribution, rebinning the FFT by averaging W frequency bins and further averaging M individual FFTs will result in a statistical  distribution of power according to a $\chi{^2}$ distribution with 2WM dof.\\

\subsection{Fast oscillation upper limits}
Table~\ref{log_obs} lists the statistical results extracted for the mean FFT that covers the full observation for our full sample. The table includes the number of averaged individual 102.4 s FFT (M-FFT), the maximum power observed ($P_{max}$) with the corresponding frequency ($f_{max}$), the statistical indicators $P_{exceed}$ (see above) and $P_{detect}$, evaluated here at a 99\% confidence level (or 2.6$\sigma$ equivalent for a normal distribution). Here, $P_{detect}$ is defined as the power level that only has a 1\% probability of being exceeded by the noise level (see eq. 3.7 in 
\citetads{1988tns..conf...27V}). %vdk88
\\ 
No significant peaks in the FFTs were found in any of the sources,  since the maximum power value $P_{max}$  is always lower than the detection limit at 2.6 $\sigma$, $P_{detect}$  (see Table~\ref{log_obs}).
The last column gives the rms upper limit (in percent) in the range (0.1-5) Hz, deduced from the corresponding FFT statistical parameters, $P_{max}$ and $P_{exceed}$ computed here for a 99\% significance. The limit amplitude has been also corrected for the binning effect and the finite size window effect assuming a typical signal frequency of 1 Hz (eq. 4.7 and 6.4 in 
\citetads{1988tns..conf...27V}). %vdk88
\\
The limits in the range (0.1- 2) Hz are not significantly different.\
The upper limits range from 6.8 to 71.1\% depending on the statistical quality of the observation directly linked to  the source counting rate. We stress that the upper limits derived here following a strict statistical analysis are in general more conservative than the ones previously reported in the literature.\\
\\
Because some sources show  a strong orbital modulation (see Figs.~\ref{lightcurves_1} to ~\ref{lightcurves_CDInd}), fast oscillations were also searched in intervals restricted to the bright phase only. Table~\ref{limit_bright}\, lists the phase limits used to define the bright phase, together with the eclipse range excluded in each case and the corresponding statistical results for the FFTs. No positive results are found. As the higher counting rate is somewhat compensated for by the lower statistic of the smaller time sample, the upper limits are not significantly different from the total observation.\\
To check for possible transitory fast oscillations, we also searched for significant peaks in FFTs summed in typical consecutive 27.3 min intervals for each source. No positive results were obtained among the 433 different intervals analysed. \\

%\subsection{High frequency (5-50 Hz) fast oscillation searches}
Because some of the sources of our sample were observed in the specific XMM timing mode that provides a higher (0.03 ms) resolution, fast oscillations were also searched at higher frequencies using accumulated 0.01s light curves. Table~\ref{limit_hires} lists the upper limits derived in the (5-50) Hz frequency range. Except for the two sources with lower statistics (HU Aqr and VV Pup), the typical limits for fast oscillations are around 10\% to 20\%.

\subsection{QPO upper limits from simulated data}
The upper limits for the amplitudes derived above are obtained by assuming
a pure sinusoidal modulation, which is present along the whole observation.
These statistical limits have been checked by simulated data in a fake observation.
We used the V834 Cen observation, corresponding to one of the longest runs and highest counting rates, to test our upper limits.
 To the 0.1s light curve, we have added an artificial sinusoidal signal of
increasing amplitude and determined the level at which a signal
is detectable by eye in the FFT. We obtained an eye detection at a power level of $\sim 2.4$ for a relative amplitude of $\sim 5\%$,
in accordance with the 99\% P$_{detect}$ level for this observation (see Table~\ref{log_obs}).
This value is significantly lower than the rms upper limit of 9\% given in Table~\ref{log_obs} that takes the different corrective factors described above into account, and gives therefore a more conservative limit.

To give more realistic limits for broad QPOs, we also simulated a fake
signal by adding the contribution of 101 sine curves with frequencies in the range [0.25-0.75 Hz] centred on
0.5 Hz with amplitudes distributed according to a  0.2 Hz FWHM Lorentzian curve, as typically observed for optical QPOs.
The peak amplitude at which signal is detected 
by eye gives a stronger constraint than the pure sinusoidal modulation, 
corresponding to a relative amplitude value of $\sim$2.5\%.
This is lower than in the pure sinusoidal case owing to the accumulative contribution of signals with frequencies spaced by 0.002Hz  into the finite width of the frequency bins 0.0097Hz of the power spectrum. 
Thus the observed amplitude limits reported in Tables 1 and 3 are in fact overestimated
for quasi-periodic oscillations. \rm

\section{Discussion}
\subsection{The QPO regime}
The data presented here offer the first systematic search for fast quasi-periodic oscillations in the X-ray flux of 
polars in a significant and representative sample of polars. Though fast QPOs are predicted in theoretical models 
to develop under specific physical conditions in the accretion column, none are observed here from 24 different sources.
An oscillating shock is so far the most promising process that can give birth to QPOs, and because a significant fraction of the 
gravitational energy, is released in the column in the X-ray range, significant QPOs are expected  in X-rays.  
The absence of detectable QPOs in our XMM sample covering a wide range of system parameters allows to derive 
interesting constraints on the models.\\
As underlined by different works
 (\citeads{1979ApJ...234L.117L}, %Lamb-Master 1979
\citeads{1999MNRAS.310..677S}, %Saxton 99
\citeads{2005ApJ...626..373M}, %Mignone 2005
see 
 \citeads{2000SSRv...93..611W} %Wu 00
 for a review), the stability of the accretion column in polars
is mainly governed by the balance between the bremsstrahlung and the cyclotron cooling of the post-shock region.
Numerical simulations  (see Paper II) show that, when a significant fraction of the energy is released in cyclotron, the 
shock oscillations are strongly damped, suppressing the QPOs.
A first order of the limit between the QPO and non-QPO regime can therefore be determined by examining the ratio
of the bremsstrahlung to cyclotron cooling time  $\epsilon_{s}= {t_{br}}/{t_{cy}}$ at the shock. 
The condition $\epsilon_{s} \ll 1$ will indicate a bremsstrahlung dominated  shock favouring QPOs, while $\epsilon_{s} > 1$ indicates
a cyclotron-dominated shock with an expected damping of the QPOs.\\

Following  
\citetads{1999PhDT........13S}, %Saxton 99b
$\epsilon_{s}$ can be expressed as a function of relevant parameters of the system, 
$\rm V_{ff}$ the free-fall velocity,  $\rho$ the flow density above the shock, B the magnetic field, and A the cylindrical column cross-section as 
%-----------
\begin{equation}
%\rm \epsilon_{s} = 1.640\,10^{-4} (A_{16})^{-\frac{17}{40}}
\rm \epsilon_{s} = 7.396\,10^{-5}\, (A_{16})^{-17/40}\,(B_{7})^{57/20}\,(\rho_{8})^{-37/20}\,(V_{8})^{4}
\end{equation}
%-----------
where $ \rm A_{16} = A/10^{16}\, cm^{2}$,  $ \rm B_{7} = B/10^{7}\, G$,  $ \rm \rho_{8} = \rho /10^{-8}\, g.cm^{-3}$, and 
$ \rm V_{8} = V_{ff}/10^{8}\, cm.s^{-1}$. Here a pure-hydrogen plasma is considered with a Gaunt factor of 1.0 and a typical ratio
of electron and ion partial pressures of 1.
Assuming the matter is captured far from the white dwarf,  $\rm V_{ff}$ can be expressed as $\rm (2GM/r)^{1/2}$ with  $\rho = (\rm \dot{M}/A.V_{ff})$,  yielding the dependency of $\epsilon_{s}$ on the system primary parameters:
%-----------
\begin{equation}
\rm \epsilon_{s} = 1.086 \, (A_{16})^{57/40}\,(B_{7})^{57/20}\,( \dot{M}_{16})^{-37/20}\,(M_{o})^{117/40}\,(R_{9})^{-117/40}
 \label{equation6}
\end{equation} 
%-----------
where $ \rm \dot{M}_{16} = \dot{M}/10^{16}\, g.s^{-1}$,  $ \rm M_{o} = M/M_{\odot}$, and $ \rm R_{9}= R/10^{9}$\,cm 
with M and R the mass and radius of the white dwarf.\\
%Imposing a bremsstrahlung dominated shock, the condition for QPOs can be set as $\rm \epsilon_{s} < 1$.
To insure a bremsstrahlung-dominated shock with significant QPOs will require $\rm \epsilon_{s} < 1$. Actually, simulations show that 
QPOs are already significantly damped at a lower value of $\rm \epsilon_{s} < 0.5$ (see Paper II).
We note that the condition $\epsilon_{s} = {t_{br}}/{t_{cy}} < 1$ is set at the shock but as the cooling efficiency of the bremsstrahlung increases
toward the white dwarf surface with respect to the cyclotron one (see Paper II), this insures that the condition is valid through the whole column.
From this expression, it follows that QPOs are favoured primarily in the case of sources showing low B and high accretion rate,
as well as for a small accretion column section and a low mass white dwarf.\\
The characteristics of the polar systems in our sample are given in Table~\ref{ephemeris}. The estimated white dwarf masses and
magnetic fields were taken from the recent review of polar physical parameters from the literature by 
\citetads{2012MNRAS.422.1601K} %Kalomeni
 and references therein.
For the magnetic field, we selected the values derived from cyclotron features and checked with the reviews by 
%Beuermann (1998)
\citetads{1998PHEAA..157..100B} %Beuermann (2000) 
and 
\citetads{2000NewAR..44...69W}, %Wickramasinghe00
 allowing for a range in case of multiple poles.  Values for V347 Pav and WW Hor are taken respectively from 
 \citetads{2000MNRAS.315..423P} %Potter00
 and 
  \citetads{2008PASP..120.1171I}. %Imamura08
\\
The net accretion rate in polars, $ \rm \dot{M}$, is not known with accuracy, but a lower limit can be obtained by considering that the total X-ray luminosity corresponds to the gravitational energy of the accreted matter 
$ \rm \dot{M} =  4 \pi  d^{2} F_{x} \,R/ G M$ where d is the source distance and F$\rm _x$ the total X-ray flux so that  \\
$ \rm \dot{M}_{16} =  8.97 \times 10^{-3} \, d_{100}^{2} \, F_{x11}\,R_{9}/M_{o}$  
where $\rm d_{100}$= d/100 pc and $\rm F_{x11}= F_{x}/10^{-11} \, erg.s^{-1}.cm^{-2}$.\\
We stress that, in systems with high magnetic field, this can only be considered as a lower limit since part of the gravitational energy is also lost in this case through the cyclotron emission in the visible and infrared. From numerical simulations, it can be shown that typical cyclotron contribution to the overall emission can reach $\sim 30\%$ for B=30 MG (see Paper II). \\
The soft+hard X-ray flux in our sample was taken from the study of the energy balance in polars by 
\citetads{2004MNRAS.347..497R}, %Ramsay04
where the bolometric unabsorbed flux is derived from spectral fits to XMM and ROSAT observations. Distances 
were taken from the compilation by 
\citetads{1999ASPC..157..180B}, %Barrett99
 and the white dwarf mass-radius relation by 
 \citetads{1972ApJ...175..417N} %Nauenberg72
  was used. \\

\begin{figure} 
  \centering
   %\includegraphics[width=8.9cm,angle=-0]{EVUma_01_lccor_ph200.pdf}
   % to exclude lower part of image
     \includegraphics*[width=8.9cm,angle=-0,trim=170 50 220 110]{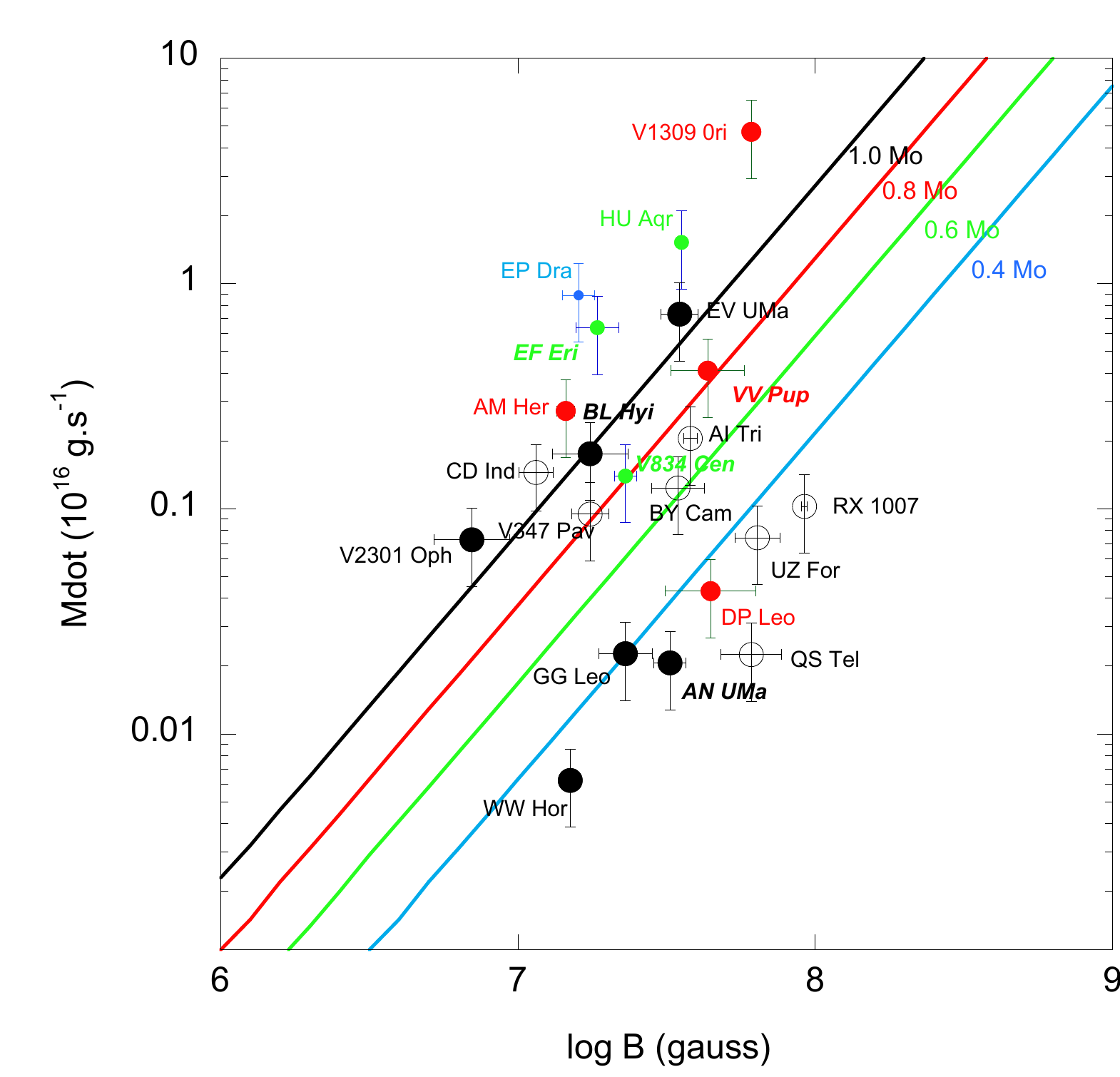}
      \caption{(B-$ \rm \dot{M}$) diagram for the polars with values from Table~\ref{ephemeris}.   
      The sources are shown by filled symbols with increasing sizes and colours corresponding to their mass in the range 0.3-0.5 (blue),
      0.5-0.7 (green), 0.7-0.9 (red), and 0.9-1.2 $\rm M_{\odot}$ (black). Sources with no mass determination are shown with open symbols and assumed at 1 $\rm M_{\odot}$. The 5 polars with known optical QPOs are shown in italics.
      The lines mark the limit of the QPO regime corresponding to $\rm \epsilon_{s}$= 1 (see text), assuming a representative column cross-section of $10^{14}\, \rm cm^{2}$ and for different white dwarf mass values shown by labels and with the above colour convention. QPOs are expected above the associated line.
                    }
       \label{B-Mdot}
   \end{figure}

Figure~\ref{B-Mdot} shows the (B-$ \rm \dot{M}$) diagram for the sample of our sources with available X-ray luminosities. 
The sources are shown according to their mass. When no mass determination was available, 
the mass was assumed to be 1$\rm M_{\odot}$. 
The polars of our list cover roughly a decade in B with values from $\sim$ 10 to 100 MG and more than two decades in accretion rate from $\sim$ 0.01 to 5 $\times 10^{16} \rm g.s^{-1}$.\\
Also shown in Fig.~\ref{B-Mdot} is the B-$ \rm \dot{M}$ relation corresponding to $\rm \epsilon_{s}$=1 as derived from Eq.~\ref{equation6}, for different white dwarf masses. 
We note that the dependency of $ \rm \dot{M}$ with B derived here from
\citetads{1999PhDT........13S} %Saxton 99b 
is significantly different from the one shown by Lamb \& Master (1979), who used different prescriptions to compute the bremsstrahlung and cyclotron cooling times.\\
 
The diagonal lines of Fig.~\ref{B-Mdot} are shown for a typical column cross-section of $10^{14}\, \rm cm^{2}$, which would correspond to
a column fractional area of $f = (0.7 - 2.7) \times 10^{-5}$ for a range of white dwarf mass (0.4--1.0) M$_{\odot}$. 
We chose to display in Fig.~\ref{B-Mdot} the absolute accretion rate $\rm  \dot{M}$ rather than the specific accretion rate $\rm \dot{m} = \dot{M} /A$ sometimes used, since the additional parameter A is not well known for polars. Different values of A will shift the corresponding curves according to Eq.~\ref{equation6}.\\
In Fig.~\ref{B-Mdot}, the bremsstrahlung regime favorable for QPOs is found for each source above the line corresponding to its mass.
From our sample, according to their accretion rate and magnetic field, the sources most strongly dominated by the bremsstrahlung cooling are the lowest mass systems EP Dra, HU Aqr, EF Eri, and more marginally V834 Cen. Amongst the systems with intermediate masses, V1309 Ori, AM Her, and marginally VV Pup will satisfy the criterion, while among high mass systems only EV UMa, V2301 Oph, and marginally BL Hyi  will be in the acceptable range. All these systems should be the best candidates for the presence of QPOs. Remarkably, out of the five systems showing optical QPOs, four are over or close to the QPO criterion. Only one, AN UMa, appears to be far from the bremsstrahlung regime following this criterion. According to its magnetic field, this source could be dominated by the bremsstrahlung cooling only with either a very unlikely column cross-section smaller than $10^{14}\, \rm cm^{2}$ or a mass  lower than 0.4 M$_{\odot}$. amongst the lowest observed for polars. We note, however, that only a lower limit for the distance exists for this source, so the accretion rate might be significantly underestimated (see Sect 5.4.3 for further discussion).
The other sources with mass determination that are probably not dominated by bremsstrahlung cooling are systems like WW Hor, DP Leo, and GG Leo, which are therefore not expected to show significant QPOs.
Several systems -- EU Lyn, HS Cam, and SDSS 2050 -- for which no reliable accretion luminosity exists, are not shown in Fig.~\ref{B-Mdot}. \\

%______________________________________________ FIGURE 5
%
%______________________________________________________________

\subsection{The model predictions}
In the context of our project POLAR to reproduce the physical conditions of an accretion column in the laboratory through adapted scaling laws
\citepads{2012HEDP....8....1F}, %Falize12
we have developed a  2D hydrodynamical code, HADES
\citepads{2011Ap&SS.336..175M}, %Michaut11
to perform numerical simulations of the accretion column evolution. Full description of the code and detailed predictions and results from variable sources parameters are given in Paper II (Busschaert et al. 2015). Here, we discuss the main results concerning the expected QPO amplitude and frequency and the influence of the magnetic field.
The model solves the hydrodynamics equations in Eulerian coordinates and includes radiative losses via a cooling function appearing as a source term.

\begin{figure} 
  \centering
    % to exclude lower part of image
     \includegraphics*[width=8.9cm,angle=-0,trim=190 55 200 125]{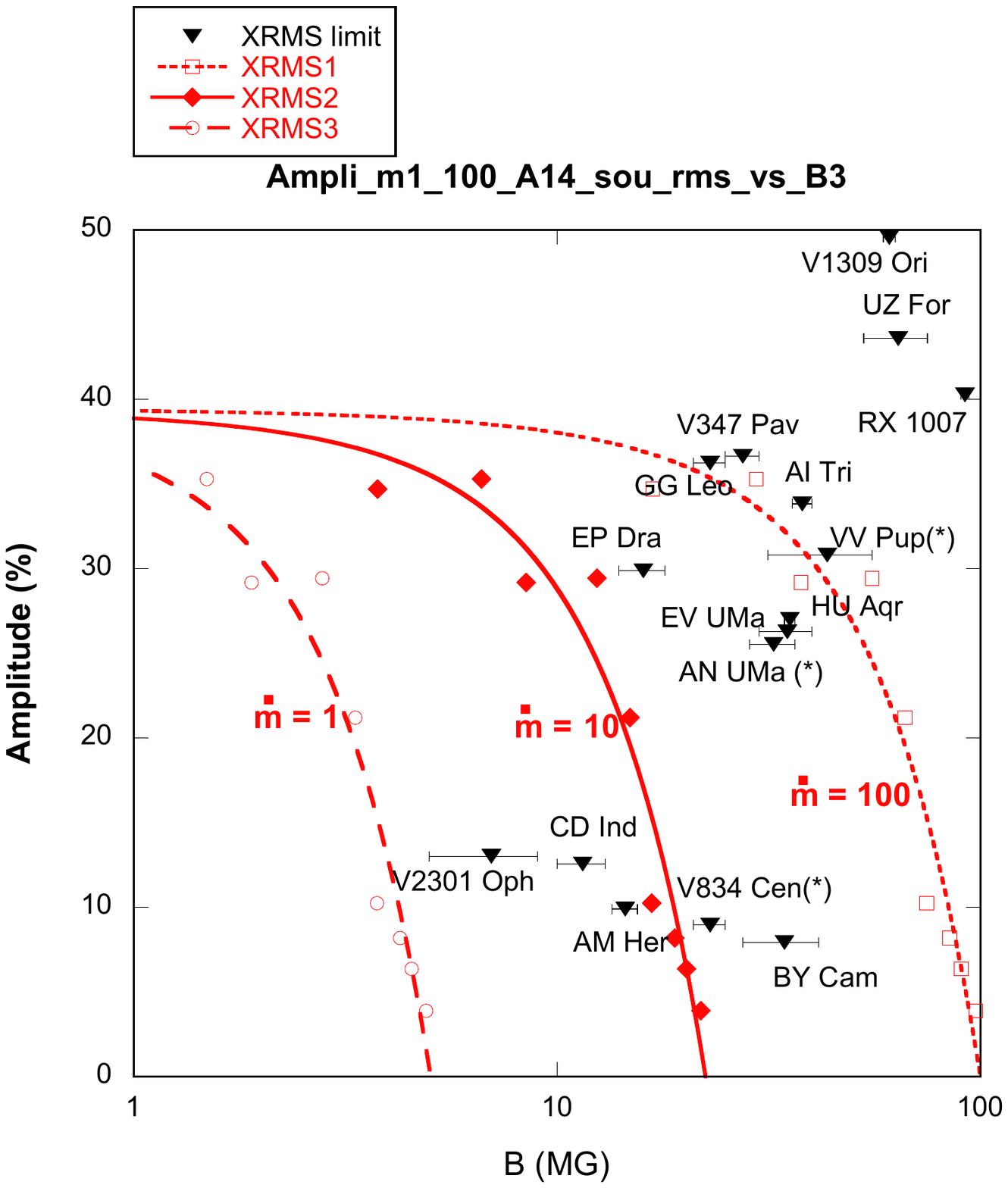}
      \caption{Upper limits on polar (0.1-5 Hz) oscillation amplitudes in  the  X-ray (0.5-10 keV) flux as a function of magnetic field strength.       Only sources with upper limits ($<50\%$) are shown. Sources with known optical QPOs are denoted by an asterisk. Also displayed are the QPO amplitudes predicted by representative numerical simulations of the shock instability (shown by symbols). Curves are polynomial fits through the individual measurements and are shown  for a typical WD mass of 0.8 M$_{\odot}$, a column cross-section of  $10^{14}\, \rm cm^{2}$ and for different values of the specific accretion rate 
      (dashed curve $\dot{m}$ = 1.0, full curve $\dot{m}$ = 10 and dotted curve $\dot{m}$ = 100 g.cm$^{-2}$.s$^{-1}$).   
                    }
       \label{QPO_model_mdot}
   \end{figure}

\begin{figure} 
  \centering
   % to exclude lower part of image
     \includegraphics*[width=8.9cm,angle=-0,trim=170 50 220 130]{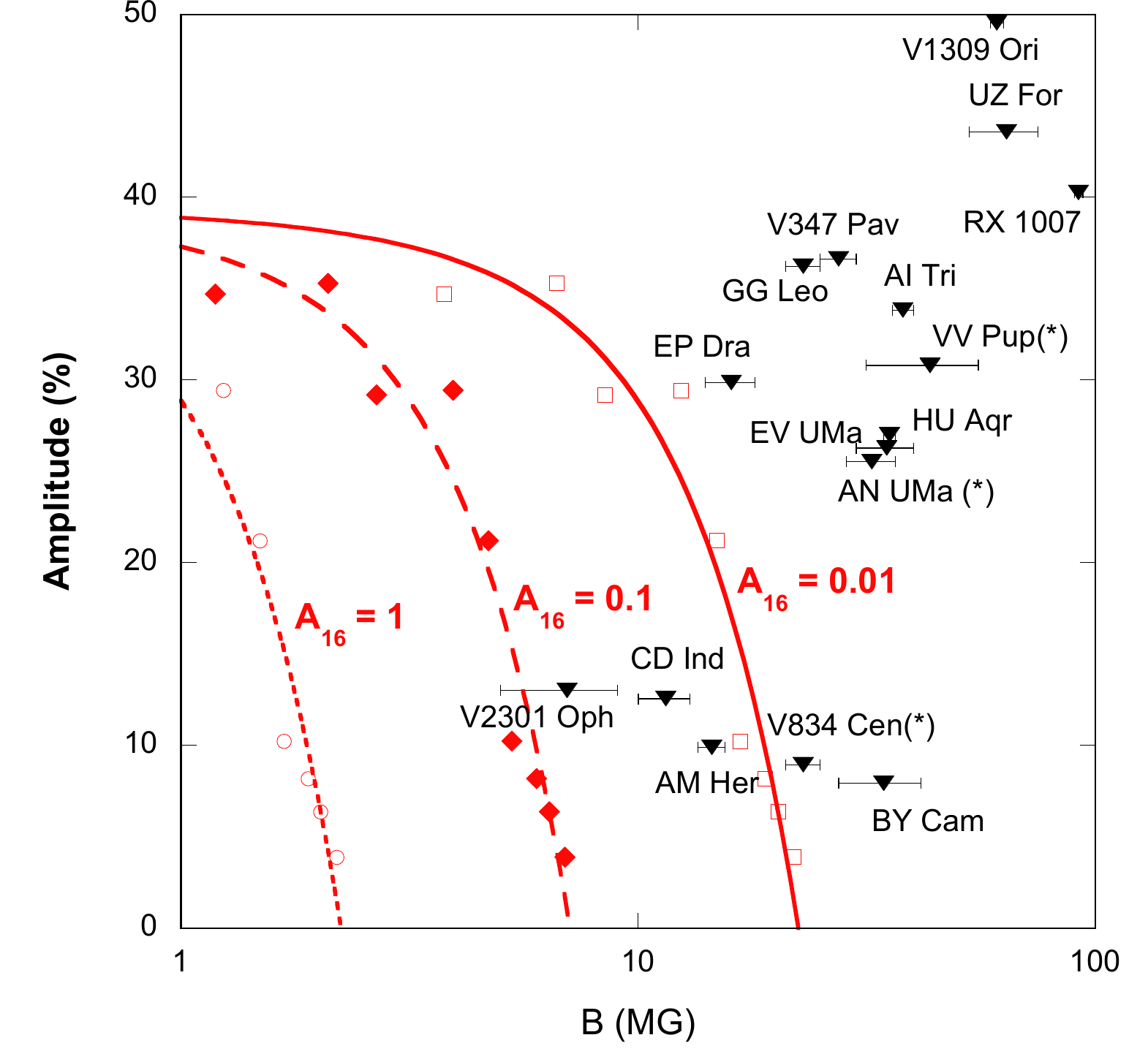}
      \caption{Upper limits on polar (0.1-5 Hz) oscillation amplitudes in  the  X-ray (0.5-10 keV) flux for the same representative model as in Fig.~\ref{QPO_model_mdot} ( $\dot{m}$=10,  M = 0.8 M$_{\odot}$) and for different values of the accretion column cross-section $ \rm A_{16} = 0.01$ (full line), $ \rm A_{16} = 0.1$  (dashed line), and $ \rm A_{16} = 1.0$ (dotted line), keeping the total accretion rate constant at $ \rm \dot{M}_{16}$ =0.1.
                    }
       \label{QPO_model_A}
   \end{figure}

\begin{figure} 
  \centering
   % to exclude lower part of image
     \includegraphics*[width=8.9cm,angle=-0,trim=170 50 220 130]{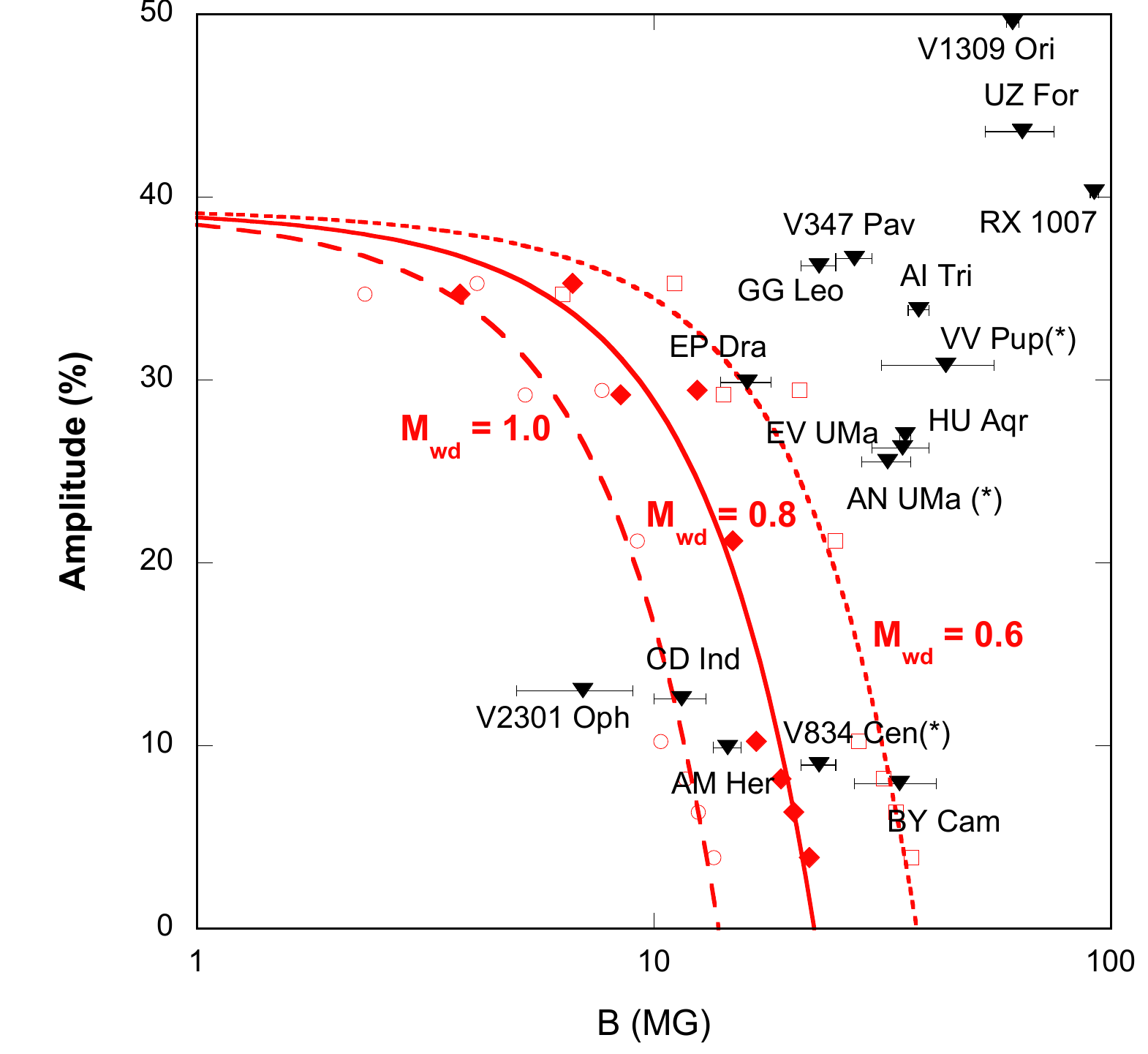}
      \caption{Upper limits on polar (0.1-5 Hz) oscillation amplitudes in  the  X-ray (0.5-10 keV) flux for the same representative model as in Fig.~\ref{QPO_model_mdot} ($\dot{m}$=10 and $ \rm A_{16} = 0.01$) and for different values of the WD mass
      ($ \rm M_{wd}= 1.0$\,M$_{\odot}$  dashed line,  0.8\,M$_{\odot}$ full line,  0.6\,M$_{\odot}$  dotted line).   
                    }
       \label{QPO_model_Mwd}
   \end{figure}

We use this model to predict the expected QPOs in different regimes.
As a first order, we used a plane-parallel geometry to reproduce the shock variations under constant homogeneous accretion. 
The numerical simulations demonstrate that under various conditions the shock front oscillates with time as a consequence of cooling instability, and this results in spontaneous quasi-periodic oscillations. This compares well with previous studies
(\citeads{1982ApJ...261..543C}, %Chevalier 1982
\citeads{2005ApJ...626..373M}, %Mignone 2005
\citeads{1984ApJ...276..667I}). %Imamura 1984
As shown above, with the given hypothesis on the cooling processes, the post-shock region is governed by four parameters: the magnetic field B, the accretion rate  $\rm  \dot{M}$, the accretion column cross-section A, and the white dwarf mass M (with radius R that can be related by an approximate  M-R relation 
 \citepads{1972ApJ...175..417N}). %Nauenberg72
\\
We produced a grid of models varying the intensity of the magnetic field in our simulations to predict the QPO amplitudes in the range of field strength observed for polars. In each run, the emission was integrated through the accretion column to compute the total bremsstrahlung luminosity in the (0.5-10 keV) energy range to compare with observations. The total  cyclotron emission was also extracted.  Temporal variations were studied that exclude the onset of the shock (usually a few seconds) over a range of time sufficient to insure that the oscillations were in a non-transitory stabilised regime. The oscillation temporal characteristics were then extracted by standard Fourier techniques to provide the typical amplitudes and frequencies present in the X-ray and cyclotron flux. When a significant power was split into different frequencies, the quadratic sum of the amplitudes was computed
 (see Paper II for full details). \\
Representative results are shown in Fig.~\ref{QPO_model_mdot} as a function of the magnetic field strength. The models considered here are for a typical mass M=0.8 M$_{\odot}$ and a column cross-section of $10^{14}\, \rm cm^{2}$. For this cross-section, the specific accretion rate range $\dot{m}$ = (1-100) g.cm$^{-2}$.s$^{-1}$  corresponds to the observed  $\dot{M}$ = $(0.01-1)\, 10^{16}$\ g.s$^{-1}$ total accretion rate range (see Fig.~\ref{B-Mdot}). 
Typical oscillation amplitudes up to $\sim 40\%$ are predicted with the amplitude decreasing very steeply with B owing to the cyclotron damping. The cut-off of the amplitude varies significantly with the specific accretion rate, because it is shifted to higher B values with increasing accretion rate as the result of a more significant bremsstrahlung contribution. For extremely high values of the specific accretion rate, sources with magnetic field up to 70 MG can still show significant QPOs.\\
As noted above, the detailed models will also depend significantly  on the column cross-section A and the white dwarf (WD) mass M. To illustrate the influence of these parameters, we show in Figs.~\ref{QPO_model_A}  and~\ref{QPO_model_Mwd}, the predicted amplitudes for the $\dot{m}$=10 reference model, varying A and M in the expected range for polars. The location of the curves in Fig.~\ref{QPO_model_mdot} are therefore only indicative for a specific model. To predict the exact amplitude for a given source will of course require taking the specific parameters into account.
To achieve this, we produced a grid of values and used interpolations according to the dependency in the different parameters given in Formula~\ref{equation6} (see Paper II for the full discussion of amplitudes and frequencies). The accretion column cross-section is the least well known parameter for polars, which is usually not easily determined by other means. The QPO limits therefore provide interesting constraints on this parameter.\\
The model predictions can be compared with the derived QPO upper limits for the polars in our observational sample that are also shown in Figs.~\ref{QPO_model_mdot} to~\ref{QPO_model_Mwd}, where parameters have to be adapted for each individual source.  
Notably, only $\sim 30\%  $ of the polars  have a magnetic field lower than B=20 MG that could insure QPOs with a reasonable accretion rate. Above B=20 MG, only a high accretion rate could maintain QPOs, a condition usually not satisfied by most of the individual sources. 
Among the low-field polars, three sources, AM Her, CD Ind, and V2301 Oph also have among the lowest upper limits ($<\sim 10\%$)  for X-ray QPOs from our sample and yield interesting constraints. We discuss them separately below, together with the polars showing optical QPOs.
%
%______________________________________________ FIGURE 6
%

%______________________________________________ FIGURE 7
%
%______________________________________________________________
%

%______________________________________________ FIGURE 8
%
%______________________________________________________________

\subsection{X-ray QPOs}
\subsubsection{AM Her}
The magnetic field for AM Her is reasonably well constrained with a value B=14-15 MG. Its distance (d=$85\pm5$\,pc) has also been well defined, including parallax measurements 
 \citepads{2012NewA...17..154G}. %B-Gariety12
The white dwarf mass is, however, still quite uncertain with a possible range of 0.77-0.97 M$_{\odot}$,  and a mean value of M=0.87 M$_{\odot}$ will be assumed. With these values and considering
$ \rm \dot{M}_{16}$ =0.27 from the overall mean X-ray luminosity, QPOs of amplitude $\sim 32\%$ are expected from interpolation through our grid of simulations when assuming a typical column cross-section of $10^{14}\, \rm cm^{2}$.
Conversely, our observed QPO upper limit of $\sim 10\%$ will impose a cross-section higher than $\sim 4.5\,10^{14}\, \rm cm^{2}$, which corresponds to a specific accretion rate lower than $\dot{m}  \lesssim 6$ g.cm$^{-2}$.s$^{-1}$. Allowing for the possible range in mass of (0.77-0.97 M$_{\odot}$) imposes A $> (2.8-6.3)\,10^{14} \rm cm^{2}$ and  an accretion rate $\dot{m} < (4.5-9)$ g.cm$^{-2}$.s$^{-1}$.
Thus for AM Her, the absence of QPOs at a significant level over 10\% can only be explained if the specific accretion rate is kept below this value. 

\subsubsection{CD Ind}
The B value has been determined by 
\citetads{1997A&A...326..195S} %Schwope97
 from a fit to the cyclotron spectrum with a rather narrow range of B=9-13 MG. No reliable mass estimation is available yet, and only a lower limit of the distance (250 pc) is derived from an assumed secondary spectral type.  With this lower limit and  the X-ray total luminosity quoted by 
% \citetads{2004ASPC..315..106R}, %Ramsay04
 \citetads{2004MNRAS.347..497R}, %Ramsay04
a minimum mass accretion rate of $ \rm \dot{M}_{16}$ =0.145 is implied if  a 1 M$_{\odot}$ white dwarf is considered. 
In this case, the predicted QPO amplitude from the simulations are $\sim 31\%$ for A= $10^{14}\, \rm cm^{2}$. Our more constraining upper limit of $\sim 13\%$ will impose A larger than $\gtrsim\,1.4\,10^{14} \rm cm^{2}$ , which will imply a specific accretion rate  $\dot{m}  \lesssim 10$  g.cm$^{-2}$.s$^{-1}$.

\subsubsection{V2301 Oph}
V2301 Oph is the weakest field polar with a value B= 7 MG
 \citepads{1995MNRAS.273...17F}, %Ferrario95
making therefore the system a prime candidate for QPOs. The white dwarf mass is not strictly determined, but an upper limit of M $< 1.2$ M$_{\odot}$ is found from the eclipse modelling and it is suggested that an acceptable range is M = 1.0-1.1 M$_{\odot}$.
 \citepads{2007MNRAS.379.1209R}. %Ramsay07
Using a distance of 150 pc, an unabsorbed bolometric X-ray luminosity of L$\rm _x$= $2 \times 10^{32}$ erg.s$^{-1}$  is derived from the spectral analysis of the bright phase during the XMM observations by 
 \citetads{2007MNRAS.379.1209R}. %Ramsay07
The same authors also reported an upper limit of (19-24) \% for QPOs in the restricted 2-10 keV range, derived by simply adding a fake signal without giving any significance level. An 99\% upper limit of  4.1\% to 7.6\% was also reported from RXTE data obtained in 1997 in a similar energy range by 
 \citetads{1999ApJ...515..404S}. %Steiman-Cameron99
\\
Our limit of 13\% is similar but for a wider (0.5-10) keV range and for a 99\% upper limit obtained with rigorous statistical criteria.
From our simulations, the expected QPO amplitude is 21\% for $\dot{M}_{16}$ = 0.07 as derived from the X-ray luminosity and assuming A= $10^{14}\, \rm cm^{2}$. This is slightly over our observed limit and our non-detection will therefore impose a higher cross section with  A $\ge 1.3\, 10^{14}\, \rm cm^{2}$ (corresponding to $\dot{m}  \lesssim 5.4$  g.cm$^{-2}$.s$^{-1}$).
Despite its low field, V2301 Oph is therefore still below detection level mainly because of its low accretion rate. 

\subsection{Optical  QPO sources}
\begin{figure} 
  \centering
   % to exclude lower part of image
     \includegraphics*[width=8.9cm,angle=-0,trim=170 50 220 130]{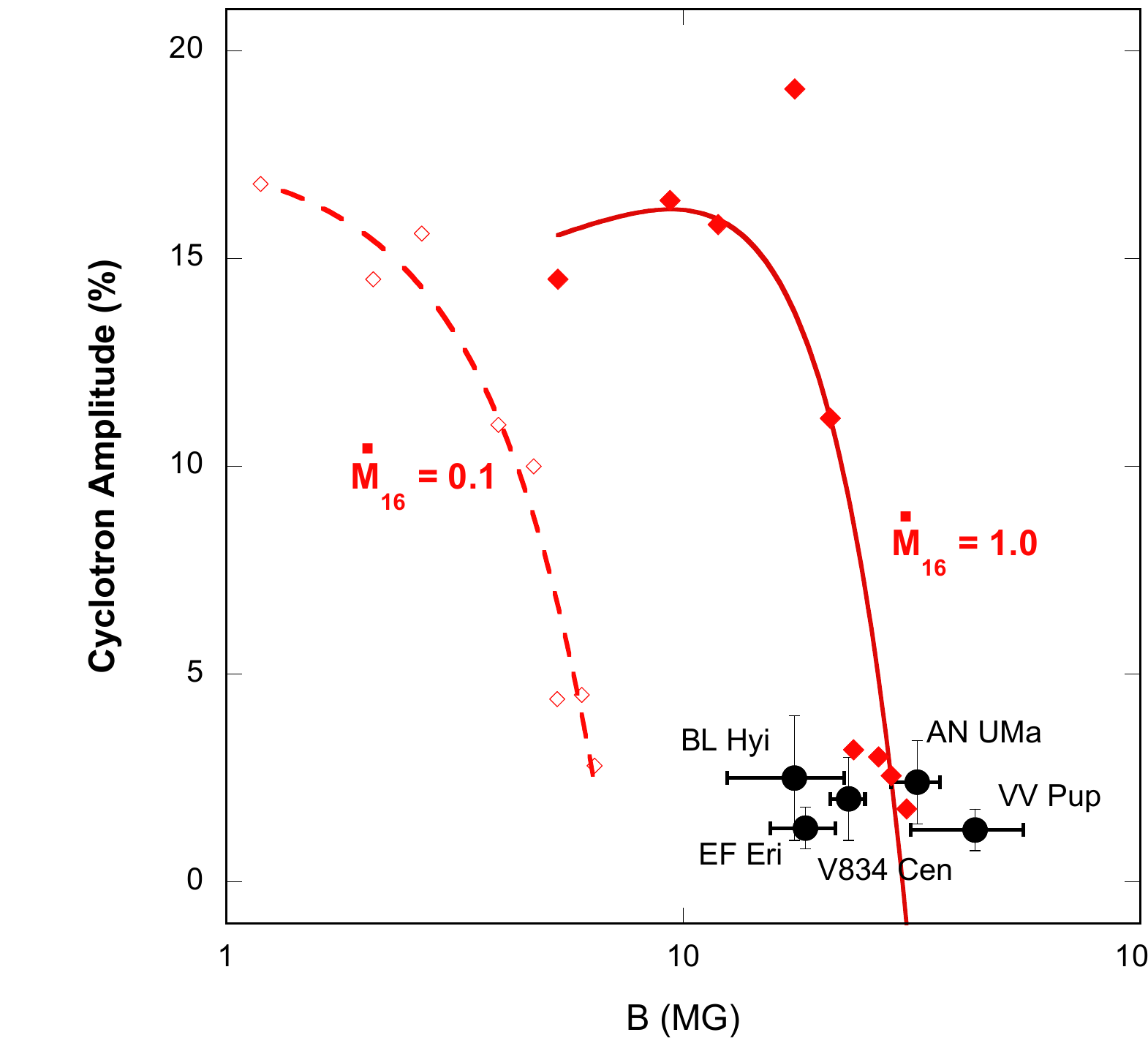}
      \caption{Measured optical oscillation amplitudes detected among polars as a function of magnetic field strength. 
   Also displayed are the QPO amplitudes predicted by numerical simulations of the shock instability (see text) onto a 0.8 M$_{\odot}$ WD for different values of the accretion rate 
      ($\dot{M}_{16}$ = 1.0 red line curve,  $\dot{M}_{16}$ = 0.1 red dotted curve) and assuming a column cross-section of  $10^{14}\, \rm cm^{2}$ .   
                    }
       \label{QPO_model_cyc}
   \end{figure}

From our numerical simulations, both the X-ray and cyclotron luminosity variations can be followed through the cycle of the shock instability, and the expected X-ray and optical QPO amplitudes can be computed (see Paper II).  Figure~\ref{QPO_model_cyc} shows the predicted cyclotron amplitudes for the same representative numerical model as in Fig.~\ref{QPO_model_mdot}, with the sources with measured optical QPO amplitudes also shown. We note here that the amplitudes from the simulations are relative to the cyclotron total flux, while the measured amplitudes are given with respect to the source total optical flux so that a dilution factor may have to be considered and is discussed below.
%
%______________________________________________ FIGURE 9
%
%______________________________________________________________
%

If attributed to the shock instability, the detected optical QPOs provide independent constraints that can be combined with the ones derived from upper limits in X-rays. Unfortunately, because of the low level of the sources during the XMM observations, the limits derived for BL Hyi ( 71.1\%) and EF Eri (58.1\%) are not very constraining. We discuss here  the  three other sources V834 Cen, AN UMa, and VV Pup in more detail. 

\subsubsection{V834 Cen}
V834 Cen is an interesting source since its characteristics are determined reasonably
well with B = 23 MG, a distance of d= $(116 \pm 8)$ pc,  and a mass of M= $0.66 \pm 0.05$ M$_{\odot}$. Previous searches for X-ray QPOs from RXTE data provided an upper limit of $\sim14\%$ at a 90\% confidence level in the range 0.2-1.2 Hz
 \citepads{2000PASP..112...18I}. %Imamura00
Similar searches in GINGA data also yielded an upper limit of 18\% at a 95\% statistical level in the range (0.4-0.8 Hz)
 \citepads{1997MNRAS.286...77B}, %Beardmore97
using the same statistical analysis as in this paper. 
Earlier simultaneous X-ray and optical observations have shown a possible hint of X-ray pulsations at twice the frequency as observed in simultaneous optical data but could not be confirmed
 \citepads{1985ESASP.236..155B}. %Bonnet-Bidaud85
 \\  
From the long XMM observations (43 ksec), the upper limit for persistent QPOs  is significantly better here (9\%). Assuming L$\rm _x$= $1.5 \times 10^{32}$ erg.s$^{-1}$
 \citepads{2004MNRAS.347..497R} %Ramsay04
gives $\dot{M}_{16}$ = 0.14, and for a typical A=$10^{14}\, \rm cm^{2}$, the predicted amplitude from the simulations is $\sim$27\%. Therefore QPOs could have been detected.  Our upper limit in X-rays at a  lower level of 9\% implies that the cross-section has to be higher than  $2.2\, 10^{14}\, \rm cm^{2}$.\\
Considering the optical range, QPOs are detected at a level of $\sim2\%$ 
\citepads{1985A&A...145L...1L}, %Larsson 85
which without any dilution, will be in accordance with the model prediction at A $\sim 2.5\, 10^{14}\, \rm cm^{2}$ (corresponding to $\dot{m} \sim 5.6$). Both conditions in X-ray and optical therefore agree for this source. The presence of optical QPOs and their absence in X-rays impose a relatively small  column cross-section A $\sim 2.5\, 10^{14}\, \rm cm^{2}$. We note that if the observed optical flux is not only due to cyclotron but is also combined with any additional source, then this dilution will impose an even smaller cross-section A (and higher $\dot{m}$). For instance, the minimum value A $\sim 2.2\, 10^{14}\, \rm cm^{2}$ imposed by the X-ray limit would produced a higher cyclotron amplitude of $\sim4\%$ that can still be in accordance with the 2\% observed amplitude if the dilution reaches 50\%.\\

\subsubsection{VV Pup}
VV Pup is a close-by source at $(146 \pm 5)$ pc  with a WD mass determined at M= $0.73 \pm 0.18$ M$_{\odot}$. The source's magnetic field is more complex than a simple dipole, and different values of the magnetic field have been derived from the two poles at 31 and 54 MG, respectively, though the dominant pole for cyclotron appears to be at 31 MG
\citepads{2006AJ....131.2216H}. %Howell06
The bolometric X-ray luminosity L$\rm _x$= $5.3 \times 10^{32}$ erg.s$^{-1}$
 \citepads{2004MNRAS.347..497R} %Ramsay04
implies a total accretion rate  $\dot{M}_{16}$ = 0.41. To our knowledge, no limits have been published yet on X-ray QPOs. Since the source is very soft, despite a significant accretion rate in the 2007 high state, the source counting rate in the XMM (0.5-10 keV) range is rather low, and the upper limit for QPOs is only 31\%.
When compared to numerical simulations with a typical A=$10^{14}\, \rm cm^{2}$, an amplitude of $\sim 24\%$ would be expected for the above source parameters, therefore below our detection limit.\\
On the other hand, the optical QPOs detected for this source at a level of $(1-1.5\%)$ 
\citepads{1989A&A...217..146L} %Larsson 89
impose a cross-section of  A$\sim 2.1\,10^{14}\, \rm cm^{2}$  from our simulations, insuring a significant specific accretion rate of $\dot{m} \sim 19$ g.cm$^{-2}$.s$^{-1}$. With these values, X-ray QPOs will be expected but with a low amplitude of only $ \sim 3.3\%$, much lower than our upper limit. Owing to the high magnetic field, the QPOs therefore appear significantly damped.

\subsubsection{AN UMa}
AN  UMa has a reasonably well-determined magnetic field in the range B = (29-36) MG from cyclotron study
\citepads{1989MNRAS.236P..29C}. %Cropper89
Its mass determination is, however, still uncertain with a commonly quoted value of 1.0 M$_{\odot}$, though other values as low as (0.4-0.6) M$_{\odot}$ have also been reported
\citepads{1996A&A...306..199B}. %BB96
Its distance is also not accurately known with only a lower limit of d $\ge 120$ pc. For this distance, the observed X-ray luminosity will correspond to a total accretion rate of $\dot{M}_{16}$ $\ge$ 0.021. With this accretion rate and assuming a typical cross section of A=$10^{14}\, \rm cm^{2}$ and a WD mass of 1M$_{\odot}$, the numerical models yield no significant X-ray QPOs, i.e. at a level less than 1\%. \\
The optical QPOs detected at a level of  (1-4)\% 
\citepads{1996A&A...306..199B} %BB96
impose independent constraints. From our numerical simulations and with the source parameters, M=1\,M$_{\odot}$ and $\dot{M}_{16}$ = 0.021, optical QPOs at this level will require an unrealistically very small cross-section with a value at A $\sim 1.9\,10^{12}\, \rm cm^{2}$. Only a significant lowering of the WD mass would increase the cross section while keeping the optical QPOs at the same level. However, even for the minimum mass of 0.4M$_{\odot}$, the requested cross section will  still be rather low A$\sim 0.5\,10^{14}\, \rm cm^{2}$, corresponding to a very small fraction of the WD surface f = $3.4\,10^{-6}$. To increase the cross section further, while keeping the QPOs level, can only then be achieved by increasing the total accretion rate and therefore by assuming a larger distance for the source. We computed that to achieve a more realistic minimum value of A = $10^{14}\, \rm cm^{2}$, the source will have to be at a minimum distance of d $\ge 155$ pc if the mass is kept at the minimum value of 0.4M$_{\odot}$ (respectively $\ge 255$ pc for a more median value of 0.6M$_{\odot}$).
Once again, these values do not take a possible dilution into account that will decrease the observed amplitudes with respect to the true cyclotron relative amplitudes. Doubling the amplitudes will, for instance, slightly increase the minimum distance  d~$\ge 164$ pc for a  0.4M$_{\odot}$ WD. 
\\
\subsection{The column parameters}
The upper limits derived for the amplitude of X-ray QPOs, together with the predicted cyclotron QPO amplitudes for sources showing optical oscillations, are a powerful tool for constraining the source parameters. The presence and the amplitudes of QPOs are mainly governed by the value of the cooling parameter, $\rm \epsilon_{s}$,  which is a measure of the cyclotron cooling efficiency with respect to bremsstrahlung. The higher the cooling parameter, the lower the QPO amplitudes because the cyclotron cooling will increase and damp the plasma thermal instabilities efficiently. 
As shown by Eq.~\ref{equation6}, the cooling parameter only depends on four source parameters: $\rm \dot{M}$ the total accretion rate, A the column cross-section (assumed here cylindrical), B the surface magnetic field strength, and M the WD mass.
Since $\rm \dot{M}$ has to be derived from the observed source luminosity, the distance to the source is an additional parameter. \\
Numerical simulations performed for a grid of parameters demonstrate that the QPO amplitudes are a monotonic function of  $\rm \epsilon_{s}$ so can be used to derive the relevant source parameters.
In most sources, the three parameters (B, M, and $\rm \dot{M}$) can be evaluated from independent observations, but the column cross-section is very poorly constrained, and only an approximate evaluation can be derived in the case of  column eclipses (see below). QPOs are therefore a powerful tool to efficiently constrain this critical parameter. \\
From our X-ray survey, the QPO upper limits  provide only a lower limit on the column cross section, leading to values of the order of  $\gtrsim1.3\,10^{14}\,\rm cm^{2}$ for V2301 Oph and $\gtrsim4\,10^{14}\,\rm cm^{2}$ for AM Her for the lower field sources, in accordance with what is expected for typical accretion columns. Contrary to what can be naively expected, even for a relatively low field, significant X-ray QPOs will therefore not be present unless the accretion flow is highly concentrated inside a very narrow column.\\
In the same way, detected optical QPOs and measured amplitudes this
time provide a direct measure of the column size. The two sources V834 Cen and VV Pup are found to be consistent with a column cross section of  $\sim2.5\,10^{14}\,\rm cm^{2}$ and $\sim2.1\,10^{14}\,\rm cm^{2}$, respectively, when not taking a possible dilution effect
into account.\\
For the source AN UMa, according to the simulations, no cyclotron QPO should be produced for the commonly adopted source parameters. Keeping the column section at a minimum value $10^{14}\,\rm cm^{2}$, the presence of the optical QPOs implies jointly a low mass WD (M<0.6M$_{\odot}$) and a higher accretion rate that will place the source at a significantly higher distance. The anomaly of a "high mass AN UMa" showing QPOs is already visible in Fig.~\ref {B-Mdot} where the source location is well below the bremsstrahlung regime, as discussed above.\\
The dimensions of the column are hard to constrain by direct observation. One of the best evaluations is provided by 
 \citetads{2006MNRAS.372..151O} %O'Donoghue06
 for the eclipsing polar FL Cet by way of optical high time resolution modeling of the ($\sim1-2$ s) ingress/egress of the white dwarf eclipse by the secondary.  Two close, diametrically opposed hot spots were located with typical projected dimensions  of $\sim (10-12)\, 10^{16}\, \rm cm^{2}$ and $\sim (5-6.5)\, 10^{16}\, \rm cm^{2}$ where the range allows for the uncertainty in the WD mass (0.5-0.7 M$_{\odot}$). 
 A $\sim50\%$ size variability was also observed from cycle to cycle. The mapping of the emission regions on the WD surface has also been attempted by modelling the soft X-ray and optical light curves of the polar ST LMi
 \citepads[see][]{1994MNRAS.267..481C}. %Cropper-Horne94
The resulting fraction area is at least $\sim 0.1\%$ and $\sim 0.6\%$,  respectively, for the X-ray and optical regions, corresponding to typical surfaces of 0.7 and 4 $10^{16}\, \rm cm^{2}$ for a WD of mass  0.76M$_{\odot}$ in ST LMi.
We stress, however, that the hot spots traced by the soft X-ray and optical light curves or the WD eclipses will correspond not only to the column cyclotron emission but also partly to the X-ray irradiated fraction of the WD surface, whose dimensions are likely to be much greater that the column cross-section 
\citepads[see][]{1995ASPC...85...21K}.  %King95
In fact, the typical surfaces of the soft X-ray component of blackbody type in polars are commonly evaluated in the range $\sim 10^{16}\, \rm cm^{2}$. The change in location of the spots on the WD surface due to variable capture regions can also contribute to the increase in the apparent emitting zone. The possibility also exists of a more complex column structure with a narrow dense core surrounded by a more diffuse cyclotron halo 
\citepads{1992MNRAS.256...80A}. %Achilleos
 In this case, more detailed 2D numerical simulations will have to be considered that are beyond the scope of this paper. 
The significant difference, of more than an order of magnitude, between the size of the apparent emission region and the typical dimension of the accretion column implied by the optical QPOs may, however, be a concern for interpreting these QPOs as a result of the thermal instability of the column.

\subsection{QPOs at high frequency}
One of the main results of the numerical simulations is the variable frequencies of the oscillations that depend on the specific accretion rate and the magnetic field strength (see Paper II). For a typical 0.8M$_{\odot}$ WD, the main QPO frequencies shift from $\sim$ 0.07 to 7 Hz with $\dot{m}$ increasing from 0.1 to 10 g.cm$^{-2}$.s$^{-1}$.  Other modes of oscillations are also shown to be activated with several overtones of even higher frequencies present up to 25 Hz (see Paper II). In our study, the model amplitude was computed using the quadratic sums of all frequencies, therefore providing an upper limit for the amplitude at any given frequency in the covered range. Owing to the data available, only five sources in our sample could be searched for frequencies higher that 5 Hz, so the possibility still exists that higher frequency QPOs can be present in the rest of our sample.\\
The oscillation frequency is an additional parameter predicted by the models, and it is mainly constrained by the specific accretion rate (and the associated column cross-section) and the WD mass. We note that for the sources showing optical QPOs, the range of the source magnetic fields  and the detected amplitudes require that their typical column sections should be in the range  A $\sim (2-2.5)\,10^{14}\, \rm cm^{2}$ with a corresponding specific accretion rate $\dot{m} \sim (6-20)$ g.cm$^{-2}$.s$^{-1}$. Therefore, for a typical 0.6M$_{\odot}$ WD, QPO frequencies in X-rays and optical would be expected in a typical range of (6-20 Hz)  (see Formula 21 in Paper II). This range is significantly excluded at least for V834 Cen  in our X-ray survey, and no high frequency (> 1 Hz) optical QPOs have been detected yet. From the shock oscillation model, the measured optical frequencies at $\sim(0.3-1)$Hz should instead indicate a much lower specific accretion rate but one that in this case leads to negligible optical QPO amplitudes. Both the amplitudes and the frequencies therefore do not appear to consistently be reproduced by the models. \\
\subsection{Alternatives to shock oscillations}
Because no X-ray QPOs have been detected yet that will definitively relate the QPO phenomena to the thermal instability process,  the origin of the optical QPOs can still be questioned further. Alternative explanations have already been proposed in terms of magnetospheric oscillations by Alfven waves to explain typical 20-60\,s pulsations
\citepads{1981ApJ...245..183T} %Tuohy 81
or predicted 0.1-10\,s excited magneto-acoustic waves in the white dwarf
\citepads{1995MNRAS.275L..11L}. %Lou 95
One other possibility that may have been overlooked so far is the radial pulsations of the white dwarf itself. The expected typical timescale of the WD radial oscillations is $\tau \sim R/v_{ff} \sim [2R^{3}/GM]^{1/2}$, corresponding to a range of periods of $\sim (4-18)$\,s for a range of masses (1.2-0.4) M$_{\odot}$. In fact, this process was the first one proposed to explain the early discovered pulsars with periods of a few seconds.  White dwarfs are expected to show p-mode radial pulsations at these fundamental frequencies and also with numerous high overtones down to the period range (0.1-1)\,s
(\citeads{1983ApJ...265..982S}, %Saio83
\citeads{2008ARA&A..46..157W}). %Winget08
Though a wide variety of non-radial g-modes oscillations have already been discovered in DA, DB,  and DO white dwarfs and more recently in accreting WDs (see 
\citeads{2008ARA&A..46..157W}, %Winget08
\citeads{2006ApJ...643L.119A}, %Arras06
\citeads{2007ApJ...667..433M}), %Mukadam07
none of the predicted radial pulsations have been discovered yet, possibly due to a somewhat still limited survey (see 
\citeads{2011A&A...525A..64S}, %Silvotti11
\citeads{2014MNRAS.437.1836K}). %Kilkenny14
Acoustic (p-mode) oscillations are usually expected at rather low amplitudes (<1\%), but higher amplitudes of a few percentage points cannot be excluded
\citepads{1983ApJ...265..982S}. %Saio83
There might still therefore be a possibility that the optical QPOs observed amongst polars originate in the white dwarf itself.\\

It is worth noticing that the negative X-ray QPO search reported in this paper is somewhat similar to the situation found amongst classical T Tauri stars. For these systems, shock wave oscillation models have also been invoked to explain their soft X-ray luminosities, and stability studies from hydrodynamic simulations have shown the expected presence of quasi-periodic instabilities with a wide range of possible frequencies
(\citeads{2008MNRAS.388..357K}, %Koldoba08
\citeads{2008A&A...491L..17S}). %Sacco08
Present searches for these QPOs  in X-ray and UV-optical  bands have been negative up to now
(\citeads{2009ApJ...703.1224D}, %Drake09
\citeads{2010A&A...518A..54G}). %Gunther10
 We note, however, that the typical QPO timescales in this case cover a much wider interval, ranging from 0.02 - 0.2 s 
 \citepads{2008MNRAS.388..357K} %Koldoba08
 to 10 min 
 \citepads{2008A&A...491L..17S}, %Sacco08
 as a consequence of different cooling and density assumptions. This
therefore makes the search more complex. With only two systems being examined so far, observational constraints are still much more limited than for polars.  Owing to a different balance between the thermal and magnetic pressure, the accretion structure is also expected to be quite different (see Paper II for further discussion).
%______________________________________________________________

\section{Conclusions}
The most complete search for X-ray QPOs among polars presented here did not succeed in a positive detection.  Despite the  largest ever sample studied systematically, no (0.5-10 keV) X-ray QPOs were detected with upper limits ranging from $\sim$7 to 70\%. From detailed 1D numerical simulation, thermal instabilities in the accretion column are expected to produce  quasi-oscillations with amplitude reaching 
$\sim40$\% in the bremsstrahlung (0.5-10 keV) X-ray regime and  $\sim20$\% in the optical cyclotron regime. These amplitudes steeply decrease with increasing magnetic field value,  but this decrease can be potentially balanced if the source reaches a higher specific accretion rate owing to a column with a smaller cross-section. However, in our sample, even for low B (<20MG) sources, the specific accretion rate is apparently not sufficient to maintain the predicted instabilities.\\
The question that arises is therefore "why so few polars show QPOs" and "why instabilities are suppressed so efficiently" despite the wide range of accretion and WD parameters covered in our sample. One explanation may be that the observation limitations mean that the searched frequencies are only at the low end of possible values, and therefore higher frequencies QPOs remain undetectable. The QPOs may also be highly transitory. Although the numerical simulations show a stabilized regime after the onset of the oscillations, the presence and the influence of a secondary shock, as clearly demonstrated by our simulations, may play a role in the long term by destroying the instability process and therefore reducing the efficient time during which QPOs are present. The conditions for QPOs may also appear difficult to achieve if a lower B field strength is conjugated with a wider accretion  column resulting in a poor capture and funnelling efficiency.\\
More problematic is the apparent inconsistency of the numerical predictions with the observed characteristics of the optical QPOs when the  source parameters are reasonably well known as for V834 Cen and VV Pup. The X-ray upper limits and the optical amplitudes can be reproduced by the thermal instability model only in the case of a high specific accretion rate of $\dot{m} \sim (6-20)$ g.cm$^{-2}$.s$^{-1}$ through a narrow accretion column. However, the QPO frequency will then be expected at much higher frequencies than observed.\\

The ultimate check for the nature of the optical QPOs relies in the measure of the polarisation that will definitely sign the cyclotron nature. However, since the overall polarization is usually quite low ($\lesssim 10\%$), the fractional light in polarized QPOs will be $\sim 0.1-0.3\%$, requiring large telescopes. Further searches for X-ray QPOs will also require greater sensitivity and will have to await the next generation of X-ray satellites such as the future ESA Athena project. Laser laboratory experiments planned in the near future with high power lasers will offer other interesting prospects for investigating the influence of the different physical parameters on the development of shocks in accretion columns in detail.
\\

\begin{acknowledgements}
We wish to thank an anonymous referee for fruitful comments and suggestions for improving the paper.
\end{acknowledgements}

%\listofobjects

%
% if biblio files
\bibliographystyle{aa}         %% aa.bst = A&A bibliography style
\bibliography{xmmqpo}    %% example.bib = Bibtex entries from ADS

\end{document}